\documentclass{aa}  
\usepackage{graphicx}
\usepackage{txfonts}
\usepackage{natbib}
\bibpunct{(}{)}{;}{a}{}{,}

\usepackage{multirow}

\usepackage{lipsum}

\usepackage{xcolor}

\usepackage{longtable}

\begin{document}

   \title{Organic chemistry in the protosolar analogue HOPS-108: Environment matters}

   \author{L. Chahine
          \inst{1,2}  
          \and
          A. L\'{o}pez-Sepulcre\inst{1,3} \and
          R. Neri\inst{1} \and
          C. Ceccarelli \inst{3} \and
          S. Mercimek \inst{4,5} \and
          C. Codella \inst{4,3} \and
          M. Bouvier \inst{3} \and
          E. Bianchi \inst{3} \and
          C. Favre \inst{3} \and
          L. Podio \inst{4} \and
          F. O. Alves \inst{6} \and
          N. Sakai \inst{7} \and
          S. Yamamoto \inst{8}
          }

   \institute{Institut de Radioastronomie Millim\'{e}trique (IRAM), 300 rue de la Piscine, 38406 Saint-Martin-d'Hères, France \\ e-mail: chahine@iram.fr 
         \and
              \'{E}cole doctorale de Physique, Universit\'{e} Grenoble Alpes, 110 Rue de la Chimie, 38400 Saint-Martin-d'Hères, France 
          \and 
              Universit\'{e} Grenoble Alpes, CNRS, IPAG, 38000 Grenoble, France 
            \and INAF, Osservatorio Astrofisico di Arcetri, Largo E. Fermi 5, I-50125, Firenze, Italy 
            \and
            Universit\`{a} degli Studi di Firenze, Dipartimento di Fisica e Astronomia, Via G. Sansone 1, 50019 Sesto Fiorentino, Italy 
            \and
            Center for Astrochemical Studies, Max Planck Institute for Extraterrestrial Physics, Garching, 85748, Germany
            \and
            The Institute of Physical and Chemical Research (RIKEN), Saitama 351-0198, Japan
            \and
            Department of Physics, The University of Tokyo, Bunkyo-ku, Tokyo 113-0033, Japan }

   \date{Received 16 July 2021; Accepted 13 October 2021}

 
  \abstract 
   {Hot corinos are compact regions around solar-mass protostellar objects that are very rich in interstellar Complex Organic Molecules (iCOMs). How the abundance of these molecules is affected by the environmental physical conditions is still an open question. More specifically, addressing this point is key to understand our own chemical origins since the Solar System formed in a large cluster of low- to high-mass stars and was therefore subject to external heating and ultraviolet irradiation which may have shaped the chemistry of its early formation stages.}
   {The goal of this high resolution study is to determine the abundance ratios of iCOMs in HOPS-108, which is a Class 0 protostar and a hot corino candidate located in the nearest Solar System analogue, the protostellar cluster OMC-2\,FIR\,4, in Orion. We aim to compare the abundance ratios to those found in other hot corinos, which  are all located in less crowded environments, in order to understand the impact of environmental conditions hot corinos' chemistry.}
   {We observed the OMC-2\,FIR\,4 proto-cluster using the Band 6 of the Atacama Large (sub-)Millimetre Array (ALMA) in Cycle 4 with an angular resolution of $\sim$0.28$"$ (110 au). We determined the abundances and temperature of the species using local thermodynamic equilibrium (LTE) and non-LTE analysis. }
   { Our results reveal a rich organic chemistry towards HOPS-108, asserting that it is a hot corino where the following iCOMs are detected: CH$_{\rm 3}$OH, HCOOCH$_{\rm 3}$, CH$_{\rm 3}$OCH$_{\rm 3}$, CH$_{\rm 3} ^{\rm 18}$OH, CH$_{\rm 2}$DOH, CH$_{\rm 3}$COCH$_{\rm 3}$, CH$_{\rm 3}$CHO, CH$_{\rm 3}$CN, $^{\rm 13}$CH$_{\rm 3}$CN, C$_{\rm 2}$H$_{\rm 5}$CN, and NH$_{\rm 2}$CHO. Remarkably, we find a possible enhancement in the HCOOCH$_{3}$ abundance with respect to other known hot corinos. Indeed, the [CH$_{3}$OCH$_{3}$]/[HCOOCH$_{3}$] abundance ratio in this source is $\sim$0.2 and, within the uncertainties, it deviates from the known correlation marginally where [CH$_{3}$OCH$_{3}$]/[HCOOCH$_{3}$]$\sim$1. A relatively low [CH$_{2}$DOH]/[CH$_{3}$OH] abundance ratio of $\sim$0.02 is also obtained, which is in agreement with that found in another Orion source, HH212, suggesting a higher gas temperature during the early phases of ice mantle formation.}
   {The [CH$_{3}$OCH$_{3}$]/[HCOOCH$_{3}$] and [CH$_{2}$DOH]/[CH$_{3}$OH] abundance ratios in HOPS-108 might result from different physical conditions in the Orion molecular complex compared to other regions. The former ratio cannot be reproduced with current chemical models, highlighting the importance of improving the chemical networks with theoretical calculations. More hot corinos located in heavily clustered regions such as Orion should be targeted in order to measure these ratios and evaluate whether they are an environmental product or whether HOPS-108 is an exceptional hot corino overall.}


   \maketitle
%

\section{Introduction}
\label{intro}

The first stages of solar-mass star formation are known for their molecular richness. Indeed, at the centre of the protostellar envelopes, some protostars may host compact ($\leq$ 100 au), hot ($\geq$ 100 K) and dense ($\geq$ 10$^{7}$ cm$^{-3}$) regions called hot corinos \citep{Ceccarelli2004,Ceccarelli2007,Caselli2012}. These are very rich in interstellar Complex Organic Molecules (iCOMs), that is, C-bearing molecules containing at least six atoms \citep{Herbst2009,Ceccarelli2017,Jorgensen2020}. Their large molecular complexity is probably inherited and reprocessed in the subsequent stages throughout the formation of a Sun-like star \citep{Caselli2012}. Hence, understanding the chemistry of hot corinos is a crucial key to shed light on the conditions that favoured the emergence of life on our planet.
\newline \indent
So far, most of the studied hot corinos are either isolated or were born in loose protoclusters. In contrast, our Sun was born in a crowded cluster in the vicinity of massive stars rather than in isolation \citep{Adams2010,Pfalzner2015}. It is also known that the Sun was subject to internal irradiation from energetic particles (>10 MeV), whose imprint is seen in the products of short-lived radionuclides in meteoritic material today \citep{Gounelle2013}. However, it is not yet understood whether these conditions affected the early chemistry of the proto-Sun and its surroundings. From this standpoint, delving into the wide diversity of iCOMs in Solar-System-like environments is therefore fundamental to decipher the secret of our own origins. One of the best targets for this aim is OMC-2\,FIR\,4, a protostellar cluster located in the Orion Molecular Cloud 2, at a distance of 393$\pm$25 pc \citep{Grossschedl2018}. It has a total mass of 30 $M_\odot$ approximately \citep{Mezger1990, Crimier2009} and a bolometric luminosity $\leq$ 1000 $L_\odot$ \citep{Crimier2009, Furlan2014}. Due to various properties, it is thought to be the closest analogue to the environment in which the Sun may have been born. To begin with, OMC-2\,FIR\,4 is a young protocluster in which both low- and intermediate-mass protostars are forming \citep{Shimajiri2008, Lopez-Sepulcre-and-taquet2013, Kainulainen2017, Tobin2019}. It is located near the Trapezium OB star cluster, which bathes it in far ultraviolet (FUV) radiation ($\sim$ 1500 $G_{0}$ where $G_{0}$ is the FUV radiation field in Habing units; \citep{Loepz-Sepulcre-and-kama2013}). 
Finally, a stream of high-energy cosmic-ray-like particles pervades the protocluster, ionising the surrounding envelope at a pace of 4000 times higher than cosmic rays (CRs) in our Galaxy \citep{Ceccarelli2014, Fontani2017, Favre2018, Ceccarelli2019}. On that premise, OMC-2\,FIR\,4 is the perfect analogue in pursuing a chemistry similar to that of our Solar System's early days. Particularly, the hot corino candidate HOPS-108 \citep{Tobin2019} located within this protocluster can provide us with important hints about the chemistry of the gas and dust envelope surrounding our proto-Sun.
\newline \indent 
In this work, we present the first dedicated chemical study on the iCOMs of HOPS-108. This was performed using high-resolution (110 au) Atacama Large (sub-)Millimetre Array (ALMA) data at 1.2 mm.
\newline \indent
The article is structured as follows: In Sect. \ref{Obs_and_lines} we describe the observations.\ In Sect. \ref{Results} we present the continuum and molecular line maps, together with the line identification and the main results of the analysis. In Sect. \ref{Discussions} we discuss the results, and, finally, in Sect. \ref{Conclusions} we summarise the conclusions.

\section{Observations}

\label{Obs_and_lines}

OMC-2\,FIR\,4 was observed with ALMA during its Cycle 4 operations, between 25 October 2016 and 5 May 2017, as part of the project 2016.1.00376.S (PI: Ana López-Sepulcre). The observations were taken with Band 6 in the ranges 218-234 GHz and 243-262 GHz. The phase-tracking centre was $\alpha_{ICRS} = 05^{h}35^{m}26^{s}.97$, $\delta_{ICRS} = -05^{\circ}09'54''.50$ and the systemic velocity was set to $V_{LSR}$ = 11.4 km s$^{-1}$. For the lower-frequency, 45 antennas of 12 m were used in the C-5 configuration of the array, probing angular scales from 0$\farcs$19 to 11$\farcs$2. The integration time on source was $\sim$9 min, and the primary beam size is $\sim$27$\farcs$1. For the higher-frequency spectral configuration, 42 antennas of 12 m were used in the C-5 configuration of the array, probing angular scales from 0$\farcs$21 to 8$\farcs$2. The integration time on source was $\sim$19 min and the primary beam size is $\sim$25$\farcs$6. For both spectral configurations, J0510+1800 and J0522-3627 were used for bandpass and flux calibration, and J0607-0834 and J0501-0159 were used for phase calibration. The absolute flux calibration uncertainty is estimated to be $<$10\%. In the present work, focussed on iCOMs, we have used the five spectral windows (spws) listed in Table \ref{spw}. 

The data calibration was performed using the standard ALMA calibration pipeline with the Common Astronomy Software Applications package (CASA\footnote{\url{https://casa.nrao.edu/}}, \citealp{McMullin2007}), while self-calibration, imaging, and data analysis were performed using the IRAM-GILDAS software package\footnote{http://www.iram.fr/IRAMFR/GILDAS/}. The continuum images were produced by averaging line-free channels from the 1.875 GHz spectral windows at 232 GHz and 246 GHz in the visibility plane. The remaining effective bandwidths are 458.9 MHz and 218.8 MHz, respectively. Phase self-calibration was then performed on the continuum emission and the gain solutions were applied to the line cubes. Continuum subtraction was performed on the cubes in the visibility plane, before line imaging. The resulting synthesised clean beam and channel root-mean-square (rms) noise level for each spectral window are summarised in Table \ref{spw}.

{\footnotesize\setlength{\tabcolsep}{3pt}
\begin{table*}[t]
  \centering
  \caption{Analysed ALMA spectral windows}
  \renewcommand{\arraystretch}{1.2}
  \begin{tabular}{l c c c c c c c}
  
    \hline
        \hline
         
         $\mathrm{Spectral \,window \,label}$ & 
         $\mathrm{Central \, Frequency} $ &
         $\mathrm{Bandwidth}$ &
        $\mathrm{Velocity \,  Resolution} $ &
         $\mathrm{Beam}$ &
         $\mathrm{Beam \, P.A.} $  &
         $\mathrm{Chan. \, rms} $  &
         \\  
               &
         (GHz) &
         (MHz) &
          (km s$\rm ^{-1}$) & ($"$)&
        ($\rm ^{\circ}$)& (mJy beam$\rm ^{-1}$) \\         \hline  
  
Continuum high frequency & 246.2 & 1875 & 1.2 & 0.32 $\times$ 0.27 & 107 & 1.3 \\  
Continuum low frequency & 232.2 & 1875 & 1.3 & 0.51 $\times$ 0.27 & 109 & 2.0 \\ 
NH$_{2}$CHO 12$_{1,12}$ $-$ 11$_{1,11}$ & 243.5 & 58.59 & 0.15 & 0.32 $\times$ 0.28 & 106 & 4.5 \\ 
CH$_{3}$CN 14$_{0}$ $-$ 13$_{0}$ & 257.5 & 58.59 & 0.14 & 0.31 $\times$ 0.26 & 109 & 4.9 \\
HCOOCH$_{3}$ 24$_{1,24}$ $-$ 23$_{1,23}$ & 259.3 & 58.59 & 0.14 & 0.31 $\times$ 0.26 & 106 & 3.7 \\
  \hline
  \end{tabular}
  \label{spw}
\end{table*}}


\section{Analysis and results}
\label{Results}

\subsection{Continuum emission}
\label{continuum}

\begin{figure}
    \centering
    \includegraphics[width=0.48\textwidth]{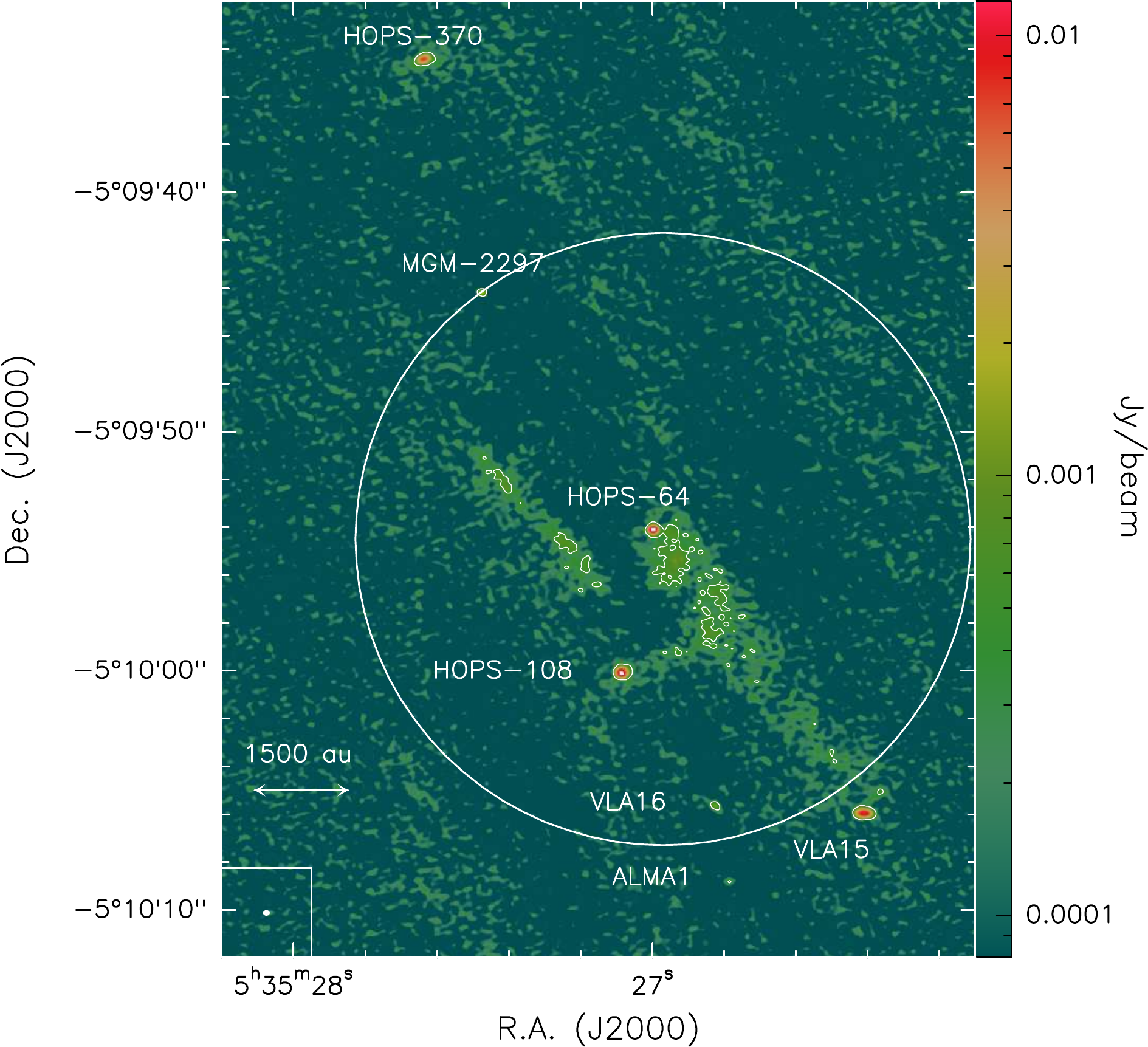}
    \caption{Dust continuum emission map of OMC-2\,FIR\,4 at 1.2 mm. Contours at 5$\sigma$ are shown with $\sigma$ = 95.7 $\mu$Jy beam$^{-1}$. The different cores are labelled. The synthesised beam is depicted in white in the lower left corner. The primary beam at full width half maximum (25$\farcs$6) is depicted with a white circle.}
    \label{cont-im}
\end{figure}

Figure \ref{cont-im} shows the map of continuum emission at 246.2 GHz (1.2 mm). This map is not corrected for the primary beam attenuation, but all the flux densities used in the analysis have been corrected accordingly. We detected seven compact cores that are labelled as follows: the sources names HOPS-370, HOPS-64, and HOPS-108 refer to the sources identified in the \textit{Herschel Orion Protostar Survey} (HOPS, \citealp{Furlan2016}); VLA16 and VLA15 refer to the sources identified in \cite{Osorio2017}; MGM-2297 refers to \cite{Megeath2012}; and ALMA1 refers to the source recently identified in \cite{Tobin2019}. In addition, we detected some extended emission tracing the envelope of the protocluster. The source HOPS-370 is associated with the OMC-2\,FIR\,3 region, MGM-2297 is probably a foreground source and has no direct association with OMC-2\,FIR\,3 \citep{Tobin2019}, while the other five cores belong to the OMC-2\,FIR\,4 region. As we report in the next section, among the five cores in OMC-2\,FIR\,4, only HOPS-108 displays rich iCOM emission at these wavelengths, making it the only clear hot corino present in the protocluster. In this context, our work is only focussed on this source. As shown in the map, the emission of HOPS-108 is compact at the resolution of our observations. We fitted it with a circular Gaussian function in the uv plane, and we obtained coordinates of RA (J2000) $ = 05^{h}35^{m}27^{s}.084$ and Dec (J2000) $= -05^{\circ}10'00''.07$ consistent with those derived by \cite{Tobin2019}, a source size of 0$\farcs$14 ($\sim$ 55 au), and a flux density of  (20.0 $\pm$ 0.1) mJy beam$^{-1}$.
 
\subsection{Line identification and molecular maps}

\begin{figure*}
    \centering
    \includegraphics[width=\textwidth]{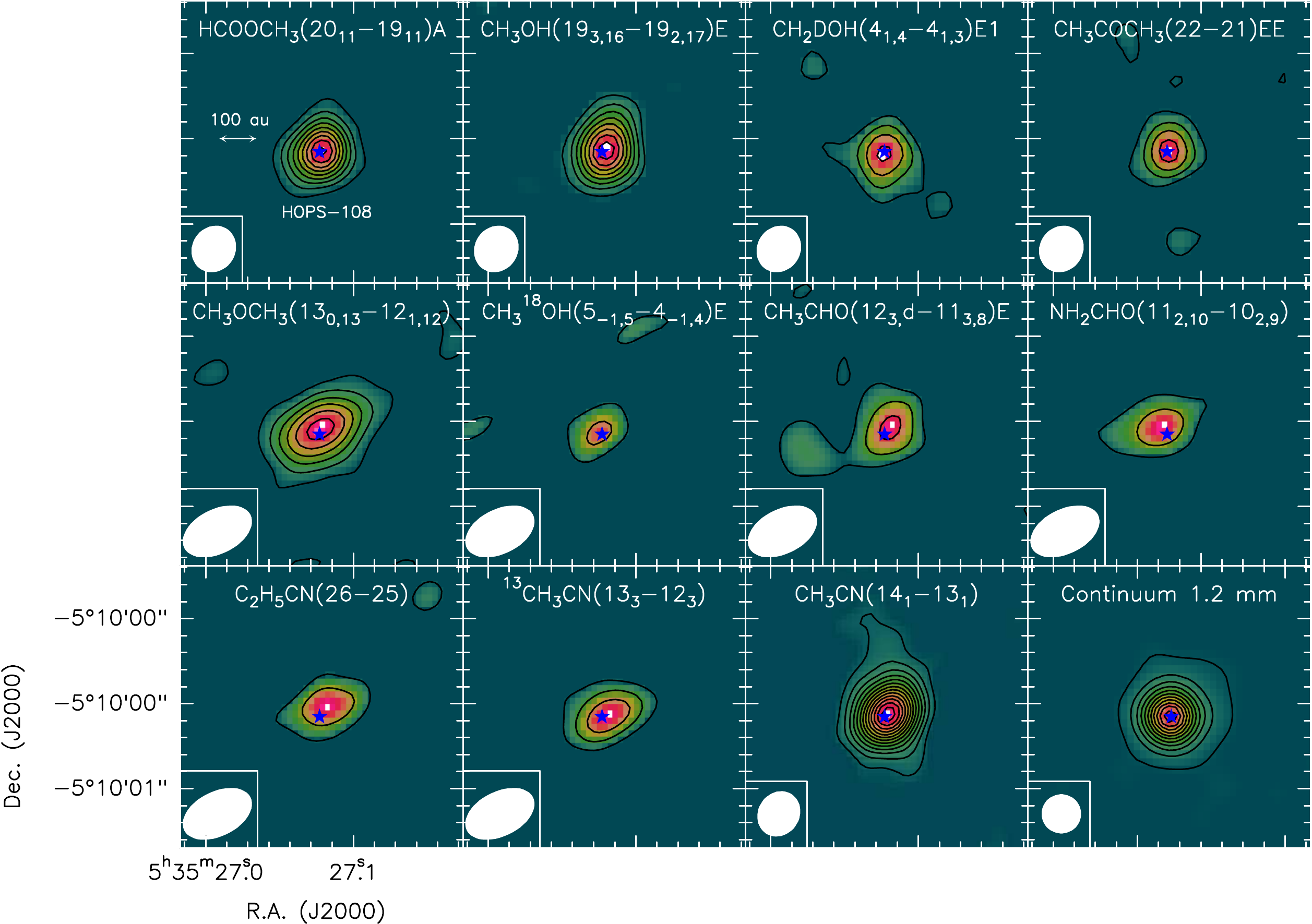}
    \caption{Moment 0 maps of each detected molecule and of continuum emission towards HOPS-108. For CH$_{3}$OH and CH$_{3}$CN, contours start at 5$\sigma$ and increase by stpdf of 5$\sigma$ with $\sigma$ = 0.6 mJy km s$^{-1}$ beam$^{-1}$. For the other molecules, contours start at 3$\sigma$ and increase by stpdf of 3$\sigma$ with $\sigma$ = 0.6-1.1 mJy km s$^{-1}$ beam$^{-1}$. For the continuum emission, contours start at 5$\sigma$ and increase by stpdf of 15$\sigma$ with $\sigma$ = 95.7 $\mu$Jy km s$^{-1}$ beam$^{-1}$. The synthesised beam is depicted in white in the lower left corner. The position of the continuum emission peak is depicted with a blue star. }
    \label{mom-0}
\end{figure*}

The line identification and analysis was performed using the extension WEEDS in the CLASS software of the GILDAS package \citep{Maret2011} using the Jet Propulsion Laboratory (JPL\footnote{\url{https://spec.jpl.nasa.gov}}, \citealt{Pickett1998}) and the Cologne Database for Molecular Spectroscopy (CDMS\footnote{\url{https://cdms.astro.uni-koeln.de}}, \citealt{Muller2001,Muller2005}) databases. The spectra were extracted from the central pixel corresponding to the position of the continuum emission peak. The line detection limit is set to 3$\sigma$ at the intensity peak. 
\newline \indent In terms of the line number, our spectra are dominated by HCOOCH$_{\rm 3}$, with more than 80 lines detected. We also detect between two and nine lines of CH$_{\rm 3}$OH, CH$_{\rm 3}$OCH$_{\rm 3}$, CH$_{\rm 3} ^{\rm 18}$OH, CH$_{\rm 2}$DOH, CH$_{\rm 3}$COCH$_{\rm 3}$, CH$_{\rm 3}$CHO, CH$_{\rm 3}$CN, $^{\rm 13}$CH$_{\rm 3}$CN, C$_{\rm 2}$H$_{\rm 5}$CN, and NH$_{\rm 2}$CHO. The detected transitions span over a large upper energy range (E$_{\rm up}$=49-537 K). The molecules are listed in Table \ref{molecules}. The line profiles were fitted with a Gaussian function with CLASS. The derived line parameters are summarised in Appendix \ref{appA}, and a sample of the observed lines for each detected molecule are shown in Figs. \ref{mf-spectra}-\ref{meth-mc-spec}. Moment 0 maps of the detected molecules are shown in Fig. \ref{mom-0}. The emission is spatially compact. For all iCOMs, it peaks at the same position of the continuum peak (See Sect. \ref{continuum}).
\newline \indent The source size $\theta_{s}$ was obtained by stacking the uv tables of three HCOOCH$_{3}$ bright lines with the same velocity and full width at half-maximum (FWHM) in order to get a high signal-to-noise ratio (S/N) and then by fitting the emission in the stacked file with a circular Gaussian function in the visibility plane. The obtained value is $\theta_{s} = 0\farcs12 \pm 0\farcs02 $ ($\sim$ 45 au), which is compatible with the value measured from continuum emission. The derived size was fixed to this value for the other molecules.

{\footnotesize\setlength{\tabcolsep}{4pt}
\begin{table*}[t]
    \caption{Detected molecules towards HOPS-108 in this work and their derived parameters.} 
    \begin{center}
    
    \renewcommand{\arraystretch}{1.5}
    \begin{tabular}{l l c c c c c}
        
        \hline
        \hline
         
          \multicolumn{1}{l}{Molecule} & 
          \multicolumn{1}{l}{$N_{\rm lines}$ $^a$} &
          \multicolumn{1}{c}{$E_{\rm up}$} &
          \multicolumn{1}{c}{$T ^b$}  &
          \multicolumn{1}{c}{$N_{\rm tot}$} &
          \multicolumn{1}{c}{$[X]/[CH_{3}OH] ^e$}  &
          \multicolumn{1}{c}{$[X]/[HCOOCH_{3}] ^e$}
         
         \\  
          \multicolumn{1}{l}{}    &
          \multicolumn{1}{l}{}    &
         \multicolumn{1}{c}{(K)} &
          \multicolumn{1}{c}{(K)} & \multicolumn{1}{c}{(cm$\rm ^{-2})$}&
            \multicolumn{1}{c}{($\%$) } & \multicolumn{1}{c}{ ($\%$) }\\ 
         
         \hline  
         \multicolumn{7}{c}{LTE analysis}\\
         \hline
         
$\mathrm{HCOOCH_{3}}$ & 28 [83] & 49-366 & 96$_{-3}^{+3}$ & 6.7$_{-0.5}^{+0.5}$ $\times$ 10$^{17}$ & 4.8-13.4 & =100 \\

$\mathrm{C_{2}H_{5}CN}$ & 6 [11] & 161-285 & 145$_{-33}^{+62}$ & 4.0$_{-1.5}^{+2.4}$ $\times$ 10$^{15}$ & 0.03-0.08 & 0.6 $_{-0.2}^{+0.4}$ \\ 

$\mathrm{CH_{2}DOH}$ & 1 [3] & 113 & 100-150  $^c$ & (1.5-2.5) $\times$ 10$^{17}$ & 1.0-5.0 & / \\

$\mathrm{CH_{3}COCH_{3}}$ & 3 [10] & 145-150 & 100-150 $^c$ & (2.9-4.5) $\times$ 10$^{16}$ & 0.2-0.9 & 5.5 $_{-1.3}^{+1.3}$\\ 

$\mathrm{NH_{2}CHO}$ & 2 [2] & 79 & 100-150 $^c$ & (1.8-2.5) $\times$ 10$^{15}$ & 0.01-0.05 & 0.3 $_{-0.1}^{+0.1}$ \\  

$\mathrm{CH_{3}OCH_{3} \, ^{d}}$ & 3 [4] & 81-174 & 100-150 $^c$ &  (0.7-2) $\times$ 10$^{17}$ &  0.5-4.0 &  19.8 $_{-10.2}^{+10.2}$ \\

$\mathrm{CH_{3}CHO \, ^{d}}$ & 2 [3] & 93 & 100-150 $^c$ &  (1.1-1.6) $\times$ 10$^{16}$ & 0.08-0.32 & 2.0 $_{-0.4}^{+0.4}$\\

\hline  
         \multicolumn{7}{c}{non-LTE analysis}\\
         \hline

$\mathrm{CH_{3}OH \, }$ & 1 [9] & 190-537 & $\geq$120 & 0.5-1.4 $\times$ 10$^{19}$  & =100 & / \\

$\mathrm{CH_{3}^{18}OH \, ^{d}}$ & 5[6] & 39-81 & $\geq$120 &  0.9-2.5 $\times$ 10$^{16}$ & =0.18 & / \\
 
$\mathrm{CH_{3}CN}$ & 2 [3] & 93-100 & 120$_{-30}^{+30}$ & 2.0$_{-1.0}^{+1.0}$ $\times$ 10$^{16}$ & 0.1-0.4 & 3.4$_{-1.7}^{+1.7}$ \\         

$\mathrm{^{13}CH_{3}CN}$ & 3 [5] & 78-142 & 120$_{-30}^{+30}$ & 4.0$_{-2.0}^{+2.0}$ $\times$ 10$^{14}$ & / & / \\

\hline
\noalign{\vskip 2mm}
       \end{tabular}

\end{center}

\small{\textbf{Notes}. \newline $^a$ Number of lines used in the analysis. The total number of detected transitions (i.e. including blends) is reported in brackets (see text).\newline $^b$ $T$ is the rotational temperature $T_{\rm rot}$ for $\mathrm{HCOOCH_{3}}$ and $\mathrm{C_{2}H_{5}CN}$, $T_{\rm ex}$ is that for the rest of molecules analysed using LTE, and $T_{\rm kin}$ is that for molecules analysed using non-LTE. \newline $^c$ Adopted range of excitation temperatures (see Sect. \ref{lte}). \newline $^d$ Line flux measured after subtracting the small contribution from $\mathrm{HCOOCH_{3}}$ and/or $\mathrm{C_{2}H_{5}CN}$ (see Sects. \ref{non-lte} and \ref{lte}). \newline $^e$ Abundance ratios of molecules marked as `/' yield no meaningful information.}

    \label{molecules}
\end{table*}}

\subsection{Non-LTE analysis}
\label{non-lte}

The collisional rates of CH$_{3}$OH and CH$_{3}$CN are available in the literature \citep{Rabli2010,Green1986}. To derive the physical properties of the gas emitting them, we performed a non-local thermodynamic equilibrium (non-LTE) analysis using the large velocity gradient (LVG) code grelvg \citep{Ceccarelli2003}. 

For CH$_{3}$CN and $^{13}$CH$_{3}$CN, we used the collisional coefficients with H$_{2}$ computed between 20 and 140 K by \cite{Green1986} and provided by the LAMDA database \citep{Schoier2005}, which includes an extrapolation of the values for $T>$140 K. We assumed the $^{12}$C$/^{13}$C ratio equal to 50 as obtained in the Orion OMC-2 region by \cite{Kahane2018}. 
In our analysis we included two CH$_{3}$CN lines, as the third is strongly blended with a HCOOCH$_{3}$ line, and three $^{13}$CH$_{3}$CN lines, as the remaining two are also strongly blended with HCOOCH$_{3}$ lines.

We ran a large grid of models ($\geq10000$) for both CH$_{3}$CN and $^{13}$CH$_{3}$CN, covering a range in column density $N_{\mathrm{CH_{3}CN}}$ from 5$\times$10$^{15}$ to 3$\times$10$^{17}$ cm$^{-2}$, which corresponds to a range of column density $N_{\mathrm{^{13}}CH_{3}CN}$ from 1$\times$10$^{14}$ to 6$\times$10$^{15}$ cm$^{-2}$, a H$_{2}$ density $n_{\mathrm{H_{2}}}$ from 3$\times$10$^{5}$ to 6$\times$10$^{8}$ cm$^{-3}$, and a kinetic temperature $T_{\rm kin}$ from 50 to 200 K. 
Then, we fitted the $^{13}$CH$_{3}$CN and CH$_{3}$CN lines simultaneously, leaving $N_{\mathrm{CH_{3}CN}}$, $N_{\mathrm{^{13}CH_{3}CN}}$, $n_{\mathrm{H_{2}}}$, $T_{\rm kin}$ and the emitting size $\theta$ as free parameters. We assumed the line width equal to 2.5 km s$^{-1}$ as measured (see Table. \ref{transitions}), and we included the calibration uncertainty (10$\%$) and the rms errors in the uncertainty of the observed line intensities.

The best fit is obtained for $N_{\mathrm{CH_{3}CN}}$ = 2$\times$10$^{16}$ cm$^{-2}$ ($N_{\mathrm{^{13}CH_{3}CN}}$ = 4$\times$10$^{14}$ cm$^{-2}$) with a reduced chi-square $\chi_{R} ^{2}$ = 0.3. The $^{13}$CH$_{3}$CN lines are found to be optically thin ($\tau \sim$0.2) while the CH$_{3}$CN lines are found to be optically thick ($\tau \sim$9). The lines of the two isotopologues are emitted by a source of 0$\farcs$18 in diameter. Their emitting size is more extended than that of HCOOCH$_{3}$. Solutions with 1$\times$10$^{16}$  $\leq N_{\mathrm{CH_{3}CN}} \leq$ 3$\times$10$^{16}$cm$^{-2}$  (2$\times$10$^{14}$  $\leq N_{\mathrm{^{13}CH_{3}CN}} \leq$ 6$\times$10$^{14}$cm$^{-2}$) are within 1$\sigma$ of confidence level. Within the errors, the temperature is (120$\pm$30) K and the H$_{2}$ density is larger than 3$\times$10$^{7}$ cm$^{-3}$ at the 1$\sigma$ confidence level.

For CH$_{3}$OH and CH$_{3}^{18}$OH, we used the collisional coefficients with H$_{2}$ computed between 10 and 200 K for the first 256 levels by \cite{Rabli2010} and provided by the BASECOL database \citep{Dubernet2013}. We assumed the $^{18}$O$/^{16}$O ratio equal to 560 \citep{Wilson1994}.
To fit the LVG predictions, we included four CH$_{3}^{18}$OH-e lines, two CH$_{3}^{18}$OH-a lines, and one CH$_{3}$OH-e line. Among the other available eight CH$_{3}$OH lines, three have high upper level energies (E$_{ \rm up} \geq$ 400 K) and they were excluded from the analysis as the collisional coefficients are not computed at these energies. The other five lines have low excitation energies; they trace extended emission and hence they were excluded to avoid contamination. Among the available CH$_{3}^{18}$OH lines, one is strongly blended with CH$_{2}$DOH and was excluded from the analysis, two are contaminated by a weaker line of HCOOCH$_{3}$, and one is contaminated by C$_{2}$H$_{5}$CN, thus their integrated intensities were estimated after removing the expected intensity of the contaminating lines (See Appendix \ref{appB}). The spectrum is shown in Fig. \ref{ch3o-18-h-spec}. For the CH$_{3}$OH line, the rms error was taken into account in the line flux. For CH$_{3}^{18}$OH, in addition to the rms error, we included the uncertainty of baseline imperfection due to rich spectra of iCOMs.
We added an extra 20$\%$ in the flux error to account for the uncertainty linked to possible non-LTE effects. 
\newline \indent
As in the case of CH$_3$CN, we ran a large grid of models ($\geq10000$) for both CH$_{3}$OH and CH$_{3}^{18}$OH, covering a range in column density $N_{\mathrm{CH_{3}OH}}$ from 1$\times$10$^{16}$ to 7$\times$10$^{19}$ cm$^{-2}$ that corresponds to a range in column density $N_{\mathrm{CH_{3}^{18}}OH}$ from 2$\times$10$^{13}$ to 1.3$\times$10$^{17}$ cm$^{-2}$, a H$_{2}$ density $n_{\mathrm{H_{2}}}$ from 8$\times$10$^{6}$ to 2$\times$10$^{9}$ cm$^{-3}$, and a temperature $T$ from 90 to 250 K. Then, we found the solution by simultaneously fitting the CH$_{3}^{18}$OH-e and CH$_{3}^{18}$OH-a lines as well as the CH$_{3}$OH-e line, leaving $N_{\mathrm{CH_{3}OH}}$, $N_{\mathrm{CH_{3}^{18}OH}}$, $n_{\mathrm{H_{2}}}$, $T_{\rm kin}$ and the emitting size $\theta$ as free parameters. We assumed the CH$_{3}^{18}$OH-e$/$ CH$_{3}^{18}$OH-a ratio equal to 1, the line width equal to 3.6 km s$^{-1}$ as measured (see Table. \ref{transitions}), and we included the uncertainties described in the previous paragraph in the observed intensities.
While the comparison between the observations and model predictions constrain the CH$_{3}$OH and CH$_{3}^{18}$OH column densities quite well (0.5$\times$10$^{19}$-1.4$\times$10$^{19}$ cm$^{-2}$ and 0.9$\times$10$^{16}$-2.5$\times$10$^{16}$ cm$^{-2}$, respectively), it does not allow one to put stringent limits on the density and temperature, other than the emitting gas temperature being larger than about 120 K and the density larger than about 10$^{6}$ cm$^{-3}$. The upper limits are even less constrained because the collisional coefficients are only available for T$\leq$200 K and the levels become LTE-populated at about 10$^{6}$ cm$^{-3}$. The emitting size of CH$_{3}$OH and CH$_{3}^{18}$OH is well constrained (0$\farcs$10- 0$\farcs$12) and it is consistent with that obtained from the analysis of the HCOOCH$_{3}$ lines. The CH$_{3}^{18}$OH lines are found to be optically thin ($\tau \sim 0.1$) while the CH$_{3}$OH lines are found to be very optically thick ($\tau \sim 21$).


\subsection{LTE analysis}
\label{lte}

For the iCOMs where collisional rates are not calculated, we assumed local thermodynamic equilibrium (LTE) conditions. Among these molecules, only two (HCOOCH$_{3}$ and C$_{2}$H$_{5}$CN) have a sufficiently large number of detected transitions spanning a large range of upper state energies, for which we could therefore build a rotational diagram (RD). In addition to the LTE assumption, to derive the column density and rotational temperature ($T_{\rm rot}$) of these two molecules, we assumed optically thin emission. The upper level column density here is $N_{u} = (8\pi k \nu^{2}/hc^{3}A_{ul}) W$, where $W = \int T_{B} \, dV$ is the line integrated intensity and $T_{B}$ is the brightness temperature.

\begin{figure}
    \centering
    \includegraphics[width=0.48\textwidth]{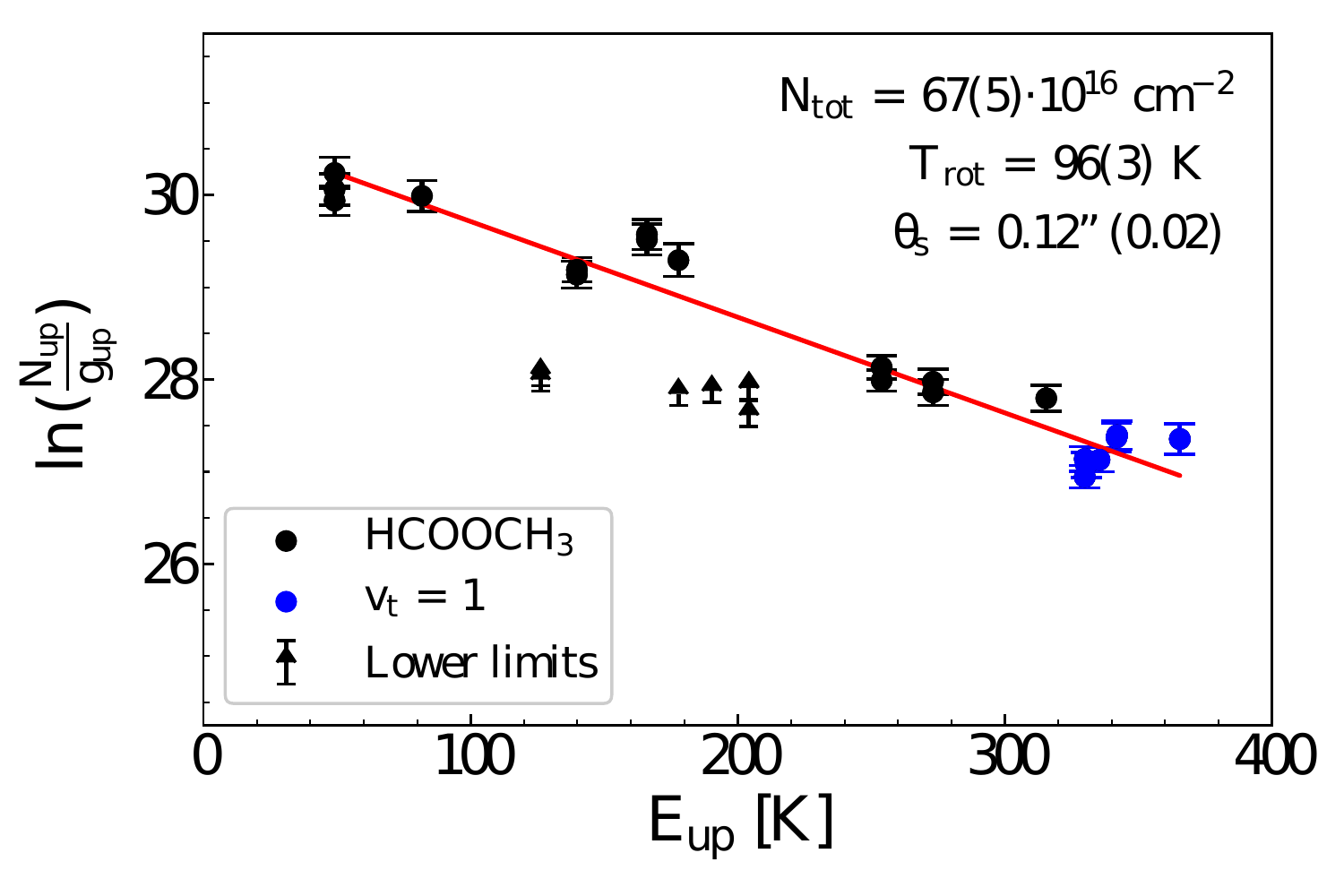}
    \includegraphics[width=0.48\textwidth]{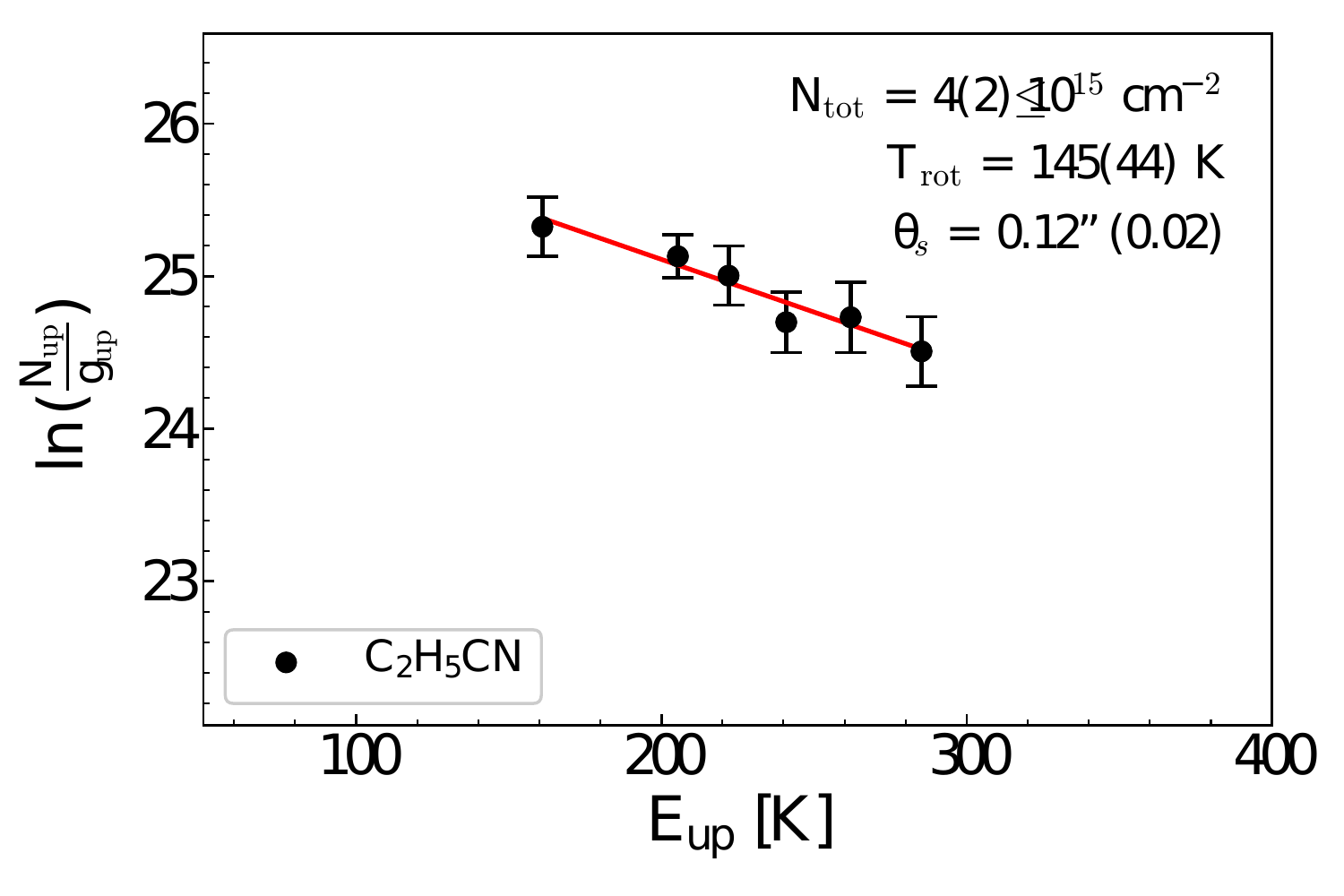}
    \caption{Rotational diagrams for HCOOCH$_{\rm 3}$ (top) and C$_{\rm 2}$H$_{\rm 5}$CN (bottom). Data points are presented in black. The HCOOCH$_{\rm 3}$ v$_{t}=1$ transitions are depicted in blue. The arrow symbols represent the lower limits that are excluded from the analysis (see Sect. \ref{lte}). The red lines correspond to the best fit to the data points.}
    \label{RDs}
\end{figure}

To build the RD, we included the isolated lines, and the lines that consist of several transitions with the same upper level energy, but different Einstein coefficients and statistical weights as done by \cite{Bianchi2019}. The blended lines were excluded from the analysis. The rms and calibration errors in the line fluxes were taken into account. The resulting rotational diagrams are shown in Fig. \ref{RDs}.

For HCOOCH$_{3}$, we obtained a rotational temperature of (96$\pm$3) K and a column density of (6.7$\pm$0.5)$\times$10$^{17}$ cm$^{-2}$. The lines with a high optical depth (>1) were excluded from the fit but are shown in the RD plot as lower limits, as a consistency check. For C$_{2}$H$_{5}$CN we obtained a rotational temperature of (145$\pm$44) K and a column density of (4.0$\pm$1.9)$\times$10$^{15}$ cm$^{-2}$.

For the other molecules, there are a few lines that cover a narrow $E_{ \rm up}$ range, hence we calculated the column densities at an excitation temperature range $T_{\rm ex}$ = 100-150 K based on the derived rotational temperature for HCOOCH$_{3}$ and the kinetic temperature for CH$_{3}$OH (see Sect. \ref{non-lte}). Also, here, we assumed LTE and optically thin line emission, and we took rms and calibration errors in the line fluxes into account.
In our analysis, we included the unblended lines. We also included the lines that are slightly blended with HCOOCH$_{3}$ and/or C$_{2}$H$_{5}$CN after subtracting the flux of the contaminant when it was estimated to be $<$20$\%$ of the total line flux (See Appendix. \ref{appB}). Finally, we excluded the lines that are blended by other molecular species or strongly blended by either HCOOCH$_{3}$ or C$_{2}$H$_{5}$CN ($>$20$\%$ of the total line flux).

More explicitly, for CH$_{2}$DOH, we included only one line as the other two are blended with CH$_{3}^{18}$OH and CH$_{3}$COCH$_{3}$. For CH$_{3}$COCH$_{3}$, we included three lines as the others were either strongly blended with HCOOCH$_{3}$, or with other molecular species. For NH$_{2}$CHO, we detected two isolated lines, which were both included in the analysis.

For CH$_{3}$OCH$_{3}$, we detected four lines with emission higher than 5$\sigma$. One of them is likely blended with CH$_{2}$DCN, and it was excluded from the analysis. Another is very slightly contaminated with two weaker lines of HCOOCH$_{3}$ and C$_{2}$H$_{5}$CN, thus its line intensity was estimated after removing the expected intensities of the contaminating lines (See Appendix \ref{appB}). The three available lines were not sufficient to build a RD, hence we derived the column density following the previously mentioned assumptions. 

For CH$_{3}$CHO, we detected three lines, one of which is strongly blended with both CH$_{2}$DOH and HCOOCH$_{3}$, and two others are slightly blended with HCOOCH$_{3}$. We derived a column density value using the latter lines after removing the expected intensities of the contaminating HCOOCH$_{3}$.
\newline
The derived column densities of these iCOMs are summarised in Table \ref{molecules}.

%


\section{Discussion}

\label{Discussions}

\subsection{The hot corino in OMC-2\,FIR\,4}

The OMC-2\,FIR\,4 region was extensively studied at large scales with the \textit{Herschel} Space Observatory \citep{Loepz-Sepulcre-and-kama2013,Kama2013,Furlan2014,Ceccarelli2014,Kama2015,Gonzalez-Garcia2016,Fontani2017,Favre2017}. At scales of $\sim$5$"$, these studies provided a rather complete overview on FIR\,4 and the surroundings. For example, it was found that the outer shell of OMC-2\,FIR\,4 is strongly irradiated by a FUV field ($\sim$1500 $G_{\rm 0}$), arising from the high-mass stars located in the nearby Trapezium OB association \citep{Loepz-Sepulcre-and-kama2013}. In addition, it was discovered that the interior envelope of FIR\,4 is subject to irradiation and ionisation by local CR-like particles with an ionisation rate of $\zeta$ = 4 $\times$ 10$^{-14}$ s$^{-1}$ \citep{Ceccarelli2014}. Further interferometric studies at millimetre and sub-millimetre wavelengths with $\sim$1$"-$3$"$ angular resolution revealed that OMC-2\,FIR\,4 is a protocluster harbouring several cores \citep{Shimajiri2008, Lopez-Sepulcre-and-taquet2013, Kainulainen2017}. They also confirmed the presence of FUV irradiation \citep{Lopez-Sepulcre-and-taquet2013, Favre2018} and that of CR-like particles \citep{Fontani2017, Favre2018}. 
At sub-arcsecond resolution, in addition to the present work, the only available study to our knowledge was performed by \cite{Tobin2019}. These authors characterised the protostellar content in OMC-2\,FIR\,4 by studying the dust continuum and methanol emission. In HOPS-108, they detected compact emission in CH$_{3}$OH, CH$_{3}$CN, and HCOOCH$_{3}$. From their three detected CH$_{3}$OH lines, they derived a rotational temperature of 140 K. While they reported that HOPS-108 is likely a hot corino, an extensive characterisation of its organic molecular content was out of their scope. In this context, our study is the most comprehensive one on iCOMs in this object. We detected a relatively large number of molecules that are typical tracers of hot corinos, including some that are more rarely detected even in such environments, such as CH$_{3}$COCH$_{3}$ and C$_{2}$H$_{5}$CN. All the detected molecules have compact emission centred on the continuum source (see Fig. \ref{mom-0}). In addition, the majority of CH$_{3}$OH, HCOOCH$_{3}$, and C$_{2}$H$_{5}$CN lines we detected are highly excited (E$_{\rm up} \geq$ 160 K with a kinetic temperature of $\sim$ 150 K and rotational temperature of $\sim$ 96 K and 145 K respectively), indicating that the emission originates from hot gas. This confirms that HOPS-108 is indeed a hot corino.
From the HCOOCH$_{3}$ lines, we found that the systemic velocity of HOPS-108 is $\sim$ +12.3 km s$^{-1}$. From CH$_{3}$OH lines we found a systemic velocity of $\sim$ +12.6 km s$^{-1}$ which is equal to that obtained by \cite{Tobin2019}, implying a redshift of $\sim$ 1.2 km s$^{-1}$ with respect to the 'bulk' velocity of the FIR\,4 protocluster. It is worth noting that \cite{Kama2013} obtained an equal velocity from highly excited CH$_{3}$OH lines ($E_{\rm up}$=450 K) at large scales with \textit{Herschel} suggesting that also those lines primarily originated in HOPS-108.


\subsection{Comparison with other hot corinos}

The detection of a hot corino in what is currently considered to be the closest analogue of the Solar System birth environment naturally raises the question of how the chemistry of the hot corino is affected by the nearby massive stars and the internal source of high-energy particles. To this end, we compared iCOM abundances in HOPS-108 to those in other known hot corinos that are either isolated or located in loose protoclusters. In this comparison, we used the abundance ratios with respect to CH$_{3}$OH and HCOOCH$_{3}$, which are the most abundant molecules in hot corinos, and included data from low-mass Class 0 and Class I hot corinos obtained with (sub-)millimetre interferometers. We report the obtained abundance ratios in Table \ref{molecules}. Figure \ref{mol-meth} shows the column densities of the detected iCOMs normalised to that of CH$_{3}$OH where the sources are separated into Class 0 and I, and colour-coded according to the cloud where they are forming.
For most of the molecules, a scatter of one order of magnitude is observed in the different abundance ratios. A larger scatter of almost two orders of magnitude is observed in the case of NH$_{2}$CHO and CH$_{3}$CHO. It is worth noting that similarly large differences in the iCOM abundance ratios across different hot corinos have been recently reported in several previous studies \citep{Bergner2019, Bianchi2019, Belloche2020, Nazari2021}. While these differences induce more complexities in the comparison of hot corinos, we can conclude that, within one order of magnitude, for most of the molecules, the abundance ratios in HOPS-108 are consistent with those found in other hot corinos, located in Perseus, Ophiuchus, and Serpens. They are also consistent with the ratios found in HH212, the other hot corino reported in Orion thus far. A different behaviour is shown by HCOOCH$_3$ and CH$_{2}$DOH, and they are described in the following subsections.

\subsubsection{Abundance of methyl formate}
\label{mf-discussion}

\begin{figure}
    \centering
    \includegraphics[width=0.48\textwidth]{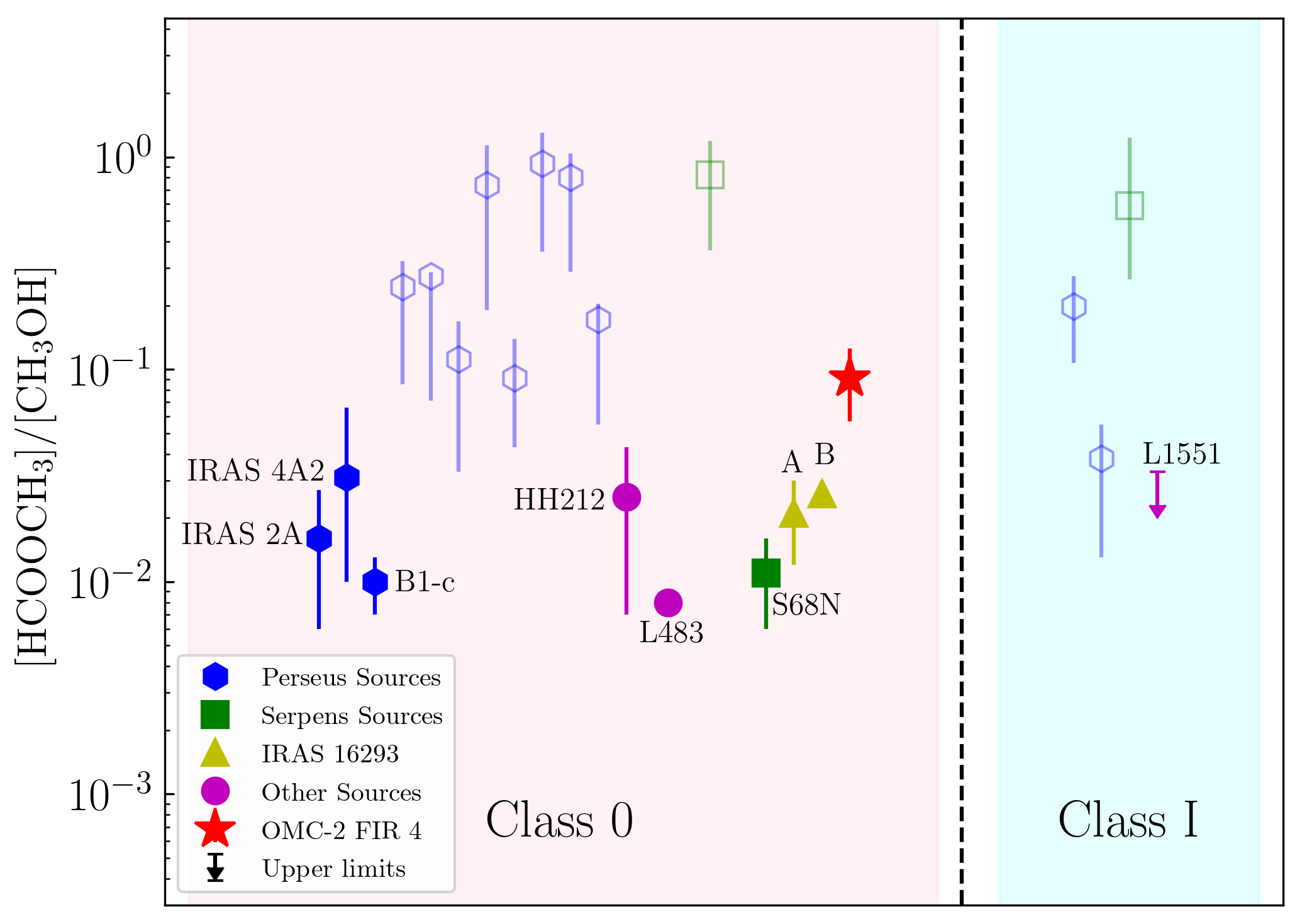}
    \includegraphics[width=0.48\textwidth]{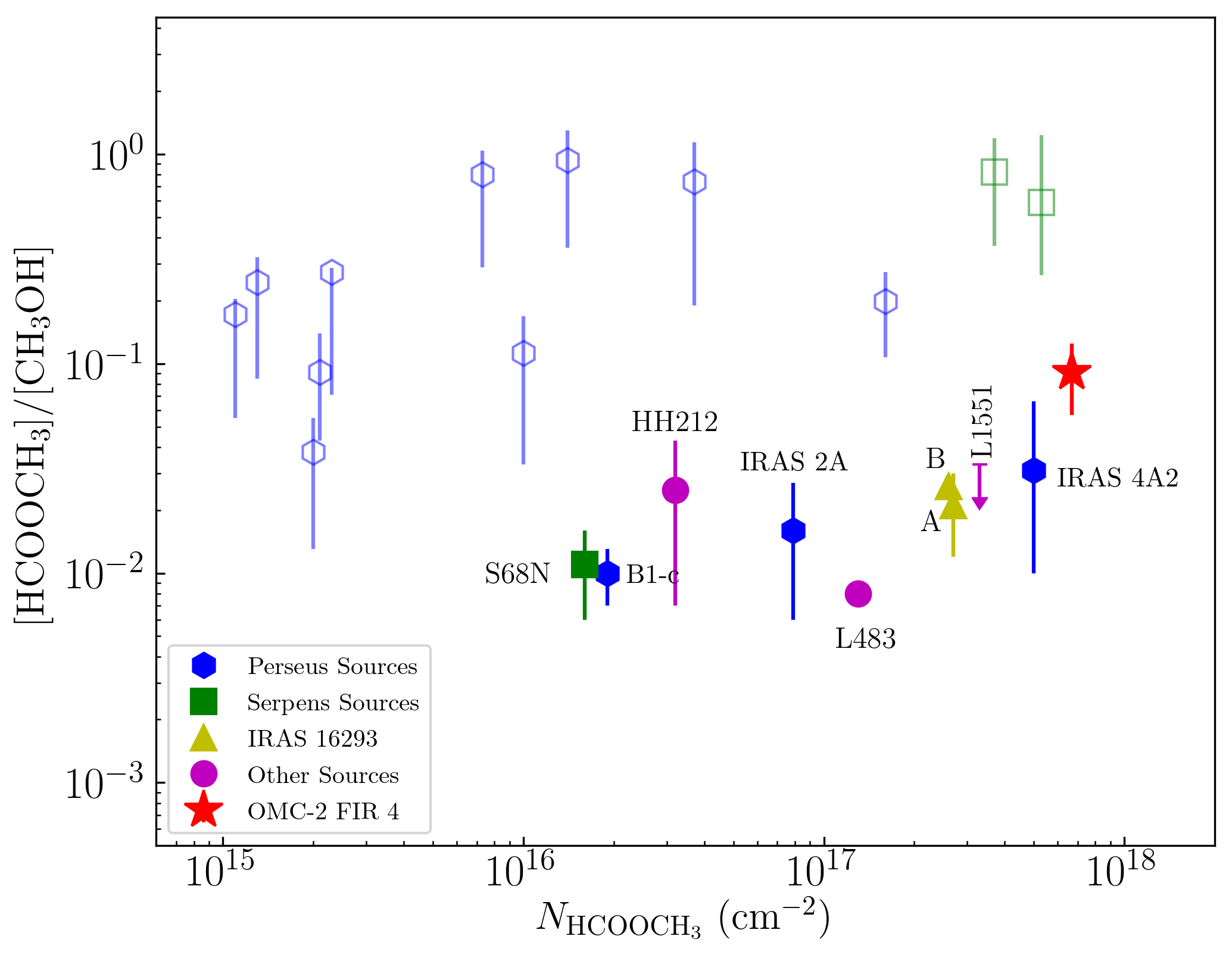}
\caption{Abundance ratio of HCOOCH$_{3}$ with respect to CH$_{3}$OH in HOPS-108 compared to other isolated hot corinos as a function of the class in no particular order within a given class (top) and HCOOCH$_{3}$ column density (bottom). The open symbols correspond to the sources where CH$_{3}$OH column density is most likely underestimated (see Sect. \ref{mf-discussion}). The Class 0 sources are: IRAS 2A, IRAS 4A2 \citep{Taquet2015,Lopez-Sepulcre2017}, B1-c, S68N \citep{vanGelder2020}, HH212 \citep{Bianchi2017, Lee2019}, L483 \citep{Jacobsen2019}, Ser-emb 1 \citep{Bergner2019}, IRAS 16293 A \citep{Manigand2020}, IRAS 16293 B \citep{Jorgensen2016, Jorgensen2018}, HOPS-108 in OMC-2\,FIR\,4 (this work), and the sources from the PEACHES survey \citep{Peaches2021}. The Class I sources are: Ser-emb 17 \citep{Bergner2019}, L1551 \citep{Bianchi2020}, and also sources from the PEACHES survey \citep{Peaches2021}.} 
    \label{mf-meth-ratio}
\end{figure}

Among all the iCOMs detected in HOPS-108, there is one that clearly stands out relative to other hot corinos, both in terms of line richness and column density: HCOOCH$_3$. Figure \ref{mf-meth-ratio} shows the abundance ratios of HCOOCH$_{3}$ normalised to those of CH$_{3}$OH. The sources are separated according to their class and colour-coded depending on the star-forming region to which they belong. At first sight, the [HCOOCH$_{3}$]/[CH$_{3}$OH] ratio in HOPS-108 seems consistent with those found in other hot corinos. However, it is worth mentioning that the CH$_{3}$OH lines can be very optically thick towards hot corinos \citep{Bianchi2020} and, therefore, in the majority of them, the column density of CH$_{3}$OH is underestimated unless it is derived from CH$_{3}^{18}$OH or the $^{13}$CH$_{3}$OH isotopologue when the latter is optically thin. If we compare the [HCOOCH$_{3}$]/[CH$_{3}$OH] ratio for the sources where the CH$_{3}$OH column density is derived using one of the secondary isotopologues (filled symbols in Fig. \ref{mf-meth-ratio}), we can notice that the HCOOCH$_{3}$ is almost a factor of 3-5 more abundant in HOPS-108 with respect to other sources, pointing to a possible enhancement of HCOOCH$_{3}$ in this source. From the comparison of iCOM abundance ratios with respect to HCOOCH$_{3}$ in all of the sources (Fig. \ref{mol-mf}), we can notice that within the errors, the N-bearing molecules ratios in HOPS-108 are lower than in other hot corinos, and those of O-bearing molecules are among the lowest, even though they are comparable to the values measured towards the IRAS 16293-2422 A and B sources. However, as previously mentioned, the large scatters in the iCOM abundance ratios across the different hot corinos make the assessment challenging \citep{ Bergner2019, Bianchi2019, Belloche2020, Nazari2021}.

\begin{figure}
    \centering
    \includegraphics[width=0.48\textwidth]{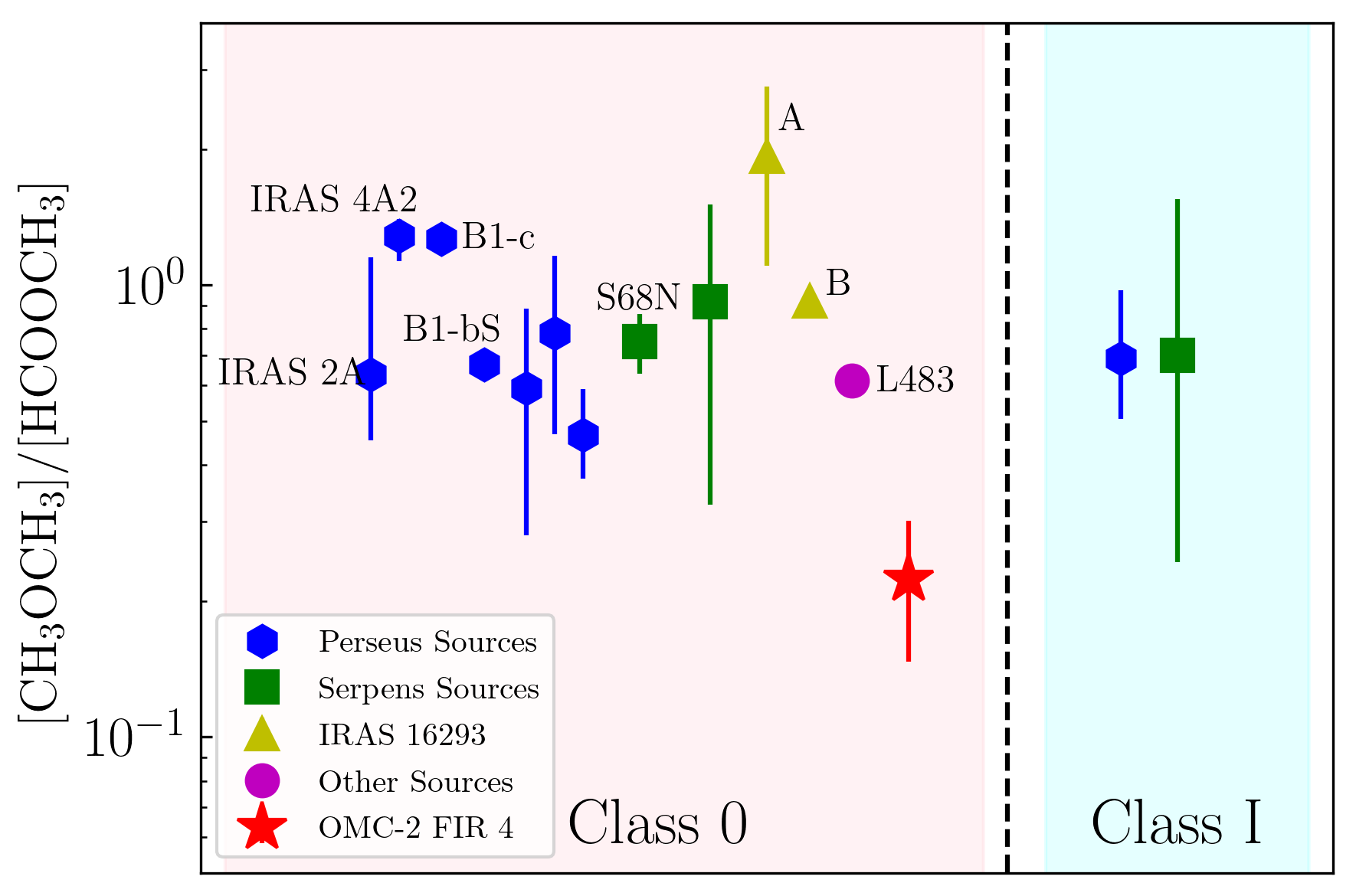}
    \includegraphics[width=0.48\textwidth]{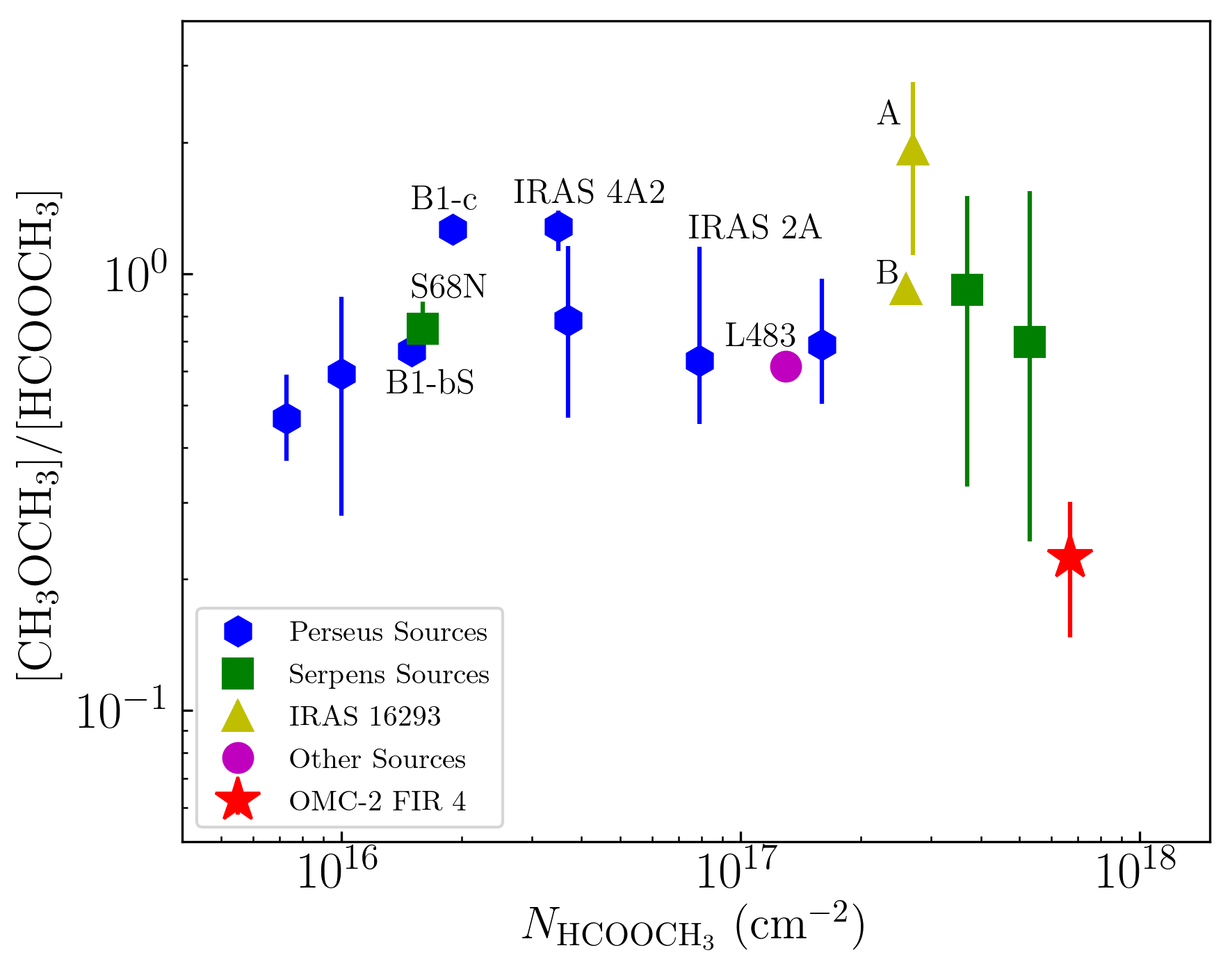}
    \includegraphics[width=0.48\textwidth]{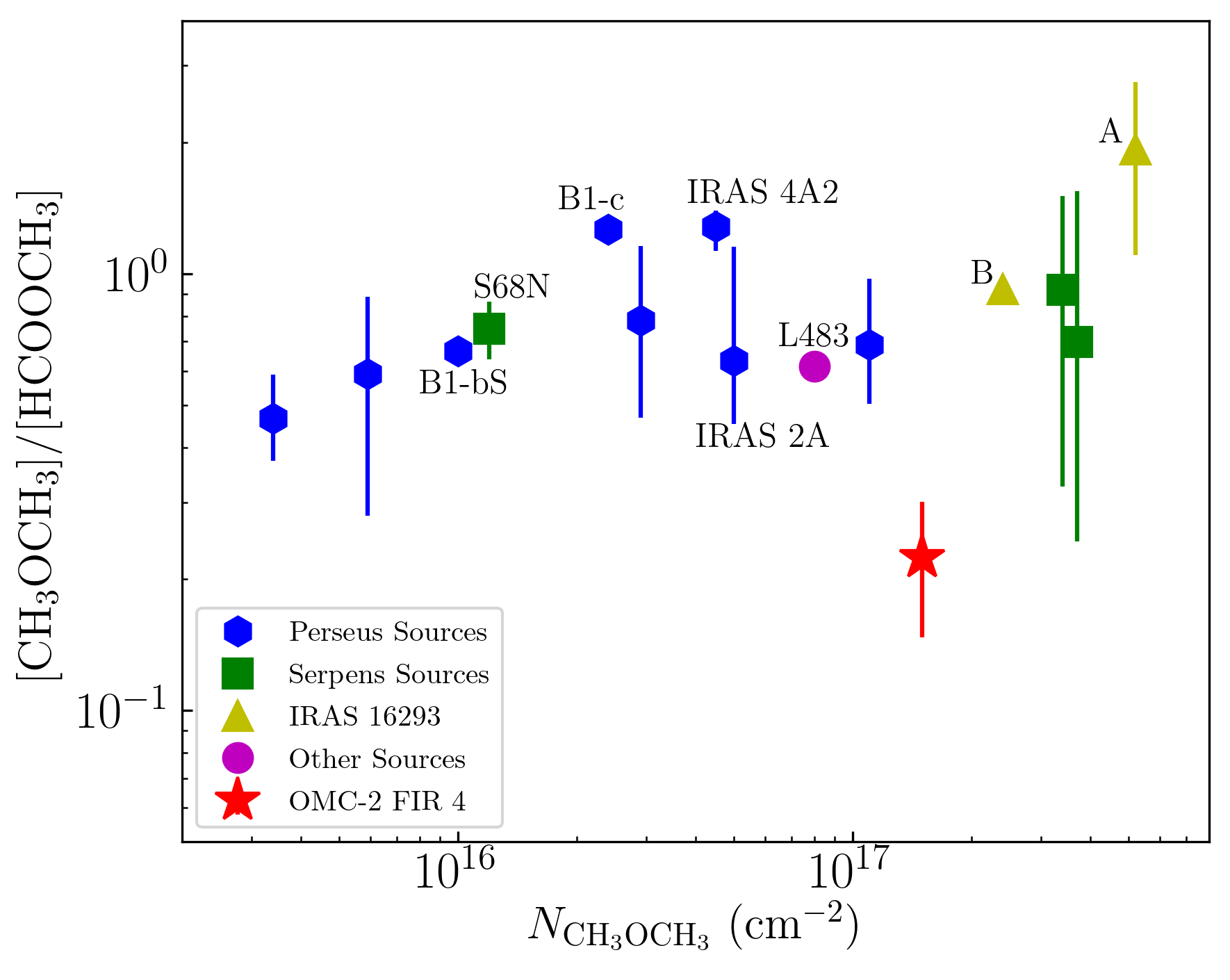}
\caption{Abundance ratio of CH$_{3}$OCH$_{3}$ with respect to HCOOCH$_{3}$ in HOPS-108 compared to other isolated hot corinos. In the top plot, the sources are separated according to their class, with no particular order within a given class. In the middle and bottom plots, the abundances are as a function of HCOOCH$_{3}$ and CH$_{3}$OCH$_{3}$ column densities, respectively. The same sources from Fig. \ref{mf-meth-ratio} are used, including B1-bS \citep{Marcelino2018}.}
    \label{mf-dme-ratio}
\end{figure}

An interesting result comes from the [CH$_{3}$OCH$_{3}$]/[HCOOCH$_{3}$] abundance ratio which is lower in our source than in the other hot corinos, implying an enhancement in HCOOCH$_{3}$. Intriguingly enough, the hot corino of OMC-2\,FIR\,4 is the only source that is marginally deviating within the errors from the well-known linear correlation between methyl formate (MF, HCOOCH$_{3}$) and dimethyl ether (DME, CH$_{3}$OCH$_{3}$), where [CH$_{3}$OCH$_{3}$]/[HCOOCH$_{3}$]$\sim$1 (Fig. \ref{mf-dme-ratio}). This correlation was observed by \cite{Jaber2014} over a range of almost five orders of magnitude, with a Pearson correlation coefficient equal to 0.95. It was also previously observed by \cite{Brouillet2013} over a smaller range. The MF-DME relation is well reproduced with the model by \cite{Balucani2015}, using gas-phase reactions in cold environments. In this model, MF is formed from DME via the oxidation of CH$_{3}$OCH$_{2}$, while DME itself forms via the effective radiative association of the radicals CH$_{3}$ and CH$_{3}$O in the gas phase (see their Fig. 1). In what follows, we discuss two possibilities to explain the DME/MF abundance ratio found in HOPS-108: (i) an enhancement of CR ionisation in the gas phase, producing more MF and destroying DME, or (ii) a formation on the grain surface 'through non-diffusive mechanisms'.

We tested the first possibility using a toy model by \cite{Balucani2015} at three values of the CR ionisation rate, $\zeta$, using the kinetic temperature, the gas density, and the emitting size obtained from the non-LTE analysis of methanol. The results are shown in Fig. \ref{crs-model}. It can be seen that the formation and destruction of MF and DME still go almost hand in hand for the three $\zeta$ values. The model actually predicts a DME/MF ratio larger than unity for CRs enhanced regions (top panel) as well as an overall lower abundance of both molecules, while our results show a relatively low ratio and a large column density of both molecules. Thus, even with an enhancement of the CR ionisation rate, if we assume formation in the gas-phase, with the present gas-phase network, the model does not yield more MF than DME. This suggests that CR-like particles do not play a role in shaping the abundance of DME relative to that of MF. Alternatively, the chemical network may not be complete concerning the destruction routes of MF and DME, where the destruction of DME is supposed to be higher to match our observations. To clarify this point, we need new theoretical and experimental chemical calculations on these pathways. For example, by studying new experimental and theoretical calculations on the destruction of MF and DME by He$^{+}$, \cite{Ascenzi2019} obtained reaction rate coefficients that are very different from those given by KIDA and UDfA databases, showing that this could impact the predicted abundances by 40$\%$ or even more depending on the physical conditions. This shows that the inclusion of correct reactions and rate coefficients in the astrochemical networks can have a crucial role in modelling and abundance predictions. 
\newline \indent
Regarding the second possibility, the model by \cite{Jin&Garrod2020} is currently the one that approaches most of our observed MF column density value, but not that of DME. This model introduces new non-diffusive mechanisms for iCOM formation on grain surfaces. The one that comes closer to our MF value is called the three-body excited formation process (3-BEF), where radicals are formed in an excited state, allowing them to overcome activation barriers to react with nearby stable molecules. In this mechanism, MF is formed through the reaction of an excited CH$_{3}$O radical with CO on the grain surface producing CH$_{3}$OCO, which is converted to HCOOCH$_{3}$ by hydrogenation. This mechanism enhances the grain-surface production of MF which results in an abundance almost two orders of magnitude higher than that of DME. Nonetheless, it is unclear why this mechanism would affect the HOPS-108 hot corino and not other sources. In addition, the model underestimates the column density of DME and over-estimates that of acetaldehyde (AA, CH$_{3}$CHO) which we measured in HOPS-108. This can be seen in our observational results, where the AA column density is almost an order of magnitude lower than that of DME. 

It is worth noting that a low DME/MF ratio was recently observed in two hot cores by \cite{Law2021}. While it was not clear if the different trend is due to a higher gas temperature or if it is specific to hot core-like environments, these authors have suggested that an additional formation pathway of HCOOCH$_{3}$ can explain their results (See their Sect. 5.2.1). Supposing that the CH$_{3}$OCH$_{3}$ is forming via ice conversion independently of the temperature, they suggested that sufficiently efficient gas-phase reactions driving the extra production of HCOOCH$_{3}$ can explain the different ratios seen in hot cores. A similar result was also observed in the hot corino BHR\,71 by \cite{Yang2020}, but no discussion on iCOM column density ratios is reported. 

In a nutshell, our DME/MF abundance ratio cannot be explained by either grain-surface or by gas-phase routes with our current knowledge. This highlights the need to theoretically revise the reaction network leading to or destroying MF and DME, and more generally iCOMs, including their rate coefficients. In addition, so far, the DME/MF ratio in hot corinos located in massive star-forming environments such as Orion have only been measured towards HOPS-108. Thus, further observations towards Orion hot corinos are needed to understand whether this low ratio is a consequence of the environmental conditions or a unique property of HOPS-108.


\subsubsection{Methanol deuteration}
\label{deuteration-discussion}
\begin{figure}
    \centering
    \includegraphics[width=0.48\textwidth]{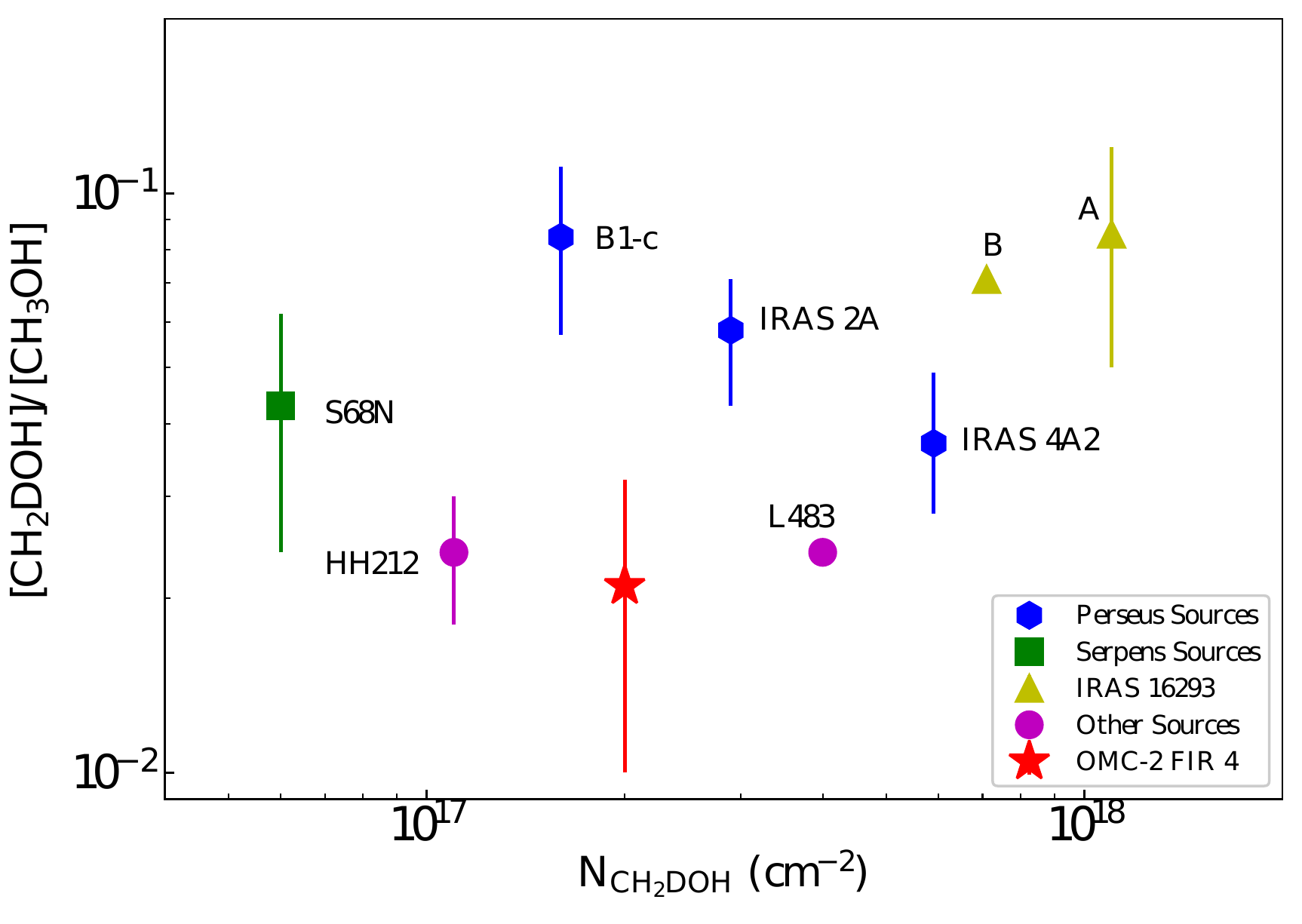}
\caption{Abundance ratio of CH$_{2}$DOH with respect to CH$_{3}$OH in HOPS-108 compared to isolated hot corinos where the CH$_{3}$OH column density was derived using one of its secondary isotopologues (see Sect. \ref{deuteration-discussion}). All the sources are Class 0: IRAS 2A, IRAS 4A2 \citep{Taquet2015,Taquet2019}, B1-c, S68N \citep{vanGelder2020}, HH212 \citep{Bianchi2017}, L483 \citep{Jacobsen2019}, IRAS 16293 A \citep{Manigand2020}, IRAS 16293 B \citep{Jorgensen2016, Jorgensen2018}, and HOPS-108 in OMC-2\,FIR\,4 (this work).}
    \label{deut-ratio}
\end{figure}

The other noteworthy result obtained is for CH$_{2}$DOH. Figure \ref{deut-ratio} shows the abundance ratios of CH$_{2}$DOH with respect to CH$_{3}$OH. In this plot, all the data belong to Class 0 sources for which the CH$_{3}$OH column density was derived using CH$_{3}^{18}$OH or $^{13}$CH$_{3}$OH. The deuteration ratios vary between $\sim$ 2$\%$ and 9$\%$. The lowest percentages ($\sim$2$\%$) belong to HH212, HOPS-108, and L483. HH212 and HOPS-108 are both located in the Orion molecular complex. On the other hand, L483 is a hybrid source \citep{Oya2017}, that is with features from both a hot corino and warm carbon chain chemistry object (WCCC; \citealp{Sakai2013}), the latter being characterised by a lower deuteration compared to hot corinos. The highest percentages are seen in IRAS 16293 A, as well as IRAS 16293 B located in the Ophiuchus star-forming region and in B1-c located in Perseus. Arguably, the [CH$_{2}$DOH]/[CH$_{3}$OH] ratio in HOPS-108 is within the typical hot corino values, even though it is at the lower end. Given that the majority of the molecules we observed in hot corinos are the result of ice mantle sublimation, the lower deuteration in HOPS-108 could be due to different physical conditions in the gas at the time when ice mantles were forming. Indeed, methanol deuteration is strongly dependent on density, temperature, and time, and on the atomic D/H ratio in the gas phase during ice formation. Hence, as for HH212, the low deuteration ratio in HOPS-108 can provide us with a hint on the past physical conditions in the gas of the parental cloud during ice formation. Being in a more active star-forming region near high-mass stars that heat the gas, the lower deuteration in HOPS-108 can be due to a higher gas temperature in the early stages that decreases the atomic D/H ratio and consequently diminishes the methanol deuteration \citep{Bianchi2017, Taquet2012, Taquet2014, Taquet2019}.
\newline \indent Nevertheless, we stress that the column density of CH$_{2}$DOH in our work was derived using a single line and is thus subject to uncertainty. Therefore, more observations are needed to constrain the value we obtained.

\section{Conclusions}

\label{Conclusions}
In this work we studied iCOMs in HOPS-108, a Class 0 protostar located in the protocluster OMC-2\,FIR\,4, using ALMA at 1.2 mm and an angular resolution of $\sim$0$\farcs$28 ($\sim$110\,au). Our main conclusions are summarised as follows:
\begin{enumerate}
    \item We detected 11 iCOMs with compact emission and high upper level energies (39 K-537 K). The molecular and continuum emission in this source are coincident and it originates from $\sim$50 au scale and the kinetic temperature is found to be $\geq$120 K, confirming that HOPS-108 is a hot corino.
    \item The kinetic temperature was derived using non-LTE analysis of CH$_{3}^{18}$OH and CH$_{3}$OH lines. From this analysis, we found that CH$_{3}$OH lines are very optically thick ($\tau\sim$21), which is probably also the case in other hot corinos. Hence, in similar studies of these sources, the column densities obtained from the main CH$_{3}$OH isotopologue are likely underestimated.
    \item The HCOOCH$_{3}$ abundance relative to CH$_{3}$OH in HOPS-108 is $\sim$3-5 times higher compared to that in other hot corinos where the CH$_{3}$OH column density was derived using one of its isotopologues. This indicates that HCOOCH$_{3}$ is likely enhanced in our source.
    \item The N-bearing molecules abundance ratio with respect to HCOOCH$_{3}$ in HOPS-108 are lower than in other hot corinos, suggesting that HCOOCH$_{3}$ might be marginally enhanced relative to this family of molecules in this source.
    \item The noticeable result in our study is the [CH$_{3}$OCH$_{3}$]/[HCOOCH$_{3}$] abundance ratio of $\sim$20$\%$ which is lower in HOPS-108 than in other hot corinos, marginally deviating from the well-known correlation between the two molecules. Interestingly, this low ratio has also been recently observed in two hot cores. This might be due to local environmental effects; however, with the current chemical network, there is no suitable model that can explain this ratio. 
    \item We obtained a [CH$_{2}$DOH]/[CH$_{3}$OH] abundance ratio of $\sim$2$\%$, which is relatively low with respect to the measured values in Perseus and Ophiuchus. This might result from higher temperatures of the parental molecular cloud at its early evolutionary stages due to high-mass star formation activity in the vicinity. 
\end{enumerate}

Our study highlights the importance of improving the chemical networks using theoretical calculations. It also suggests that more observations of CH$_{2}$DOH are needed to better constrain the deuteration in HOPS-108, and that more hot corinos located in dense clusters should be targeted to assess whether the observed abundance ratios are an environmental product or if HOPS-108 is an exceptional hot corino overall.

\begin{acknowledgements}
We thank the anonymous referee for careful reading and fruitful comments that helped to improve the paper. This project has received funding from the European Union’s Horizon 2020 research and innovation program under the Marie Skłodowska-Curie grant agreement No 811312 for the Project "Astro-Chemical Origins” (ACO). This paper makes use of the following ALMA data: ADS/JAO.ALMA 2016.1.00376.S. ALMA is a partnership of ESO (representing its member states), NSF (USA) and NINS (Japan), together with NRC (Canada), MOST and ASIAA (Taiwan), and KASI (Republic of Korea), in cooperation with the Republic of Chile. The Joint ALMA Observatory is operated by ESO, AUI/NRAO and NAOJ. We thank Paola Caselli, Francesco Fontani and Valerio Lattanzi for useful discussions. CCo and LP acknowledge the project PRIN-INAF 2016 
The Cradle of Life - GENESIS-SKA (General Conditions in Early Planetary Systems for the rise of life with SKA). CF acknowledges support from the French National Research Agency in the framework of the Investissements d’Avenir program (ANR-15-IDEX-02), through the funding of the "Origin of Life" project of the Univ. Grenoble-Alpes. ALS, CC, CCo, MB, EB and CF acknowledge the funding from the European Research Council (ERC) under the European Union’s Horizon 2020 research and innovation program, for the Project "The Dawn of Organic Chemistry" (DOC), grant agreement No 741002.  
\end{acknowledgements}

%
%

\bibliographystyle{aa} 
\bibliography{1st-chahine-hops-108.bib}

\begin{thebibliography}{72}
\expandafter\ifx\csname natexlab\endcsname\relax\def\natexlab#1{#1}\fi

\bibitem[{{Adams}(2010)}]{Adams2010}
{Adams}, F.~C. 2010, \araa, 48, 47

\bibitem[{{Ascenzi} {et~al.}(2019){Ascenzi}, {Cernuto}, {Balucani}, {Tosi},
  {Ceccarelli}, {Martini}, \& {Pirani}}]{Ascenzi2019}
{Ascenzi}, D., {Cernuto}, A., {Balucani}, N., {et~al.} 2019, \aap, 625, A72

\bibitem[{{Balucani} {et~al.}(2015){Balucani}, {Ceccarelli}, \&
  {Taquet}}]{Balucani2015}
{Balucani}, N., {Ceccarelli}, C., \& {Taquet}, V. 2015, \mnras, 449, L16

\bibitem[{{Belloche} {et~al.}(2020){Belloche}, {Maury}, {Maret}, {Anderl},
  {Bacmann}, {Andr{\'e}}, {Bontemps}, {Cabrit}, {Codella}, {Gaudel}, {Gueth},
  {Lef{\`e}vre}, {Lefloch}, {Podio}, \& {Testi}}]{Belloche2020}
{Belloche}, A., {Maury}, A.~J., {Maret}, S., {et~al.} 2020, \aap, 635, A198

\bibitem[{{Bergner} {et~al.}(2019){Bergner}, {Mart{\'\i}n-Dom{\'e}nech},
  {{\"O}berg}, {J{\o}rgensen}, {Artur de la Villarmois}, \&
  {Brinch}}]{Bergner2019}
{Bergner}, J.~B., {Mart{\'\i}n-Dom{\'e}nech}, R., {{\"O}berg}, K.~I., {et~al.}
  2019, ACS Earth and Space Chemistry, 3, 1564

\bibitem[{{Bianchi} {et~al.}(2020){Bianchi}, {Chandler}, {Ceccarelli},
  {Codella}, {Sakai}, {L{\'o}pez-Sepulcre}, {Maud}, {Moellenbrock}, {Svoboda},
  {Watanabe}, {Sakai}, {M{\'e}nard}, {Aikawa}, {Alves}, {Balucani}, {Bouvier},
  {Caselli}, {Caux}, {Charnley}, {Choudhury}, {De Simone}, {Dulieu},
  {Dur{\'a}n}, {Evans}, {Favre}, {Fedele}, {Feng}, {Fontani}, {Francis},
  {Hama}, {Hanawa}, {Herbst}, {Hirota}, {Imai}, {Isella}, {Jim{\'e}nez-Serra},
  {Johnstone}, {Kahane}, {Lefloch}, {Loinard}, {Maureira}, {Mercimek},
  {Miotello}, {Mori}, {Nakatani}, {Nomura}, {Oba}, {Ohashi}, {Okoda},
  {Ospina-Zamudio}, {Oya}, {Pineda}, {Podio}, {Rimola}, {Cox}, {Shirley},
  {Taquet}, {Testi}, {Vastel}, {Viti}, {Watanabe}, {Witzel}, {Xue}, {Zhang},
  {Zhao}, \& {Yamamoto}}]{Bianchi2020}
{Bianchi}, E., {Chandler}, C.~J., {Ceccarelli}, C., {et~al.} 2020, \mnras, 498,
  L87

\bibitem[{{Bianchi} {et~al.}(2017){Bianchi}, {Codella}, {Ceccarelli}, {Taquet},
  {Cabrit}, {Bacciotti}, {Bachiller}, {Chapillon}, {Gueth}, {Gusdorf},
  {Lefloch}, {Leurini}, {Podio}, {Rygl}, {Tabone}, \& {Tafalla}}]{Bianchi2017}
{Bianchi}, E., {Codella}, C., {Ceccarelli}, C., {et~al.} 2017, \aap, 606, L7

\bibitem[{{Bianchi} {et~al.}(2019){Bianchi}, {Codella}, {Ceccarelli}, {Vazart},
  {Bachiller}, {Balucani}, {Bouvier}, {De Simone}, {Enrique-Romero}, {Kahane},
  {Lefloch}, {L{\'o}pez-Sepulcre}, {Ospina-Zamudio}, {Podio}, \&
  {Taquet}}]{Bianchi2019}
{Bianchi}, E., {Codella}, C., {Ceccarelli}, C., {et~al.} 2019, \mnras, 483,
  1850

\bibitem[{{Brouillet} {et~al.}(2013){Brouillet}, {Despois}, {Baudry}, {Peng},
  {Favre}, {Wootten}, {Remijan}, {Wilson}, {Combes}, \&
  {Wlodarczak}}]{Brouillet2013}
{Brouillet}, N., {Despois}, D., {Baudry}, A., {et~al.} 2013, \aap, 550, A46

\bibitem[{{Calcutt} {et~al.}(2018){Calcutt}, {J{\o}rgensen}, {M{\"u}ller},
  {Kristensen}, {Coutens}, {Bourke}, {Garrod}, {Persson}, {van der Wiel}, {van
  Dishoeck}, \& {Wampfler}}]{Calcutt2018}
{Calcutt}, H., {J{\o}rgensen}, J.~K., {M{\"u}ller}, H.~S.~P., {et~al.} 2018,
  \aap, 616, A90

\bibitem[{{Caselli} \& {Ceccarelli}(2012)}]{Caselli2012}
{Caselli}, P. \& {Ceccarelli}, C. 2012, \aapr, 20, 56

\bibitem[{{Ceccarelli}(2004)}]{Ceccarelli2004}
{Ceccarelli}, C. 2004, in Astronomical Society of the Pacific Conference
  Series, Vol. 323, Star Formation in the Interstellar Medium: In Honor of
  David Hollenbach, ed. D.~{Johnstone}, F.~C. {Adams}, D.~N.~C. {Lin}, D.~A.
  {Neufeeld}, \& E.~C. {Ostriker}, 195

\bibitem[{{Ceccarelli} {et~al.}(2017){Ceccarelli}, {Caselli}, {Fontani},
  {Neri}, {L{\'o}pez-Sepulcre}, {Codella}, {Feng}, {Jim{\'e}nez-Serra},
  {Lefloch}, {Pineda}, {Vastel}, {Alves}, {Bachiller}, {Balucani}, {Bianchi},
  {Bizzocchi}, {Bottinelli}, {Caux}, {Chac{\'o}n-Tanarro}, {Choudhury},
  {Coutens}, {Dulieu}, {Favre}, {Hily-Blant}, {Holdship}, {Kahane}, {Jaber
  Al-Edhari}, {Laas}, {Ospina}, {Oya}, {Podio}, {Pon}, {Punanova}, {Quenard},
  {Rimola}, {Sakai}, {Sims}, {Spezzano}, {Taquet}, {Testi}, {Theul{\'e}},
  {Ugliengo}, {Vasyunin}, {Viti}, {Wiesenfeld}, \& {Yamamoto}}]{Ceccarelli2017}
{Ceccarelli}, C., {Caselli}, P., {Fontani}, F., {et~al.} 2017, \apj, 850, 176

\bibitem[{{Ceccarelli} {et~al.}(2007){Ceccarelli}, {Caselli}, {Herbst},
  {Tielens}, \& {Caux}}]{Ceccarelli2007}
{Ceccarelli}, C., {Caselli}, P., {Herbst}, E., {Tielens}, A.~G.~G.~M., \&
  {Caux}, E. 2007, in Protostars and Planets V, ed. B.~{Reipurth}, D.~{Jewitt},
  \& K.~{Keil}, 47

\bibitem[{{Ceccarelli} {et~al.}(2014){Ceccarelli}, {Dominik},
  {L{\'o}pez-Sepulcre}, {Kama}, {Padovani}, {Caux}, \&
  {Caselli}}]{Ceccarelli2014}
{Ceccarelli}, C., {Dominik}, C., {L{\'o}pez-Sepulcre}, A., {et~al.} 2014,
  \apjl, 790, L1

\bibitem[{{Ceccarelli} {et~al.}(2019){Ceccarelli}, {Favre},
  {L{\'o}pez-Sepulcre}, \& {Fontani}}]{Ceccarelli2019}
{Ceccarelli}, C., {Favre}, C., {L{\'o}pez-Sepulcre}, A., \& {Fontani}, F. 2019,
  Philosophical Transactions of the Royal Society of London Series A, 377,
  20180403

\bibitem[{{Ceccarelli} {et~al.}(2003){Ceccarelli}, {Maret}, {Tielens},
  {Castets}, \& {Caux}}]{Ceccarelli2003}
{Ceccarelli}, C., {Maret}, S., {Tielens}, A.~G.~G.~M., {Castets}, A., \&
  {Caux}, E. 2003, \aap, 410, 587

\bibitem[{{Crimier} {et~al.}(2009){Crimier}, {Ceccarelli}, {Lefloch}, \&
  {Faure}}]{Crimier2009}
{Crimier}, N., {Ceccarelli}, C., {Lefloch}, B., \& {Faure}, A. 2009, \aap, 506,
  1229

\bibitem[{{Dubernet} {et~al.}(2013){Dubernet}, {Alexander}, {Ba},
  {Balakrishnan}, {Balan{\c{c}}a}, {Ceccarelli}, {Cernicharo}, {Daniel},
  {Dayou}, {Doronin}, {Dumouchel}, {Faure}, {Feautrier}, {Flower}, {Grosjean},
  {Halvick}, {K{\l}os}, {Lique}, {McBane}, {Marinakis}, {Moreau}, {Moszynski},
  {Neufeld}, {Roueff}, {Schilke}, {Spielfiedel}, {Stancil}, {Stoecklin},
  {Tennyson}, {Yang}, {Vasserot}, \& {Wiesenfeld}}]{Dubernet2013}
{Dubernet}, M.~L., {Alexander}, M.~H., {Ba}, Y.~A., {et~al.} 2013, \aap, 553,
  A50

\bibitem[{{Favre} {et~al.}(2018){Favre}, {Ceccarelli}, {L{\'o}pez-Sepulcre},
  {Fontani}, {Neri}, {Manigand}, {Kama}, {Caselli}, {Jaber Al-Edhari},
  {Kahane}, {Alves}, {Balucani}, {Bianchi}, {Caux}, {Codella}, {Dulieu},
  {Pineda}, {Sims}, \& {Theul{\'e}}}]{Favre2018}
{Favre}, C., {Ceccarelli}, C., {L{\'o}pez-Sepulcre}, A., {et~al.} 2018, \apj,
  859, 136

\bibitem[{{Favre} {et~al.}(2017){Favre}, {L{\'o}pez-Sepulcre}, {Ceccarelli},
  {Dominik}, {Caselli}, {Caux}, {Fuente}, {Kama}, {Le Bourlot}, {Lefloch},
  {Lis}, {Montmerle}, {Padovani}, \& {Vastel}}]{Favre2017}
{Favre}, C., {L{\'o}pez-Sepulcre}, A., {Ceccarelli}, C., {et~al.} 2017, \aap,
  608, A82

\bibitem[{{Fontani} {et~al.}(2017){Fontani}, {Ceccarelli}, {Favre}, {Caselli},
  {Neri}, {Sims}, {Kahane}, {Alves}, {Balucani}, {Bianchi}, {Caux}, {Jaber
  Al-Edhari}, {Lopez-Sepulcre}, {Pineda}, {Bachiller}, {Bizzocchi},
  {Bottinelli}, {Chacon-Tanarro}, {Choudhury}, {Codella}, {Coutens}, {Dulieu},
  {Feng}, {Rimola}, {Hily-Blant}, {Holdship}, {Jimenez-Serra}, {Laas},
  {Lefloch}, {Oya}, {Podio}, {Pon}, {Punanova}, {Quenard}, {Sakai}, {Spezzano},
  {Taquet}, {Testi}, {Theul{\'e}}, {Ugliengo}, {Vastel}, {Vasyunin}, {Viti},
  {Yamamoto}, \& {Wiesenfeld}}]{Fontani2017}
{Fontani}, F., {Ceccarelli}, C., {Favre}, C., {et~al.} 2017, \aap, 605, A57

\bibitem[{{Furlan} {et~al.}(2016){Furlan}, {Fischer}, {Ali}, {Stutz}, {Stanke},
  {Tobin}, {Megeath}, {Osorio}, {Hartmann}, {Calvet}, {Poteet}, {Booker},
  {Manoj}, {Watson}, \& {Allen}}]{Furlan2016}
{Furlan}, E., {Fischer}, W.~J., {Ali}, B., {et~al.} 2016, \apjs, 224, 5

\bibitem[{{Furlan} {et~al.}(2014){Furlan}, {Megeath}, {Osorio}, {Stutz},
  {Fischer}, {Ali}, {Stanke}, {Manoj}, {Adams}, \& {Tobin}}]{Furlan2014}
{Furlan}, E., {Megeath}, S.~T., {Osorio}, M., {et~al.} 2014, \apj, 786, 26

\bibitem[{{Gonz{\'a}lez-Garc{\'\i}a} {et~al.}(2016){Gonz{\'a}lez-Garc{\'\i}a},
  {Manoj}, {Watson}, {Vavrek}, {Megeath}, {Stutz}, {Osorio}, {Wyrowski},
  {Fischer}, {Tobin}, {S{\'a}nchez-Portal}, {Diaz Rodriguez}, \&
  {Wilson}}]{Gonzalez-Garcia2016}
{Gonz{\'a}lez-Garc{\'\i}a}, B., {Manoj}, P., {Watson}, D.~M., {et~al.} 2016,
  \aap, 596, A26

\bibitem[{{Gounelle} {et~al.}(2013){Gounelle}, {Chaussidon}, \&
  {Rollion-Bard}}]{Gounelle2013}
{Gounelle}, M., {Chaussidon}, M., \& {Rollion-Bard}, C. 2013, \apjl, 763, L33

\bibitem[{{Green}(1986)}]{Green1986}
{Green}, S. 1986, \apj, 309, 331

\bibitem[{{Grossschedl} {et~al.}(2018){Grossschedl}, {Alves}, {Meingast},
  {Ackerl}, {Ascenso}, {Bouy}, {Burkert}, {Forbrich}, {Fuernkranz}, {Goodman},
  {Hacar}, {Herbst-Kiss}, {Lada}, {Larreina}, {Leschinski}, {Lombardi},
  {Moitinho}, {Mortimer}, \& {Zari}}]{Grossschedl2018}
{Grossschedl}, J.~E., {Alves}, J., {Meingast}, S., {et~al.} 2018, VizieR Online
  Data Catalog, J/A+A/619/A106

\bibitem[{{Herbst} \& {van Dishoeck}(2009)}]{Herbst2009}
{Herbst}, E. \& {van Dishoeck}, E.~F. 2009, \araa, 47, 427

\bibitem[{{Imai} {et~al.}(2016){Imai}, {Sakai}, {Oya}, {L{\'o}pez-Sepulcre},
  {Watanabe}, {Ceccarelli}, {Lefloch}, {Caux}, {Vastel}, {Kahane}, {Sakai},
  {Hirota}, {Aikawa}, \& {Yamamoto}}]{Imai2016}
{Imai}, M., {Sakai}, N., {Oya}, Y., {et~al.} 2016, \apjl, 830, L37

\bibitem[{{Jaber} {et~al.}(2014){Jaber}, {Ceccarelli}, {Kahane}, \&
  {Caux}}]{Jaber2014}
{Jaber}, A.~A., {Ceccarelli}, C., {Kahane}, C., \& {Caux}, E. 2014, \apj, 791,
  29

\bibitem[{{Jacobsen} {et~al.}(2019){Jacobsen}, {J{\o}rgensen}, {Di Francesco},
  {Evans}, {Choi}, \& {Lee}}]{Jacobsen2019}
{Jacobsen}, S.~K., {J{\o}rgensen}, J.~K., {Di Francesco}, J., {et~al.} 2019,
  \aap, 629, A29

\bibitem[{{Jin} \& {Garrod}(2020)}]{Jin&Garrod2020}
{Jin}, M. \& {Garrod}, R.~T. 2020, \apjs, 249, 26

\bibitem[{{J{\o}rgensen} {et~al.}(2020){J{\o}rgensen}, {Belloche}, \&
  {Garrod}}]{Jorgensen2020}
{J{\o}rgensen}, J.~K., {Belloche}, A., \& {Garrod}, R.~T. 2020, \araa, 58, 727

\bibitem[{{J{\o}rgensen} {et~al.}(2018){J{\o}rgensen}, {M{\"u}ller}, {Calcutt},
  {Coutens}, {Drozdovskaya}, {{\"O}berg}, {Persson}, {Taquet}, {van Dishoeck},
  \& {Wampfler}}]{Jorgensen2018}
{J{\o}rgensen}, J.~K., {M{\"u}ller}, H.~S.~P., {Calcutt}, H., {et~al.} 2018,
  \aap, 620, A170

\bibitem[{{J{\o}rgensen} {et~al.}(2016){J{\o}rgensen}, {van der Wiel},
  {Coutens}, {Lykke}, {M{\"u}ller}, {van Dishoeck}, {Calcutt}, {Bjerkeli},
  {Bourke}, {Drozdovskaya}, {Favre}, {Fayolle}, {Garrod}, {Jacobsen},
  {{\"O}berg}, {Persson}, \& {Wampfler}}]{Jorgensen2016}
{J{\o}rgensen}, J.~K., {van der Wiel}, M.~H.~D., {Coutens}, A., {et~al.} 2016,
  \aap, 595, A117

\bibitem[{{Kahane} {et~al.}(2018){Kahane}, {Jaber Al-Edhari}, {Ceccarelli},
  {L{\'o}pez-Sepulcre}, {Fontani}, \& {Kama}}]{Kahane2018}
{Kahane}, C., {Jaber Al-Edhari}, A., {Ceccarelli}, C., {et~al.} 2018, \apj,
  852, 130

\bibitem[{{Kainulainen} {et~al.}(2017){Kainulainen}, {Stutz}, {Stanke},
  {Abreu-Vicente}, {Beuther}, {Henning}, {Johnston}, \&
  {Megeath}}]{Kainulainen2017}
{Kainulainen}, J., {Stutz}, A.~M., {Stanke}, T., {et~al.} 2017, \aap, 600, A141

\bibitem[{{Kama} {et~al.}(2015){Kama}, {Caux}, {L{\'o}pez-Sepulcre}, {Wakelam},
  {Dominik}, {Ceccarelli}, {Lanza}, {Lique}, {Ochsendorf}, {Lis}, {Caballero},
  \& {Tielens}}]{Kama2015}
{Kama}, M., {Caux}, E., {L{\'o}pez-Sepulcre}, A., {et~al.} 2015, \aap, 574,
  A107

\bibitem[{{Kama} {et~al.}(2013){Kama}, {L{\'o}pez-Sepulcre}, {Dominik},
  {Ceccarelli}, {Fuente}, {Caux}, {Higgins}, {Tielens}, \&
  {Alonso-Albi}}]{Kama2013}
{Kama}, M., {L{\'o}pez-Sepulcre}, A., {Dominik}, C., {et~al.} 2013, \aap, 556,
  A57

\bibitem[{{Law} {et~al.}(2021){Law}, {Zhang}, {{\"O}berg}, {Galv{\'a}n-Madrid},
  {Keto}, {Liu}, \& {Ho}}]{Law2021}
{Law}, C.~J., {Zhang}, Q., {{\"O}berg}, K.~I., {et~al.} 2021, \apj, 909, 214

\bibitem[{{Lee} {et~al.}(2019){Lee}, {Codella}, {Li}, \& {Liu}}]{Lee2019}
{Lee}, C.-F., {Codella}, C., {Li}, Z.-Y., \& {Liu}, S.-Y. 2019, \apj, 876, 63

\bibitem[{{L{\'o}pez-Sepulcre}
  {et~al.}(2013{\natexlab{a}}){L{\'o}pez-Sepulcre}, {Kama}, {Ceccarelli},
  {Dominik}, {Caux}, {Fuente}, \& {Alonso-Albi}}]{Loepz-Sepulcre-and-kama2013}
{L{\'o}pez-Sepulcre}, A., {Kama}, M., {Ceccarelli}, C., {et~al.}
  2013{\natexlab{a}}, \aap, 549, A114

\bibitem[{{L{\'o}pez-Sepulcre} {et~al.}(2017){L{\'o}pez-Sepulcre}, {Sakai},
  {Neri}, {Imai}, {Oya}, {Ceccarelli}, {Higuchi}, {Aikawa}, {Bottinelli},
  {Caux}, {Hirota}, {Kahane}, {Lefloch}, {Vastel}, {Watanabe}, \&
  {Yamamoto}}]{Lopez-Sepulcre2017}
{L{\'o}pez-Sepulcre}, A., {Sakai}, N., {Neri}, R., {et~al.} 2017, \aap, 606,
  A121

\bibitem[{{L{\'o}pez-Sepulcre}
  {et~al.}(2013{\natexlab{b}}){L{\'o}pez-Sepulcre}, {Taquet},
  {S{\'a}nchez-Monge}, {Ceccarelli}, {Dominik}, {Kama}, {Caux}, {Fontani},
  {Fuente}, {Ho}, {Neri}, \& {Shimajiri}}]{Lopez-Sepulcre-and-taquet2013}
{L{\'o}pez-Sepulcre}, A., {Taquet}, V., {S{\'a}nchez-Monge}, {\'A}., {et~al.}
  2013{\natexlab{b}}, \aap, 556, A62

\bibitem[{{Lykke} {et~al.}(2017){Lykke}, {Coutens}, {J{\o}rgensen}, {van der
  Wiel}, {Garrod}, {M{\"u}ller}, {Bjerkeli}, {Bourke}, {Calcutt},
  {Drozdovskaya}, {Favre}, {Fayolle}, {Jacobsen}, {{\"O}berg}, {Persson}, {van
  Dishoeck}, \& {Wampfler}}]{Lykke2017}
{Lykke}, J.~M., {Coutens}, A., {J{\o}rgensen}, J.~K., {et~al.} 2017, \aap, 597,
  A53

\bibitem[{{Manigand} {et~al.}(2020){Manigand}, {J{\o}rgensen}, {Calcutt},
  {M{\"u}ller}, {Ligterink}, {Coutens}, {Drozdovskaya}, {van Dishoeck}, \&
  {Wampfler}}]{Manigand2020}
{Manigand}, S., {J{\o}rgensen}, J.~K., {Calcutt}, H., {et~al.} 2020, \aap, 635,
  A48

\bibitem[{{Marcelino} {et~al.}(2018){Marcelino}, {Gerin}, {Cernicharo},
  {Fuente}, {Wootten}, {Chapillon}, {Pety}, {Lis}, {Roueff}, {Commer{\c{c}}on},
  \& {Ciardi}}]{Marcelino2018}
{Marcelino}, N., {Gerin}, M., {Cernicharo}, J., {et~al.} 2018, \aap, 620, A80

\bibitem[{{Maret} {et~al.}(2011){Maret}, {Hily-Blant}, {Pety}, {Bardeau}, \&
  {Reynier}}]{Maret2011}
{Maret}, S., {Hily-Blant}, P., {Pety}, J., {Bardeau}, S., \& {Reynier}, E.
  2011, \aap, 526, A47

\bibitem[{{McMullin} {et~al.}(2007){McMullin}, {Waters}, {Schiebel}, {Young},
  \& {Golap}}]{McMullin2007}
{McMullin}, J.~P., {Waters}, B., {Schiebel}, D., {Young}, W., \& {Golap}, K.
  2007, in Astronomical Society of the Pacific Conference Series, Vol. 376,
  Astronomical Data Analysis Software and Systems XVI, ed. R.~A. {Shaw},
  F.~{Hill}, \& D.~J. {Bell}, 127

\bibitem[{{Megeath} {et~al.}(2012){Megeath}, {Gutermuth}, {Muzerolle},
  {Kryukova}, {Flaherty}, {Hora}, {Allen}, {Hartmann}, {Myers}, {Pipher},
  {Stauffer}, {Young}, \& {Fazio}}]{Megeath2012}
{Megeath}, S.~T., {Gutermuth}, R., {Muzerolle}, J., {et~al.} 2012, \aj, 144,
  192

\bibitem[{{Mezger} {et~al.}(1990){Mezger}, {Wink}, \& {Zylka}}]{Mezger1990}
{Mezger}, P.~G., {Wink}, J.~E., \& {Zylka}, R. 1990, \aap, 228, 95

\bibitem[{{M{\"u}ller} {et~al.}(2005){M{\"u}ller}, {Schl{\"o}der}, {Stutzki},
  \& {Winnewisser}}]{Muller2005}
{M{\"u}ller}, H. S.~P., {Schl{\"o}der}, F., {Stutzki}, J., \& {Winnewisser}, G.
  2005, Journal of Molecular Structure, 742, 215

\bibitem[{{M{\"u}ller} {et~al.}(2001){M{\"u}ller}, {Thorwirth}, {Roth}, \&
  {Winnewisser}}]{Muller2001}
{M{\"u}ller}, H.~S.~P., {Thorwirth}, S., {Roth}, D.~A., \& {Winnewisser}, G.
  2001, \aap, 370, L49

\bibitem[{{Nazari} {et~al.}(2021){Nazari}, {van Gelder}, {van Dishoeck},
  {Tabone}, {van't Hoff}, {Ligterink}, {Beuther}, {Boogert}, {Caratti o
  Garatti}, {Klaassen}, {Linnartz}, {Taquet}, \& {Tychoniec}}]{Nazari2021}
{Nazari}, P., {van Gelder}, M.~L., {van Dishoeck}, E.~F., {et~al.} 2021, \aap,
  650, A150

\bibitem[{{Osorio} {et~al.}(2017){Osorio}, {D{\'\i}az-Rodr{\'\i}guez},
  {Anglada}, {Megeath}, {Rodr{\'\i}guez}, {Tobin}, {Stutz}, {Furlan},
  {Fischer}, {Manoj}, {G{\'o}mez}, {Gonz{\'a}lez-Garc{\'\i}a}, {Stanke},
  {Watson}, {Loinard}, {Vavrek}, \& {Carrasco-Gonz{\'a}lez}}]{Osorio2017}
{Osorio}, M., {D{\'\i}az-Rodr{\'\i}guez}, A.~K., {Anglada}, G., {et~al.} 2017,
  \apj, 840, 36

\bibitem[{{Oya} {et~al.}(2017){Oya}, {Sakai}, {Watanabe}, {Higuchi}, {Hirota},
  {L{\'o}pez-Sepulcre}, {Sakai}, {Aikawa}, {Ceccarelli}, {Lefloch}, {Caux},
  {Vastel}, {Kahane}, \& {Yamamoto}}]{Oya2017}
{Oya}, Y., {Sakai}, N., {Watanabe}, Y., {et~al.} 2017, \apj, 837, 174

\bibitem[{{Pfalzner} {et~al.}(2015){Pfalzner}, {Davies}, {Gounelle},
  {Johansen}, {M{\"u}nker}, {Lacerda}, {Portegies Zwart}, {Testi}, {Trieloff},
  \& {Veras}}]{Pfalzner2015}
{Pfalzner}, S., {Davies}, M.~B., {Gounelle}, M., {et~al.} 2015, \physscr, 90,
  068001

\bibitem[{{Pickett} {et~al.}(1998){Pickett}, {Poynter}, {Cohen}, {Delitsky},
  {Pearson}, \& {M{\"u}ller}}]{Pickett1998}
{Pickett}, H.~M., {Poynter}, R.~L., {Cohen}, E.~A., {et~al.} 1998, \jqsrt, 60,
  883

\bibitem[{{Rabli} \& {Flower}(2010)}]{Rabli2010}
{Rabli}, D. \& {Flower}, D.~R. 2010, \mnras, 406, 95

\bibitem[{{Sakai} \& {Yamamoto}(2013)}]{Sakai2013}
{Sakai}, N. \& {Yamamoto}, S. 2013, Chemical Reviews, 113, 8981

\bibitem[{{Sch{\"o}ier} {et~al.}(2005){Sch{\"o}ier}, {van der Tak}, {van
  Dishoeck}, \& {Black}}]{Schoier2005}
{Sch{\"o}ier}, F.~L., {van der Tak}, F.~F.~S., {van Dishoeck}, E.~F., \&
  {Black}, J.~H. 2005, \aap, 432, 369

\bibitem[{{Shimajiri} {et~al.}(2008){Shimajiri}, {Takahashi}, {Takakuwa},
  {Saito}, \& {Kawabe}}]{Shimajiri2008}
{Shimajiri}, Y., {Takahashi}, S., {Takakuwa}, S., {Saito}, M., \& {Kawabe}, R.
  2008, \apj, 683, 255

\bibitem[{{Taquet} {et~al.}(2019){Taquet}, {Bianchi}, {Codella}, {Persson},
  {Ceccarelli}, {Cabrit}, {J{\o}rgensen}, {Kahane}, {L{\'o}pez-Sepulcre}, \&
  {Neri}}]{Taquet2019}
{Taquet}, V., {Bianchi}, E., {Codella}, C., {et~al.} 2019, \aap, 632, A19

\bibitem[{{Taquet} {et~al.}(2012){Taquet}, {Ceccarelli}, \&
  {Kahane}}]{Taquet2012}
{Taquet}, V., {Ceccarelli}, C., \& {Kahane}, C. 2012, \aap, 538, A42

\bibitem[{{Taquet} {et~al.}(2014){Taquet}, {Charnley}, \&
  {Sipil{\"a}}}]{Taquet2014}
{Taquet}, V., {Charnley}, S.~B., \& {Sipil{\"a}}, O. 2014, \apj, 791, 1

\bibitem[{{Taquet} {et~al.}(2015){Taquet}, {L{\'o}pez-Sepulcre}, {Ceccarelli},
  {Neri}, {Kahane}, \& {Charnley}}]{Taquet2015}
{Taquet}, V., {L{\'o}pez-Sepulcre}, A., {Ceccarelli}, C., {et~al.} 2015, \apj,
  804, 81

\bibitem[{{Tobin} {et~al.}(2019){Tobin}, {Megeath}, {van't Hoff},
  {D{\'\i}az-Rodr{\'\i}guez}, {Reynolds}, {Osorio}, {Anglada}, {Furlan},
  {Karnath}, {Offner}, {Sheehan}, {Sadavoy}, {Stutz}, {Fischer}, {Kama},
  {Persson}, {Di Francesco}, {Looney}, {Watson}, {Li}, {Stephens}, {Chandler},
  {Cox}, {Dunham}, {Kratter}, {Kounkel}, {Mazur}, {Murillo}, {Patel}, {Perez},
  {Segura-Cox}, {Sharma}, {Tychoniec}, \& {Wyrowski}}]{Tobin2019}
{Tobin}, J.~J., {Megeath}, S.~T., {van't Hoff}, M., {et~al.} 2019, \apj, 886, 6

\bibitem[{{van Gelder} {et~al.}(2020){van Gelder}, {Tabone}, {Tychoniec}, {van
  Dishoeck}, {Beuther}, {Boogert}, {Caratti o Garatti}, {Klaassen}, {Linnartz},
  {M{\"u}ller}, \& {Taquet}}]{vanGelder2020}
{van Gelder}, M.~L., {Tabone}, B., {Tychoniec}, {\L}., {et~al.} 2020, \aap,
  639, A87

\bibitem[{{Wilson} \& {Rood}(1994)}]{Wilson1994}
{Wilson}, T.~L. \& {Rood}, R. 1994, \araa, 32, 191

\bibitem[{{Yang} {et~al.}(2020){Yang}, {Evans}, {Smith}, {Lee}, {Tobin},
  {Terebey}, {Calcutt}, {J{\o}rgensen}, {Green}, \& {Bourke}}]{Yang2020}
{Yang}, Y.-L., {Evans}, Neal~J., I., {Smith}, A., {et~al.} 2020, \apj, 891, 61

\bibitem[{{Yang} {et~al.}(2021){Yang}, {Sakai}, {Zhang}, {Murillo}, {Zhang},
  {Higuchi}, {Zeng}, {L{\'o}pez-Sepulcre}, {Yamamoto}, {Lefloch}, {Bouvier},
  {Ceccarelli}, {Hirota}, {Imai}, {Oya}, {Sakai}, \& {Watanabe}}]{Peaches2021}
{Yang}, Y.-L., {Sakai}, N., {Zhang}, Y., {et~al.} 2021, \apj, 910, 20

\end{thebibliography}

\appendix

\section{Line Parameters}

\label{appA}
\onecolumn

{\footnotesize\setlength{\tabcolsep}{5pt}
    \begin{longtable}{l l c c c c c c c c}
       \caption{List of transitions and line parameters used in this work.}
       \label{transitions}
       
        \\
       
        \hline \hline
        
        $\mathrm{Molecule}$ &
       $\mathrm{Transition}$ & 
        $\mathrm{\nu} ^{a} $ &
        $\mathrm{E_{up}} ^{a}$ &
         $\mathrm{g_{up}} ^{a}$ &
        $\mathrm{A_{ij}} ^{a}$&
        $\mathrm{FWHM} ^{b}$ &
        $\mathrm{T_{peak}} ^{b}$ &
       $\mathrm{Line \; flux } ^{b}$ &
        $\mathrm {V_{peak}} ^{b}$
         \\
          
          & &
       
         (GHz) &
        (K) &
         &
        (10$^{-5}$ s$^{-1}$) &
        (km  s$^{-1}$) &
        (K) &
       (K km s$^{-1}$) &
       (km s$^{-1}$) \\  \hline \noalign{\vskip 0.05cm} 

\endfirsthead

\multicolumn{10}{c}%
{{\bfseries \tablename\ \thetable{} -- \textit{Continued}}} \\
           
\hline \hline
        
        $\mathrm{Molecule}$ &
        $\mathrm{Transition}$ & 
        $\mathrm{\nu} ^{a} $ &
        $\mathrm{E_{up}} ^{a}$ &
        $\mathrm{g_{up}} ^{a}$ &
        $\mathrm{A_{ij}} ^{a}$ &
        $\mathrm{FWHM} ^{b}$ &
         $\mathrm{T_{peak}} ^{b}$ &
        $\mathrm{Line \; flux } ^{b}$ &
        $\mathrm {V_{peak}} ^{b}$
         \\
          
          & &
       
         (GHz) &
        (K) &
         &
        (10$^{-5}$ s$^{-1}$) &
        (km  s$^{-1}$) &
        (K) &
       (K km s$^{-1}$) &
       (km s$^{-1}$) \\  \hline

\endhead        
 
 \multicolumn{10}{r}{{Continued on next page}} \\ \hline
\endfoot

\endlastfoot

 $\mathrm{HCOOCH_{3}} \, ^{c}$ &  $\mathrm{20_{17,3}-19_{17,2} \, A}$ & 245.5175 & 315.5 & 82 & 6.23 & \multirow{2}{*}{ 3.2 $\pm$ 0.4} & \multirow{2}{*}{4.29} & \multirow{2}{*}{14.6 $\pm$ 0.3}  & \multirow{2}{*}{ 12.2 $\pm$ 0.2}  \\
 \vspace{0.25cm}
 $\mathrm{HCOOCH_{3}} \, ^{c}$ &  $\mathrm{20_{17,4}-19_{17,3} \, A}$ & 245.5175 & 315.5 & 82 & 6.23 \\
  
$\mathrm{HCOOCH_{3}}$ & $\mathrm{20_{15,5}-19_{15,4} \, E}$ & 245.6567 & 273.1 & 82 & 9.83 & 2.7 $\pm$ 0.1 & 4.25 & 12.3 $\pm$ 0.4 & 12.3 $\pm$ 0.1 \\ \vspace{0.25cm}

$\mathrm{HCOOCH_{3}}$ & $\mathrm{20_{15,6}-19_{15,5} \, E}$ & 245.6729 & 273.1 & 82 & 9.84 & 2.6 $\pm$ 0.1 & 5.02 & 13.8 $\pm$ 0.5 & 12.3 $\pm$ 0.1 \\
 
 $\mathrm{HCOOCH_{3}}  \, ^{c}$ & $\mathrm{20_{14,6}-19_{14,5} \, A}$ & 245.7522 & 253.9 & 82 & 11.5 & \multirow{2}{*}{ 4.1 $\pm$ 0.1} & \multirow{2}{*}{7.54} & \multirow{2}{*}{32.8 $\pm$ 0.5} & \multirow{2}{*}{11.2 $\pm$ 0.1} \\ \vspace{0.25cm}
 $\mathrm{HCOOCH_{3}} \, ^{c}$ & $\mathrm{20_{14,7}-19_{14,6} \, A}$ & 245.7522 & 253.9 & 82 & 11.5 \\
 
 $\mathrm{HCOOCH_{3}}$ & $\mathrm{20_{14,7}-19_{14,6} \, E}$ & 245.7726 & 253.9 & 82 & 11.5  & 2.7 $\pm$ 0.1 & 6.64 & 18.9 $\pm$ 0.5 & 12.6 $\pm$ 0.1  \\
 
 $\mathrm{HCOOCH_{3}}$ & $\mathrm{21_{4,18}-20_{3,17} \, E}$ & 246.0275 & 139.8   & 86 & 2.37 & 2.6 $\pm$ 0.1 & 4.08 & 11.1 $\pm$ 0.4 & 12.4 $\pm$ 0.1 \\ 
     
$\mathrm{HCOOCH_{3}}$ & $\mathrm{21_{2,19}-20_{3,18} \, E}$ & 246.0388 & 139.8 & 86 & 2.37 & 2.5 $\pm$ 0.1 & 4.38 & 11.7 $\pm$ 0.4 & 12.5 $\pm$ 0.1  \\     \vspace{0.25cm}
 
 $\mathrm{HCOOCH_{3}}$ & $\mathrm{20_{11,9}-19_{11,8} \, E}$ & 246.2854 & 204.2 & 82      & 15.8  & 2.8 $\pm$ 0.2 & 6.05 & 18.3 $\pm$ 0.7 & 12.4 $\pm$ 0.1 \\ 

 $\mathrm{HCOOCH_{3}}  \, ^{c}$ & $\mathrm{20_{11,10}-19_{11,9} \,A}$ & 246.2951 & 204.2 & 82 & 15.8  & \multirow{2}{*}{3.2 $\pm$ 0.3} & \multirow{2}{*}{7.91} & \multirow{2}{*}{27.3 $\pm$ 1.8 }& \multirow{2}{*}{12.6 $\pm$ 0.1} \\
  \vspace{0.25cm}
   $\mathrm{HCOOCH_{3}} \, ^{c}$ & $\mathrm{20_{11,9}-19_{11,8} \,A}$ & 246.2951 & 204.2 & 82 & 15.8 \\
  
 $\mathrm{HCOOCH_{3}}$ & $\mathrm{20_{11,10}-19_{11,9} \, E}$ & 246.3082 & 204.2 & 82    & 15.8  & 2.8 $\pm$ 0.1 & 6.14 & 18.0 $\pm$ 0.7 & 12.3 $\pm$ 0.1  \\
  
$\mathrm{HCOOCH_{3}}$ & $\mathrm{22_{1,21}-21_{1,20} \,A , \nu_{t}=1 }$ & 246.4884 & 330.4 & 90 & 22.2 & 2.7 $\pm$ 0.1 & 4.86 & 13.8 $\pm$ 0.4 & 12.3 $\pm$ 0.1 \\  
  
$\mathrm{HCOOCH_{3}}$ & $\mathrm{20_{10,11}-19_{10,10} \, E}$ & 246.6231 & 190.3 & 82 & 17.4 & 2.6 $\pm$ 0.1 & 7.04 & 19.4 $\pm$ 0.7 & 12.3 $\pm$ 0.1  \\

$\mathrm{HCOOCH_{3}}$ & $\mathrm{15_{4,12}-14_{3,11} \, A}$ & 246.6834 & 81.8 & 62 & 1.17 & 2.8 $\pm$ 0.5 & 3.12 & 9.3 $\pm$ 1.3 & 12.4 $\pm$ 1.2 \\

$\mathrm{HCOOCH_{3}}$ & $\mathrm{22_{2,21}-21_{2,20} \, E , \nu_{t}=1}$ & 246.7065 & 329.9 & 90 & 22.3 & 2.7 $\pm$ 0.2 & 4.28 & 12.2 $\pm$ 0.6 & 12.4 $\pm$ 0.1 \\

$\mathrm{HCOOCH_{3}}$ & $\mathrm{22_{1,21}-21_{1,20} \,E, , \nu_{t}=1}$ & 246.7317 & 329.9 & 90 & 22.3  & 2.7 $\pm$ 0.1 & 5.22 & 14.8 $\pm$ 0.5 & 12.3 $\pm$ 0.1\\
  
$\mathrm{HCOOCH_{3}}$ & $\mathrm{10_{5,5}-9_{4,5} \, E}$ & 246.7529 & 49.1 & 42 & 1.25 & 2.2 $\pm$ 0.2 & 2.78 & 6.4 $\pm$ 0.7 & 12.3 $\pm$ 0.1 \\

$\mathrm{HCOOCH_{3}}$ & $\mathrm{19_{2,17}-18_{2,16} \, E}$ & 246.8916 & 126.2 & 78 & 21.8 & 3.3 $\pm$ 0.1 & 8.02 & 27.9 $\pm$ 0.7 & 12.4 $\pm$ 0.1  \\

$\mathrm{HCOOCH_{3}}$ & $\mathrm{19_{4,15}-18_{4,14} \, A}$ & 246.9146 & 126.2 & 78 & 21.8 & 3.0 $\pm$ 0.1 & 8.15 & 26.3 $\pm$ 0.8 & 12.3 $\pm$ 0.1  \\

$\mathrm{HCOOCH_{3}}$ & $\mathrm{10_{5,6}-9_{4,6} \, E}$ & 246.9457 & 49.1      & 42 & 1.26  & 2.4 $\pm$ 0.2 & 3.41 & 8.7 $\pm$ 0.7 & 12.2 $\pm$ 0.1 \\

$\mathrm{HCOOCH_{3}}$ & $\mathrm{20_{6,15}-19_{6,14} \,A , \nu_{t}=1}$ & 246.9852 & 335.4 & 82 & 20.7  & 2.4 $\pm$ 0.1 & 4.85 & 12.4 $\pm$ 0.6 & 12.4 $\pm$ 0.1 \\

$\mathrm{HCOOCH_{3}}$ & $\mathrm{20_{9,12}-19_{9,11} \, E}$ & 247.0636 & 177.8 & 82 & 18.2 & 2.7 $\pm$ 0.1 & 6.77 & 19.7 $\pm$ 0.9 & 12.2 $\pm$ 0.1 \\

$\mathrm{HCOOCH_{3}}$ & $\mathrm{10_{5,5}-9_{4,6} \, A}$ & 247.1242 & 49.1 & 42 & 1.82 & 3.1 $\pm$ 0.3 & 3.20 & 10.5 $\pm$ 0.5 & 12.4 $\pm$ 0.1  \\

$\mathrm{HCOOCH_{3}}$ & $\mathrm{20_{9,11}-20_{8,12} \, E}$ & 231.9552 & 177.8 & 82 & 1.56 & 2.5 $\pm$ 0.5 & 2.28 & 6.1 $\pm$ 1.0 & 12.1 $\pm$ 0.2 \\
         
$\mathrm{HCOOCH_{3}}$ & $\mathrm{19_{9,10}-19_{8,11} \, A}$ & 232.6172 & 166.0 & 78 & 1.53 & 2.2 $\pm$ 0.3 & 3.15 & 7.5 $\pm$ 0.8 & 12.2 $\pm$ 0.2 \\

$\mathrm{HCOOCH_{3}}$ & $\mathrm{19_{9,11}-19_{8,12} \, A}$ & 232.6252 & 166.0 & 78 & 1.53 & 2.4 $\pm$ 0.3 & 2.77 & 7.1 $\pm$ 0.8 & 12.4 $\pm$ 0.2 \\

$\mathrm{HCOOCH_{3}}$ & $\mathrm{19_{10,10}-18_{10,9} \,E , \nu_{t}=1}$ & 232.6839 & 365.5 & 78 & 13.8 & 2.2 $\pm$ 0.3 & 3.19 & 7.4 $\pm$ 0.8 & 12.4 $\pm$ 0.1 \\ 

$\mathrm{HCOOCH_{3}}$ & $\mathrm{19_{8,11}-18_{8,10} \,E , \nu_{t}=1}$& 232.7386 & 342.0 & 78 & 15.7 & 2.0 $\pm$ 0.3 & 4.07 & 8.7 $\pm$ 1.0 & 12.4 $\pm$ 0.1 \\

$\mathrm{HCOOCH_{3}}$ & $\mathrm{19_{8,11}-18_{8,10} \, A , \nu_{t}=1}$ & 232.8396 & 341.8 & 78 & 15.7 & 2.4 $\pm$ 0.3 & 3.32 & 8.5 $\pm$ 0.9 & 12.3 $\pm$ 0.1 \\

\hline \noalign{\vskip 0.05cm} 

$\mathrm{CH_{3}OH}$ & 18$\rm{_{3, 16}}-$  17 $\rm{_{4, 13} \, A}$ &232.7834 & 446.5 & 148 & 2.17 & 3.5 $\pm$ 0.2 & 6.94 & 26.2 $\pm$ 1.3 & 12.4 $\pm$ 0.1 \\
$\mathrm{CH_{3}OH}$ & 10$\rm{_{-3, 8}}- $11$\rm{_{-2, 10}} \, E$ & 232.9457 & 190.4 & 84 & 2.13 & 4.3 $\pm$ 0.2 & 9.50 & 43.1 $\pm$ 1.6 & 12.7 $\pm$ 0.1 \\
$\mathrm{CH_{3}OH}$ & 19$\rm{_{3, 16}}-$ 19$\rm{_{2, 17}} \, E$ & 246.8733 & 490.7 & 156 & 8.27 & 4.2 $\pm$ 0.1 & 11.24 & 49.8 $\pm$ 0.7 & 12.4 $\pm$ 0.1 \\    

\hline  \noalign{\vskip 0.05cm} 

\vspace{0.25cm}
$\mathrm{CH_{3}^{18}OH} \, ^{d}$ & 5$\rm{_{-1, 5}}-$ 4$\rm{_{-1, 4}} \, E$ & 231.7358 & 39.0 & 44 & 5.13 & 4.3 $\pm$ 0.8 & 1.75 & 8.0 $\pm$ 1.3 & 12.1 $\pm$ 0.4 \\

$\mathrm{CH_{3}^{18}OH} \, ^{c,d,*}$ & 5$\rm{_{3, 2}}-$ 4$\rm{_{3, 1}} \, E$ & 231.8013 & 81.3 & 44 & 3.42 & \multirow{2}{*}{3.8 $\pm$ 0.8} & \multirow{2}{*}{2.69} & \multirow{2}{*}{10.9 $\pm$ 1.8} & \multirow{2}{*}{11.4 $\pm$ 0.4} \\ \vspace{0.25cm}
$\mathrm{CH_{3}^{18}OH} \, ^{c,d,*}$ & 5$\rm{_{2, 4}}-$ 4$\rm{_{2, 3}} \, A$ & 231.8015 & 70.9 & 44 & 4.53 \\

$\mathrm{CH_{3}^{18}OH}$ & 5$\rm{_{1, 4}}-$ 4$\rm{_{1, 3}} \, E$ & 231.8267 & 54.1 & 44 & 5.33 & 2.7 $\pm$ 0.7 & 1.46 & 4.1 $\pm$ 0.9 & 12.3 $\pm$ 0.3 \\

$\mathrm{CH_{3}^{18}OH} \, ^{e}$ & 5$\rm{_{-2, 4}}-$ 4$\rm{_{-2, 3}} \, E$ & 231.8538 & 59.2 & 44 & 4.49 & 3.8 $\pm$ 1.1 & 1.40 & 5.6 $\pm$ 1.3 & 10.5 $\pm$ 0.4 \\

$\mathrm{CH_{3}^{18}OH}$ & 5$\rm{_{2, 3}}-$ 4$\rm{_{2, 2}} \, E$ & 231.8645 & 55.8 & 44 & 4.41 & 3.5 $\pm$ 0.8 & 1.30 & 4.9 $\pm$ 1.0 & 11.3 $\pm$ 0.4 \\

\hline \noalign{\vskip 0.05cm} 

$\mathrm{CH_{2}DOH}$ & 4$\rm{_{1, 4}}-$ 4$\rm{_{1, 3}} \, E1$ & 246.9731 & 37.7 & 9 & 2.15 & 2.9 $\pm$ 0.2 & 3.48 & 10.6 $\pm$ 0.6 & 12.4 $\pm$ 0.1 \\

\hline \noalign{\vskip 0.05cm} 

$\mathrm{CH_{3}OCH_{3}} \, ^{c,d,e}$ & 13$\rm{_{0, 13}}-$ 12$\rm{_{1, 12}} \, AA$ & 231.9878 & 80.9 & 270 & 9.15 & \multirow{4}{*}{2.9 $\pm$ 0.2} & \multirow{4}{*}{6.52} & \multirow{4}{*}{20.3 $\pm$ 1.1} & \multirow{4}{*}{12.3 $\pm$ 0.1} \\ 
$\mathrm{CH_{3}OCH_{3}}\, ^{c,d,e}$ & 13$\rm{_{0, 13}}-$ 12$\rm{_{1, 12}} \, EE$ & 231.9879 & 80.9 & 432 & 9.15 \\
$\mathrm{CH_{3}OCH_{3}} \, ^{c,d,e}$ & 13$\rm{_{0, 13}}-$ 12$\rm{_{1, 12}} \, AE$ & 231.9879 & 80.9 & 162 & 9.15 \\ \vspace{0.25cm}
$\mathrm{CH_{3}OCH_{3}} \, ^{c,d,e}$ & 13$\rm{_{0, 13}}-$ 12$\rm{_{1, 12}} \, EA$ & 231.9879 & 80.9 & 108 & 9.15 \\

$\mathrm{CH_{3}OCH_{3}} \, ^{f}$ & 17$\rm{_{5, 12}}-$ 17$\rm{_{4, 13}} \, AE$ & 259.3094 & 174.5 & 140 & 8.74 & \multirow{2}{*}{2.1 $\pm$ 0.3} & \multirow{2}{*}{3.83} & \multirow{2}{*}{8.6 $\pm$ 1.1} & \multirow{2}{*}{12.2 $\pm$ 0.1} \\ \vspace{0.25cm}
$\mathrm{CH_{3}OCH_{3}}\, ^{f}$ & 17$\rm{_{5, 12}}-$ 17$\rm{_{4, 13}} \, EA$ & 259.3097 & 174.5 & 140 & 8.74 \\

$\mathrm{CH_{3}OCH_{3}} \, ^{f}$ & 17$\rm{_{5, 12}}-$ 17$\rm{_{4, 13}} \, EE$  & 259.3119 & 174.5 & 560 & 8.76 & 1.6 $\pm$ 0.2 & 5.20 & 9.0 $\pm$ 1.0 & 12.1 $\pm$ 0.1 \\

\hline  \noalign{\vskip 0.05cm}       

$\mathrm{CH_{3}COCH_{3}}  \, ^{c}$ & 21$\rm{_{4, 17}}-$ 20$\rm{_{5, 16}} \, AE $ & 245.2965 & 145.1 & 258 & 53.9 & \multirow{4}{*}{ 2.8 $\pm$ 0.9} & \multirow{4}{*}{1.85} & \multirow{4}{*}{ 5.4 $\pm$ 1.4} & \multirow{4}{*}{12.5 $\pm$ 0.4} \\
$\mathrm{CH_{3}COCH_{3}} \, ^{c}$ & 21$\rm{_{5, 17}}-$ 20$\rm{_{4, 16}} \, AE$ & 245.2965 & 145.1 & 86 & 53.9 \\
$\mathrm{CH_{3}COCH_{3}} \, ^{c}$ & 21$\rm{_{4, 17}}-$ 20$\rm{_{5, 16}} \, EA $ & 245.2965 & 145.1 & 172 & 53.9 \\ \vspace{0.25cm}
$\mathrm{CH_{3}COCH_{3}}\, ^{c}$ & 21$\rm{_{5, 17}}-$ 20$\rm{_{4, 16}} \, EA $ & 245.2965 & 145.1 & 172 & 53.9 \\

$\mathrm{CH_{3}COCH_{3}} \, ^{c}$ & 21$\rm{_{4, 17}}-$ 20$\rm{_{5, 16}} \, AA $ & 245.4095 & 145.0 & 430 & 53.9 & \multirow{2}{*}{ 2.8 $\pm$ 0.2} & \multirow{2}{*}{2.00} & \multirow{2}{*}{ 5.9 $\pm$ 0.5} & \multirow{2}{*}{ 12.4 $\pm$ 0.1} \\ \vspace{0.25cm}
$\mathrm{CH_{3}COCH_{3}}\, ^{c}$ & 21$\rm{_{5, 17}}-$ 20$\rm{_{4, 16}} \, AA $ & 245.4095 & 145.0 & 258 & 53.9 \\ \\

$\mathrm{CH_{3}COCH_{3}} \, ^{c}$ & 22$\rm{_{3, 19}}-$ 21$\rm{_{4, 18}} \, EE $ & 246.4504 & 149.6 & 720 & 58.2 & \multirow{4}{*}{ 2.6 $\pm$ 0.1} & \multirow{4}{*}{3.64} & \multirow{4}{*}{10.3 $\pm$ 0.5} & \multirow{4}{*}{12.2 $\pm$ 0.1 }\\
$\mathrm{CH_{3}COCH_{3}} \, ^{c}$ & 22$\rm{_{4, 19}}-$ 21$\rm{_{4, 18}} \, EE $ & 246.4504 & 149.6 & 720 & 1.2 \\
$\mathrm{CH_{3}COCH_{3}} \, ^{c}$ & 22$\rm{_{3, 19}}-$ 21$\rm{_{3, 18}} \, EE $ & 246.4504 & 149.6 & 720 & 0.81 \\        
$\mathrm{CH_{3}COCH_{3}} \, ^{c}$ & 22$\rm{_{4, 19}}-$ 21$\rm{_{3, 18}} \, EE $ & 246.4504 & 149.6 & 720 & 58.6 \\         
        
\hline \noalign{\vskip 0.05cm} 

$\mathrm{CH_{3}CHO} \, ^{d}$ & 12$\rm{_{3,10}}-$ 11$\rm{_{3, 9}} \, E $ & 231.7487 & 92.5 & 50 & 39.4 & 3.9 $\pm$ 0.8 & 1.46 & 5.7 $\pm$ 0.9 & 12.4 $\pm$ 0.4 \\

$\mathrm{CH_{3}CHO} \, ^{d}$ & 12$\rm{_{3,9}}-$ 11$\rm{_{3, 8}} \, E$ & 231.8475 & 92.6 & 50 & 39.4 & 3.1 $\pm$ 0.4 & 2.26 & 7.5 $\pm$ 1.0 & 13.1 $\pm$ 0.2 \\

\hline \noalign{\vskip 0.05cm} 

$\mathrm{CH_{3}CN \, ^{f}}$ & 14$\rm{_{1}}-$ 13$\rm{_{1}}$ & 257.5224 & 99.8 & 58 & 147 & 4.6 $\pm$ 0.1 & 18.2 & 89.2 $\pm$ 2.1 & 12.3  $\pm$ 0.1 \\       

$\mathrm{CH_{3}CN \, ^{f}}$ & 14$\rm{_{0}}-$ 13$\rm{_{0}}$ & 257.5274 & 92.7 & 58 & 148 & 3.9 $\pm$ 0.1 & 19.05 & 81.8 $\pm$ 2.7 & 12.3  $\pm$ 0.1 \\     

\hline \noalign{\vskip 0.05cm} 

$\mathrm{^{13}CH_{3}CN}$ & 13$\rm{_{3}}-$ 12$\rm{_{3}}$ & 232.1949 & 142.4 & 108 & 102 & 2.5 $\pm$ 0.3 & 3.51 & 9.4 $\pm$ 1.1 & 12.5 $\pm$ 0.1  \\
$\mathrm{^{13}CH_{3}CN}$ & 13$\rm{_{1}}-$ 12$\rm{_{1}}$ & 232.2298 & 85.2 & 54 & 107 & 2.4$\pm$ 0.3 & 3.34 & 8.4 $\pm$ 1.1 & 12.2 $\pm$ 0.2 \\
$\mathrm{^{13}CH_{3}CN}$ & 13$\rm{_{0}}-$ 12$\rm{_{0}}$ & 232.2341 & 78.0 & 54 & 108 & 2.4 $\pm$ 0.4 & 3.35 & 8.5 $\pm$ 1.3 & 12.3 $\pm$ 0.2 \\

\hline \noalign{\vskip 0.05cm} 

\vspace{0.25cm}
$\mathrm{C_{2}H_{5}CN}$ & 26$\rm{_{3,24}}-$ 25$\rm{_{3, 23}}$  & 232.7900 & 161.0 & 53 & 105 & 3.3 $\pm$ 0.8 & 1.44 & 5.0 $\pm$ 1.1 & 12.3 $\pm$ 0.4 \\

$\mathrm{C_{2}H_{5}CN} \, ^{c}$ & 26$\rm{_{10,16}}-$ 25$\rm{_{10, 15}}$ & 232.9623 & 262.0 & 53 & 91.2 & \multirow{2}{*}{ 2.3 $\pm$ 0.6} & \multirow{2}{*}{1.98} & \multirow{2}{*}{4.8 $\pm$ 1.0} & \multirow{2}{*}{12.4 $\pm$ 0.2} \\ \vspace{0.25cm}
$\mathrm{C_{2}H_{5}CN} \, ^{c}$ & 26$\rm{_{10,17}}-$ 25$\rm{_{10, 16}}$ & 232.9623 & 262.0 & 53 & 91.2 \\

$\mathrm{C_{2}H_{5}CN} \, ^{c}$ & 26$\rm{_{9,17}}-$ 25$\rm{_{9, 16}}$ & 232.9675 & 240.9 & 53 & 94.2 &  \multirow{2}{*}{  2.8 $\pm$ 0.7} &  \multirow{2}{*}{1.61} &  \multirow{2}{*}{4.8 $\pm$ 1.0} &  \multirow{2}{*}{12.4 $\pm$ 0.3} \\ \vspace{0.25cm}
$\mathrm{C_{2}H_{5}CN} \, ^{c}$ & 26$\rm{_{9,18}}-$ 25$\rm{_{9, 17}}$ & 232.9675 & 240.9 & 53 & 94.2 \\

$\mathrm{C_{2}H_{5}CN} \, ^{c}$ & 26$\rm{_{11,15}}-$ 25$\rm{_{11, 14}}$ & 232.9755 & 285.2 & 53 & 87.9 &  \multirow{2}{*}{1.6 $\pm$ 1.0} &  \multirow{2}{*}{2.15} &  \multirow{2}{*}{3.7 $\pm$ 0.8} &  \multirow{2}{*}{12.3 $\pm$ 0.1} \\ \vspace{0.25cm}
$\mathrm{C_{2}H_{5}CN} \, ^{c}$ & 26$\rm{_{11,16}}-$ 25$\rm{_{11, 15}}$ & 232.9755 & 285.2 & 53 & 87.9 \\

$\mathrm{C_{2}H_{5}CN} \, ^{c}$ & 26$\rm{_{8,19}}-$ 25$\rm{_{8, 18}}$ & 232.9987 & 222.0 & 53 & 96.9 & \multirow{2}{*}{2.7 $\pm$ 0.4} & \multirow{2}{*}{2.29} & \multirow{2}{*}{6.7 $\pm$ 0.9} & \multirow{2}{*}{12.7 $\pm$ 0.2} \\ \vspace{0.25cm}
$\mathrm{C_{2}H_{5}CN} \, ^{c}$ & 26$\rm{_{8,18}}-$ 25$\rm{_{8, 17}}$ & 232.9987 & 222.0 & 53 & 96.9 \\

$\mathrm{C_{2}H_{5}CN} \, ^{c}$ &  26$\rm{_{7,20}}-$ 25$\rm{_{7, 19}}$ & 233.0693 & 205.4 & 53 & 99.4 & \multirow{2}{*}{2.7 $\pm$ 0.5} & \multirow{2}{*}{2.73} & \multirow{2}{*}{7.8 $\pm$ 1.2} & \multirow{2}{*}{12.5 $\pm$ 0.3} \\ \vspace{0.25cm}
$\mathrm{C_{2}H_{5}CN} \, ^{c}$ &  26$\rm{_{7,19}}-$ 25$\rm{_{7, 18}}$ & 233.0693 & 205.4 & 53 & 99.4 \\

\hline \noalign{\vskip 0.05cm} 

$\mathrm{NH_{2}CHO}$ & 11$\rm{_{2,10}}-$ 10$\rm{_{2, 9}}$ & 232.2736 & 78.9 & 23 & 88.2 & 3.5 $\pm$ 0.4 & 2.56 & 9.6 $\pm$ 1.0 & 12.0 $\pm$ 0.2 \\

$\mathrm{NH_{2}CHO \, ^{f}}$ & 12$\rm{_{1,12}}-$ 11$\rm{_{1, 11}}$ & 243.5210 & 79.2 & 25 & 105 & 1.7 $\pm$ 0.2 & 5.71 & 10.1 $\pm$ 0.9 & 13.0 $\pm$ 0.1 \\ 

\hline

    \end{longtable}
    
\noindent
\textit{\textbf{Notes:}} \newline $^{a}$ The frequencies and spectroscopic parameters of $\mathrm{HCOOCH_{3}}$, $\mathrm{CH_{2}DOH}$, $\mathrm{CH_{3}COCH_{3}}$, and $\mathrm{CH_{3}CHO}$ have been extracted from the JPL catalogue \citep{Pickett1998}. The frequencies and spectroscopic parameters of $\mathrm{CH_{3}OH}$, $\mathrm{CH_{3}^{18}OH}$, $\mathrm{CH_{3}OCH_{3}}$, $\mathrm{CH_{3}CN}$, $\mathrm{^{13}CH_{3}CN}$, $\mathrm{C_{2}H_{5}CN}$, and $\mathrm{NH_{2}CHO}$ have been extracted from the CDMS catalogue \citep{Muller2001,Muller2005}.  \newline $^{b}$ Parameters and uncertainties determined by Gaussian fit. \newline $^{c}$ The lines cannot be distinguished. \newline $^{d}$ Estimated after subtracting the small contribution from $\mathrm{HCOOCH_{3}}$. \newline $^{e}$ Estimated after subtracting the small contribution from $\mathrm{C_{2}H_{5}CN}$. \newline $^{f}$ Lines from narrow spws (58.59 MHz) and different channel width ($\sim$ 0.15 km s$^{-1})$ (see Table. \ref{spw}).\newline $^{*}$ The collisional coefficients have not been calculated for the E transition so we used only the A transition in the LVG. We assumed a flux value equal to 40\% of the total line flux because the A transition would be less populated with respect to the E one given that it has lower E$_{\rm up}$ and A$\rm ij$. However, the result does not change much if we assume a value between 30\% and 60\% of the total flux.}
    

\newpage
\section{Figures}

\label{appB}

For each studied molecule, we provide a sample of spectra of the detected lines. The transitions of CH$_{3}$CHO, CH$_{3}$OCH$_{3}$, and CH$_{3}^{18}$OH that were slightly contaminated by either HCOOCH$_{3}$ or C$_{2}$H$_{5}$CN (by $<$20$\%$ of the total line flux) are reported together with the expected intensities of the contaminating lines. The corresponding intensities were obtained with the LTE model of the CLASS software using the column densities and rotational temperatures obtained in our analysis, and they were removed after from the CH$_{3}$CHO, CH$_{3}$OCH$_{3}$, and CH$_{3}^{18}$OH lines using the 'subtract' command in CLASS.

\begin{figure}[hbt!]
    \centering
    \includegraphics[width=1\textwidth]{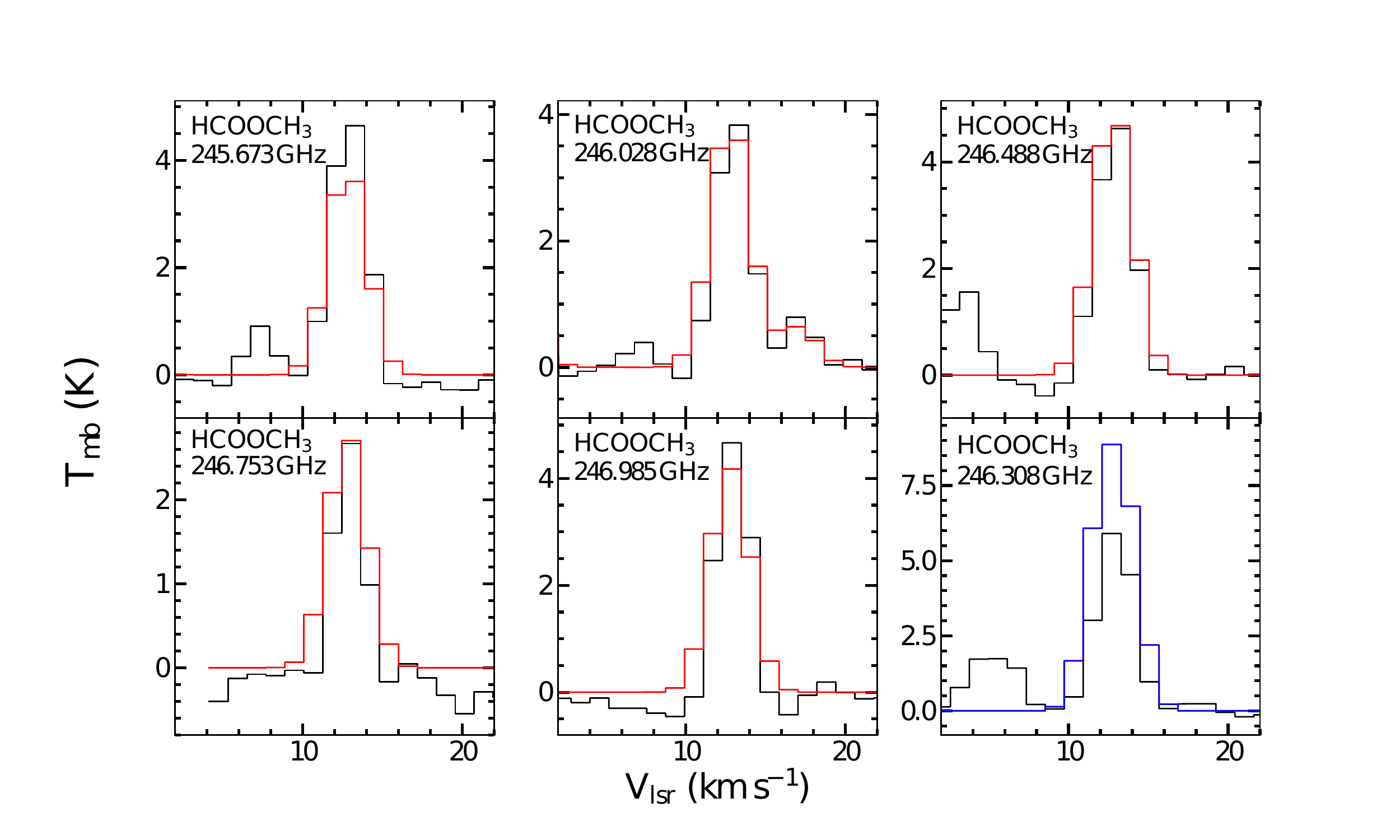}
    \caption{Sample of some observed HCOOCH$\rm_{3}$ spectral lines towards HOPS-108 (black). The spectra predicted by the best-fit LTE model are depicted in red. The blue spectrum corresponds to the best-fit model prediction of an optically thick line, which was excluded from the analysis (see Sect. \ref{lte}).}
    \label{mf-spectra}
\end{figure}

\begin{figure}[hbt!]
    \centering
    \includegraphics[width=1\textwidth]{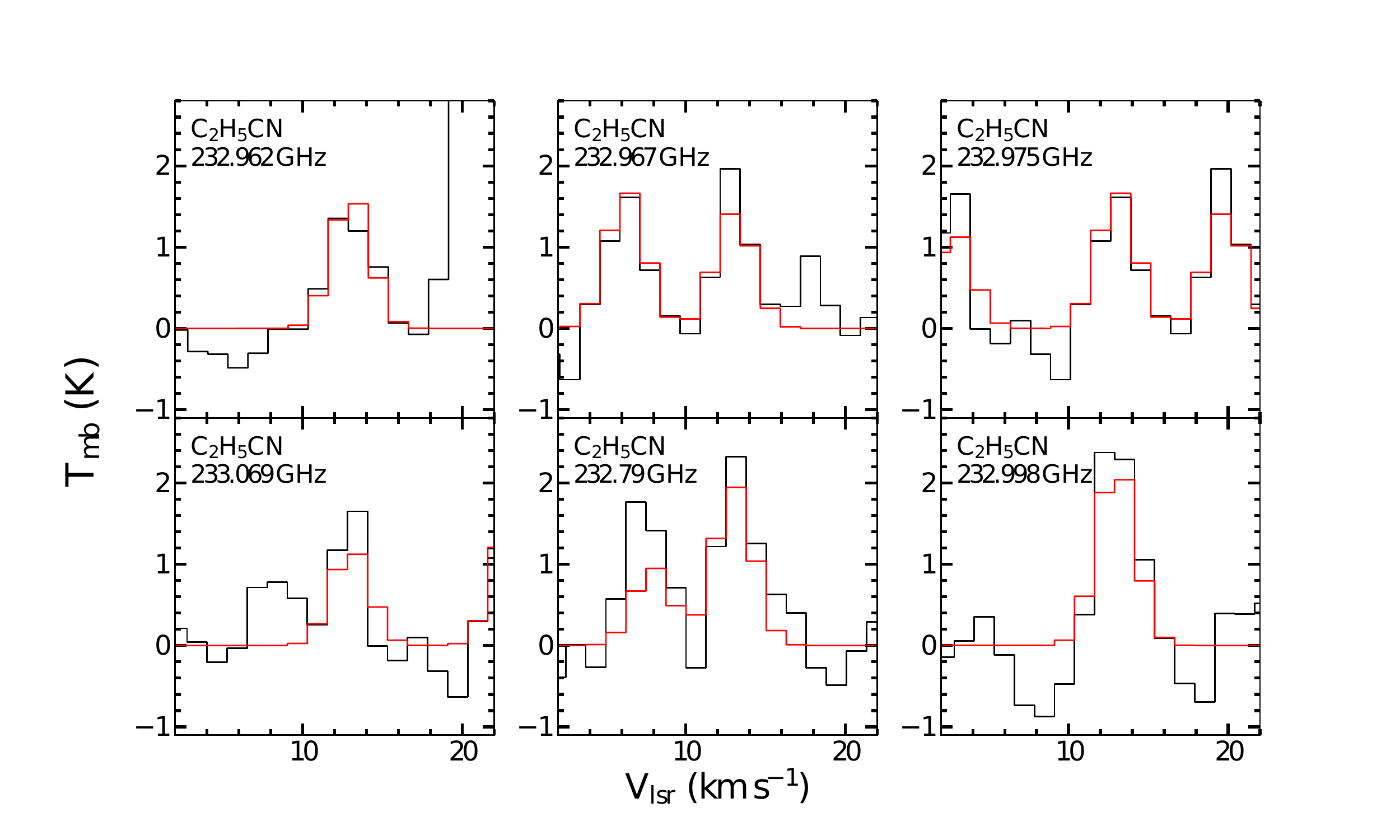}
    \caption{C$\rm_{2}$H$\rm_{5}$CN observed spectral lines (black) towards HOPS-108, and the spectra predicted by the best-fit LTE model (red).}
    \label{ec-spec}
\end{figure}

\begin{figure}[hbt!]
    \centering
    \includegraphics[width=0.25\textwidth]{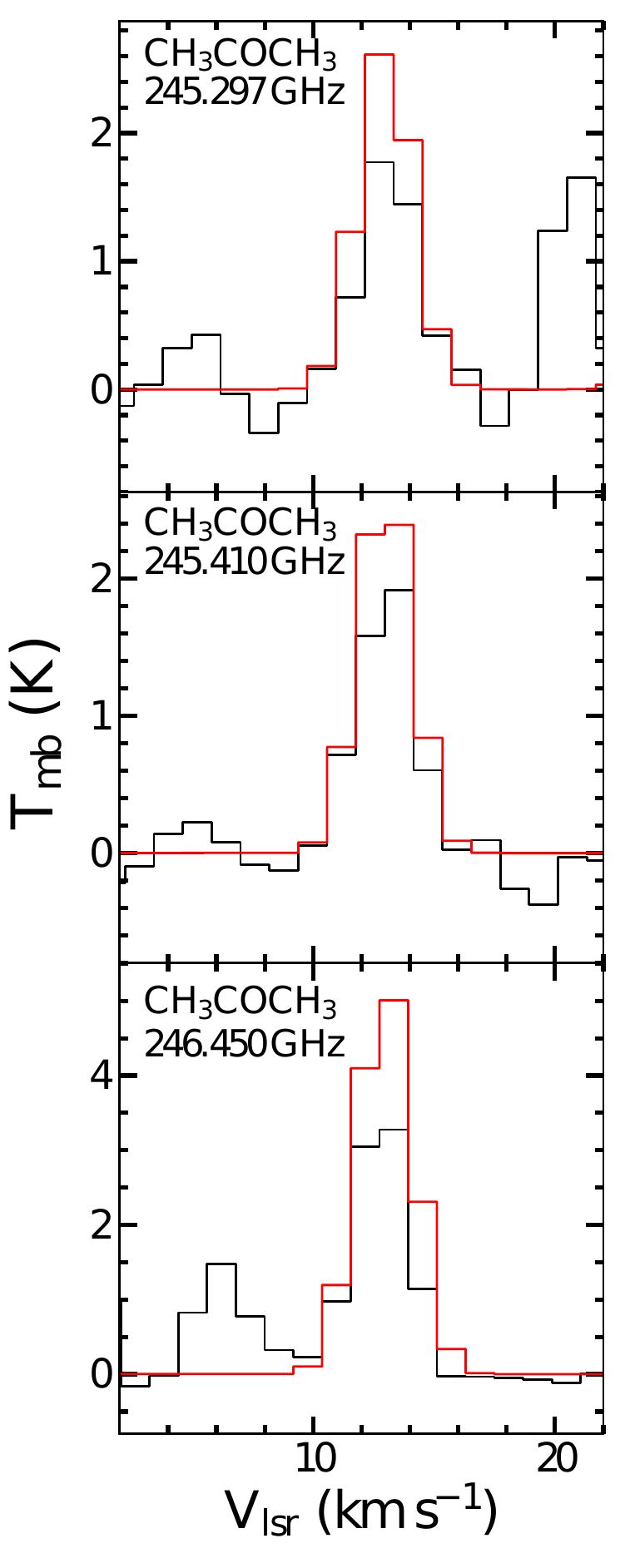}
    \includegraphics[width=0.25\textwidth]{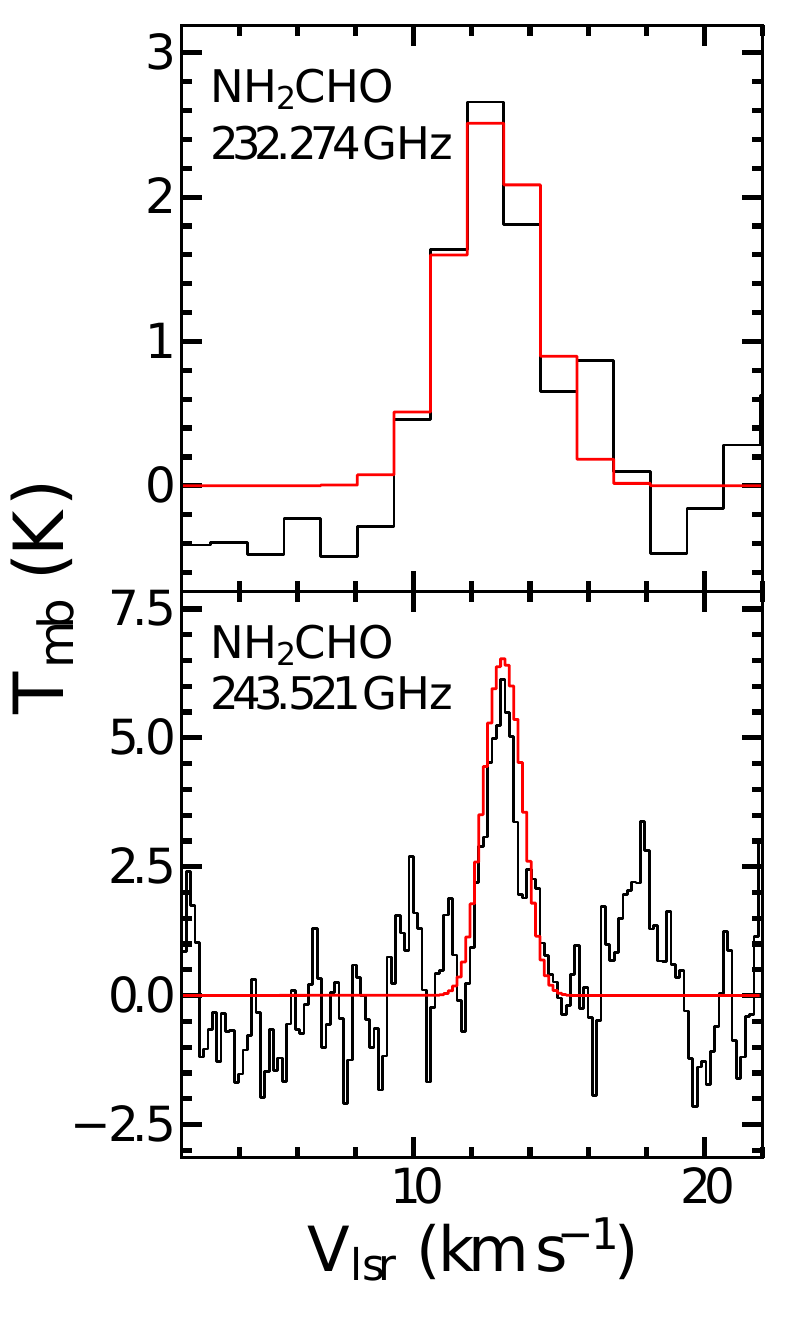}
    \includegraphics[width=0.25\textwidth]{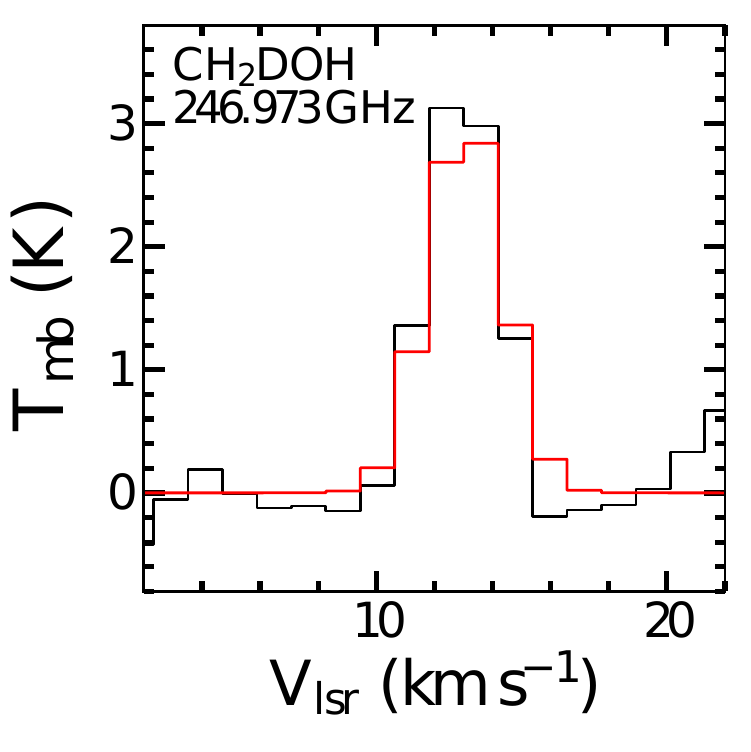}
    
    \caption{CH$\rm_{3}$COCH$\rm_{3}$, NH$\rm_{2}$CHO, and CH$\rm_{2}$DOH observed spectral lines (black) towards HOPS-108. The spectra predicted by the best-fit LTE model are depicted in red.}
    \label{other-spec}
\end{figure}

\begin{figure}[hbt!]
    \centering
    \includegraphics[width=0.3\textwidth]{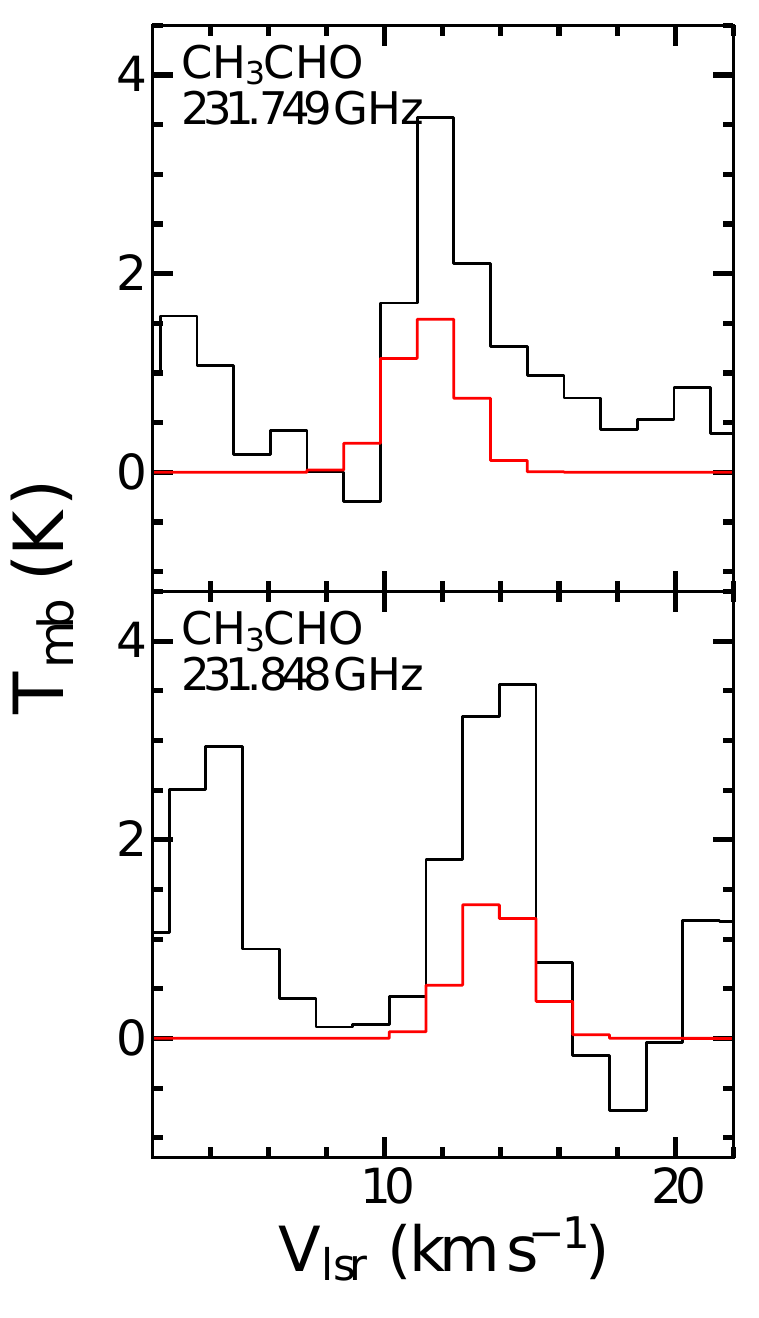}

   \caption{CH$\rm_{3}$CHO observed spectral lines (black) towards HOPS-108. The contamination from HCOOCH$_{3}$ is shown in red.}
    \label{dme-acetal}
\end{figure}

\begin{figure}[hbt!]
    \centering
    \includegraphics[width=0.8\textwidth]{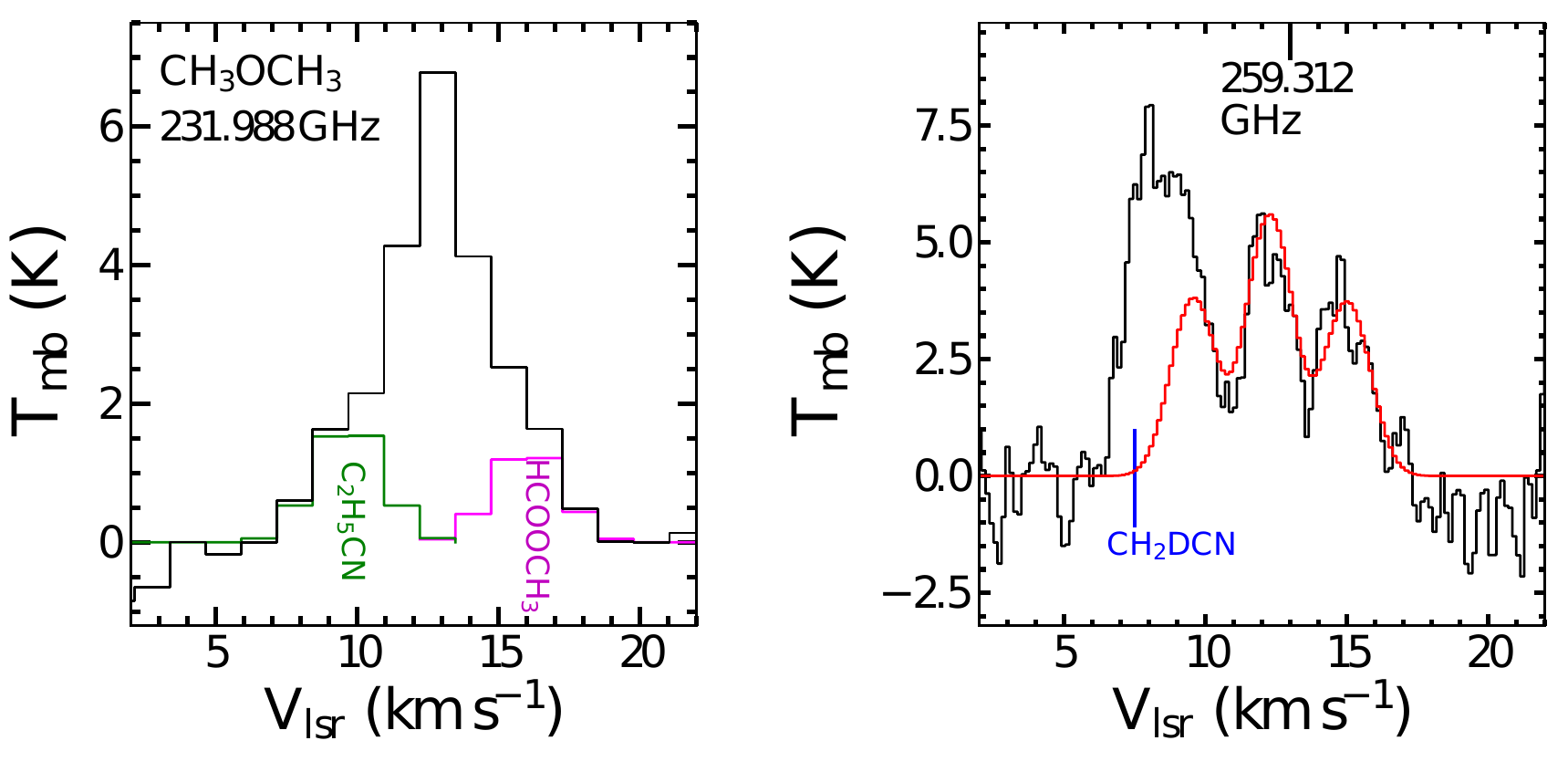}

   \caption{CH$\rm_{3}$OCH$\rm_{3}$ observed spectral lines (black) towards HOPS-108. Left: The contamination from HCOOCH$_{3}$ and C$\rm_{2}$H$\rm_{5}$CN are shown in magenta and green, respectively. Right: The spectrum predicted by the best-fit LTE model is depicted in red.}
    \label{dme-only}
\end{figure}

\begin{figure}[hbt!]
    \centering
   \includegraphics[width=0.8\textwidth]{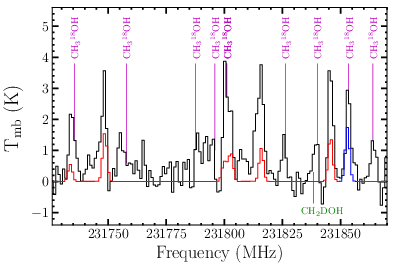}

    \caption{CH$\rm_{3}^{18}$OH observed spectral lines (black) towards HOPS-108. The contamination from HCOOCH$_{3}$ and C$\rm_{2}$H$\rm_{5}$CN are shown in red and blue, respectively. The CH$\rm_{3}^{18}$OH lines are marked in magenta.}
    \label{ch3o-18-h-spec}
\end{figure}

\begin{figure}[hbt!]
    \centering
   \includegraphics[width=0.25\textwidth]{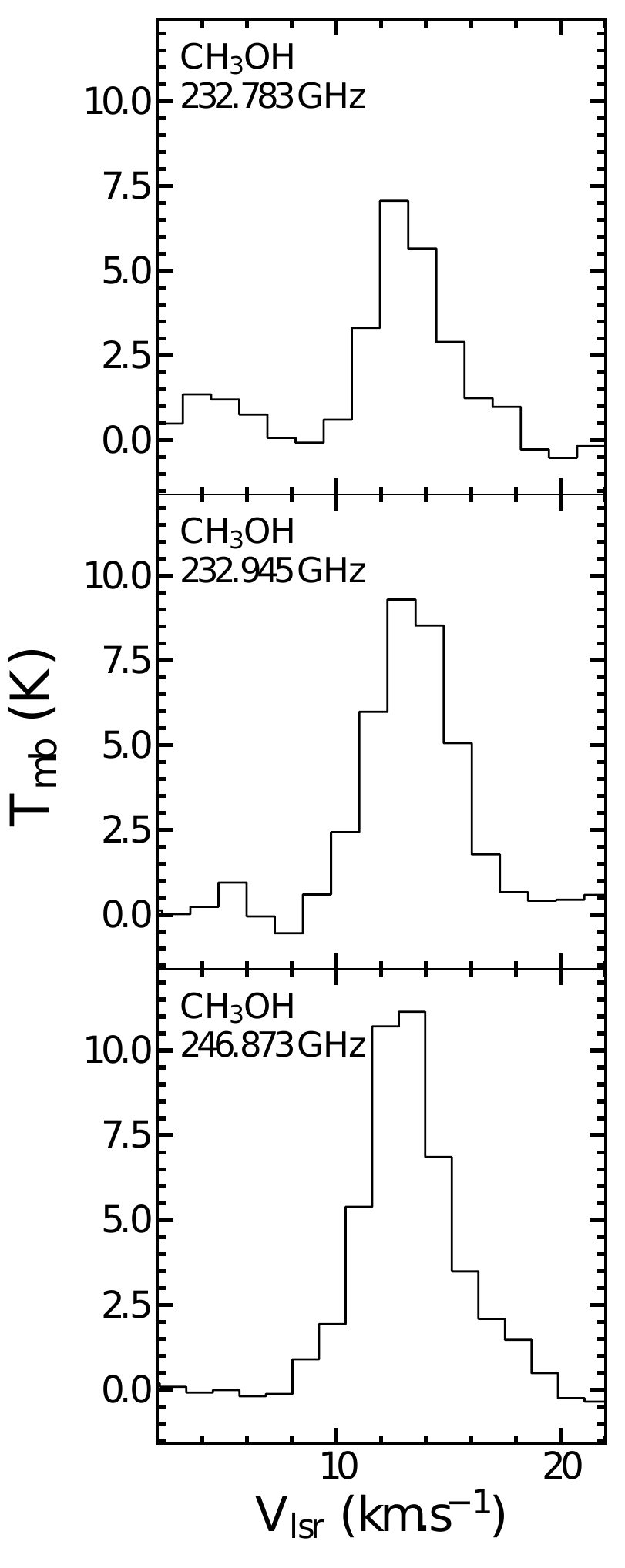}
   \includegraphics[width=0.25\textwidth]{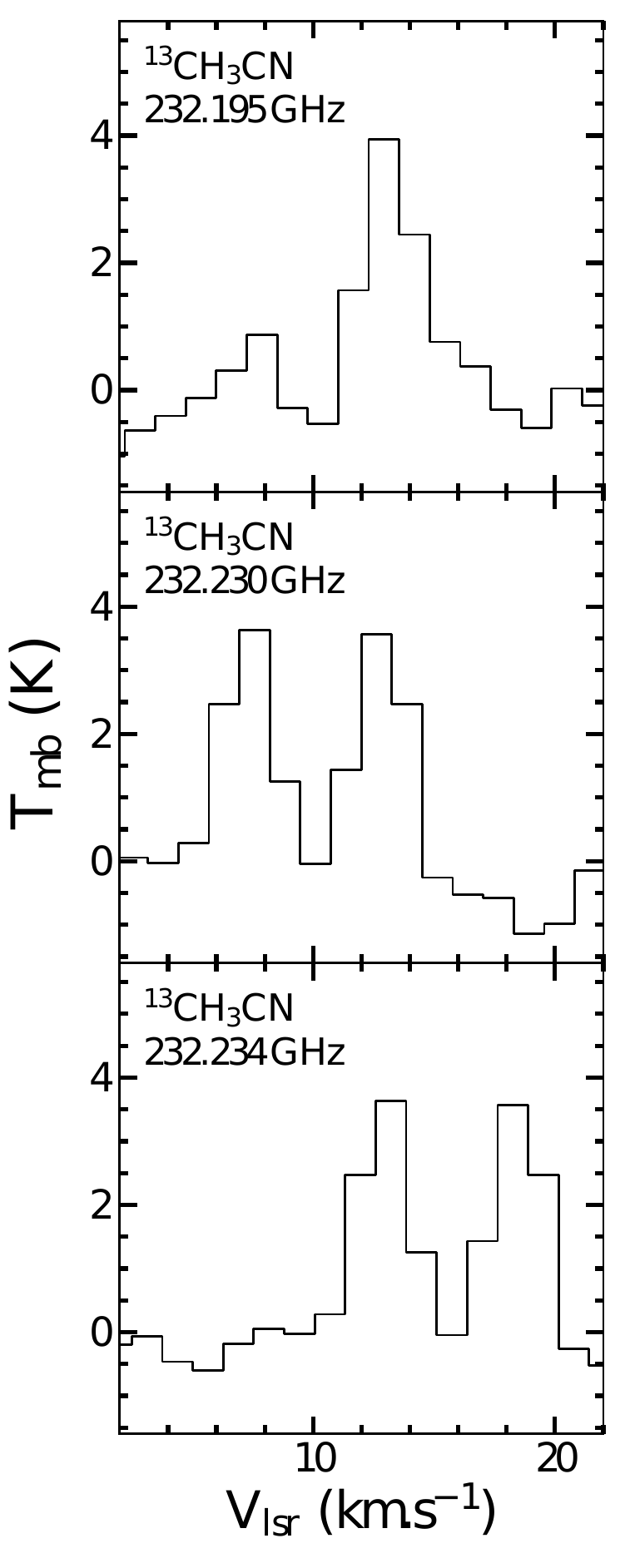}
   \includegraphics[width=0.25\textwidth]{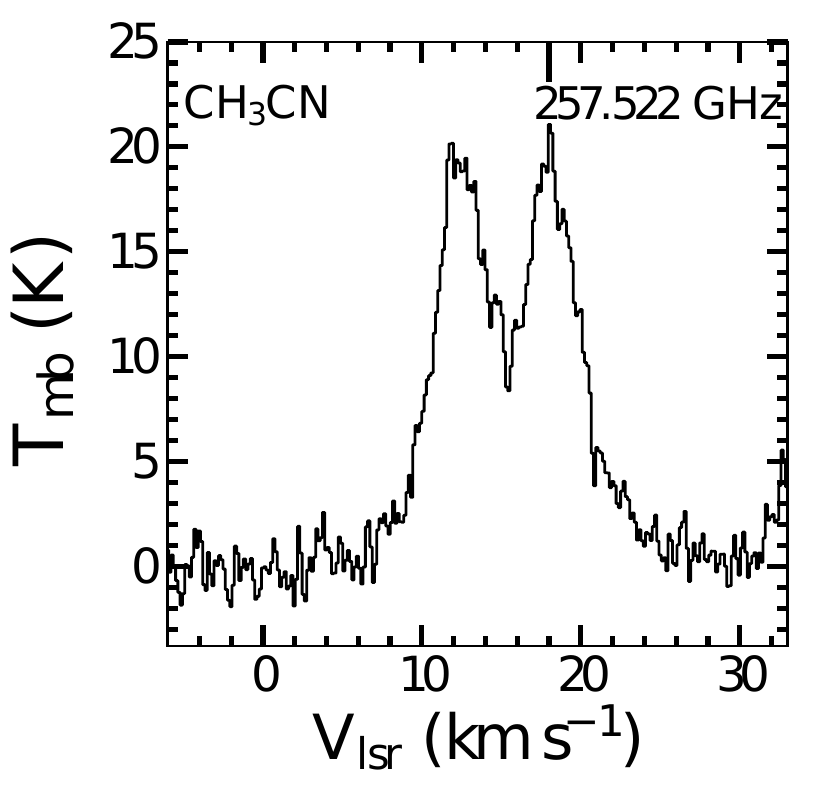}

    \caption{CH$\rm_{3}$OH, $\rm^{13}$CH$\rm_{3}$CN, and CH$\rm_{3}$CN observed spectral lines towards HOPS-108. The $\rm^{13}$CH$\rm_{3}$CN and CH$\rm_{3}$CN lines are used in the non-LTE analysis. Among the CH$\rm_{3}$OH lines, only the one at 232.783 GHz was used in the non-LTE analysis.}
    \label{meth-mc-spec}
\end{figure}

\begin{figure}[hbt!]
    \centering
    \begin{tabular}{ccc}
        \includegraphics[width=0.32\textwidth]{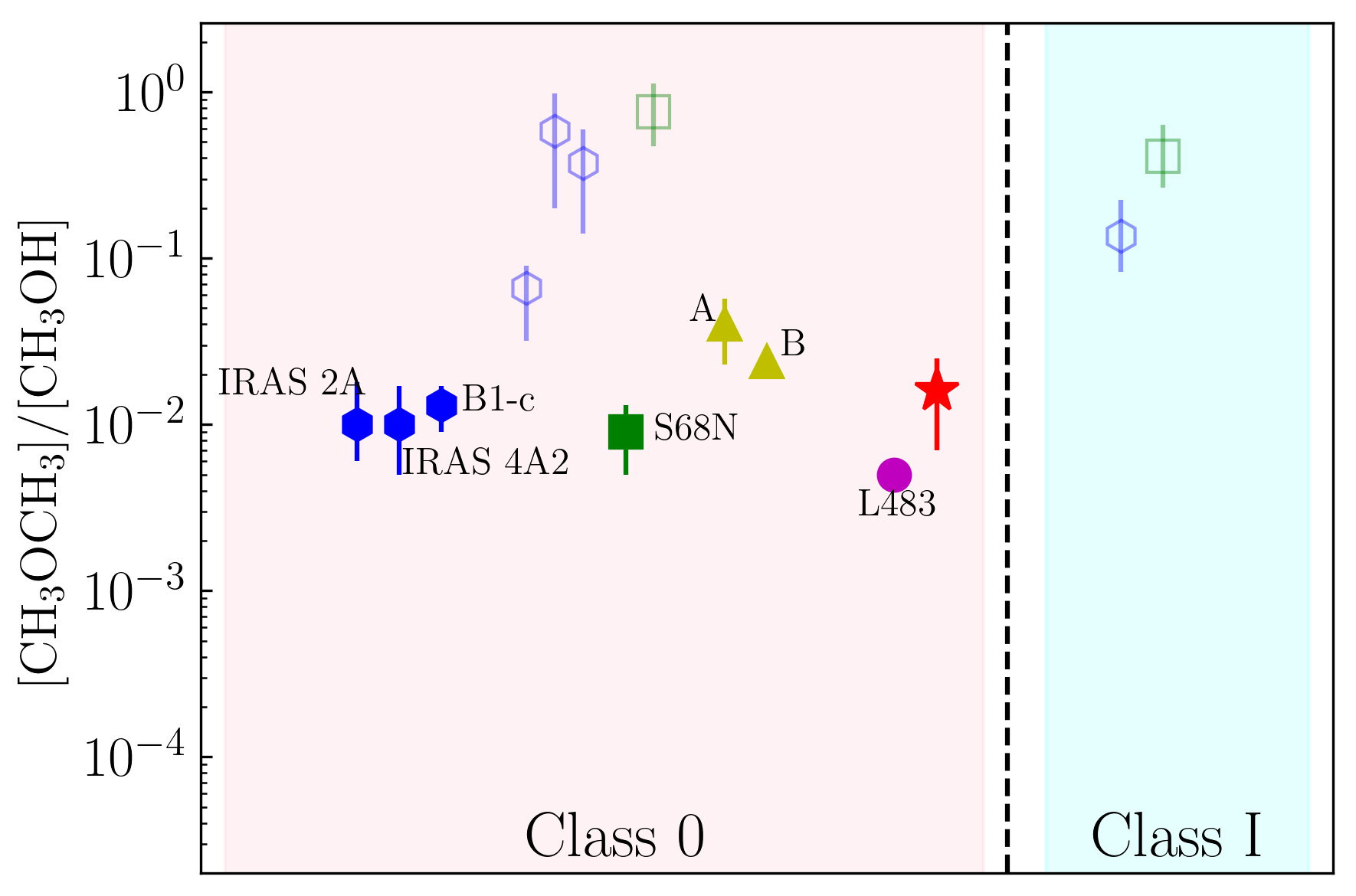} &
        \includegraphics[width=0.32\textwidth]{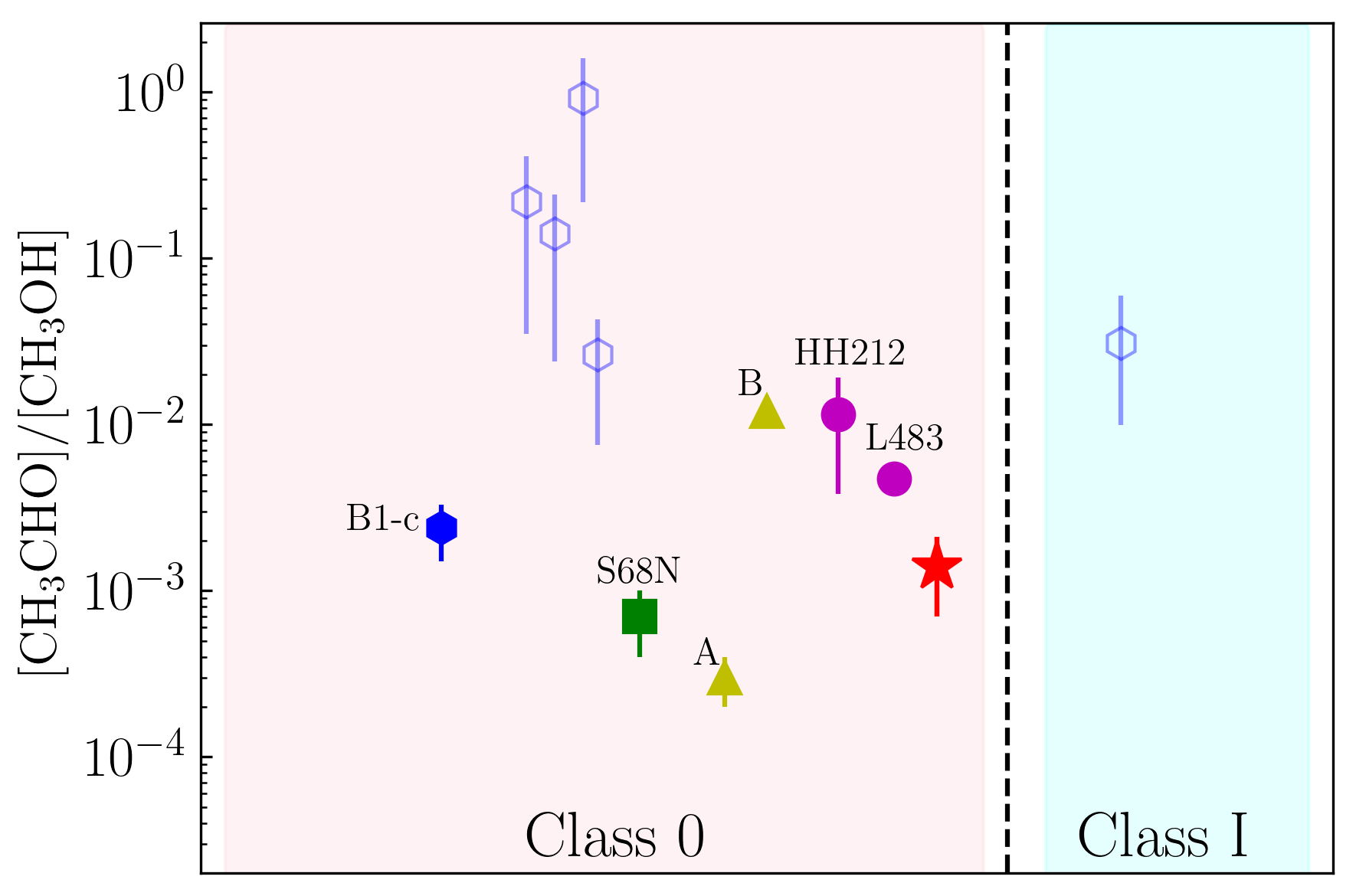} &
        \includegraphics[width=0.32\textwidth]{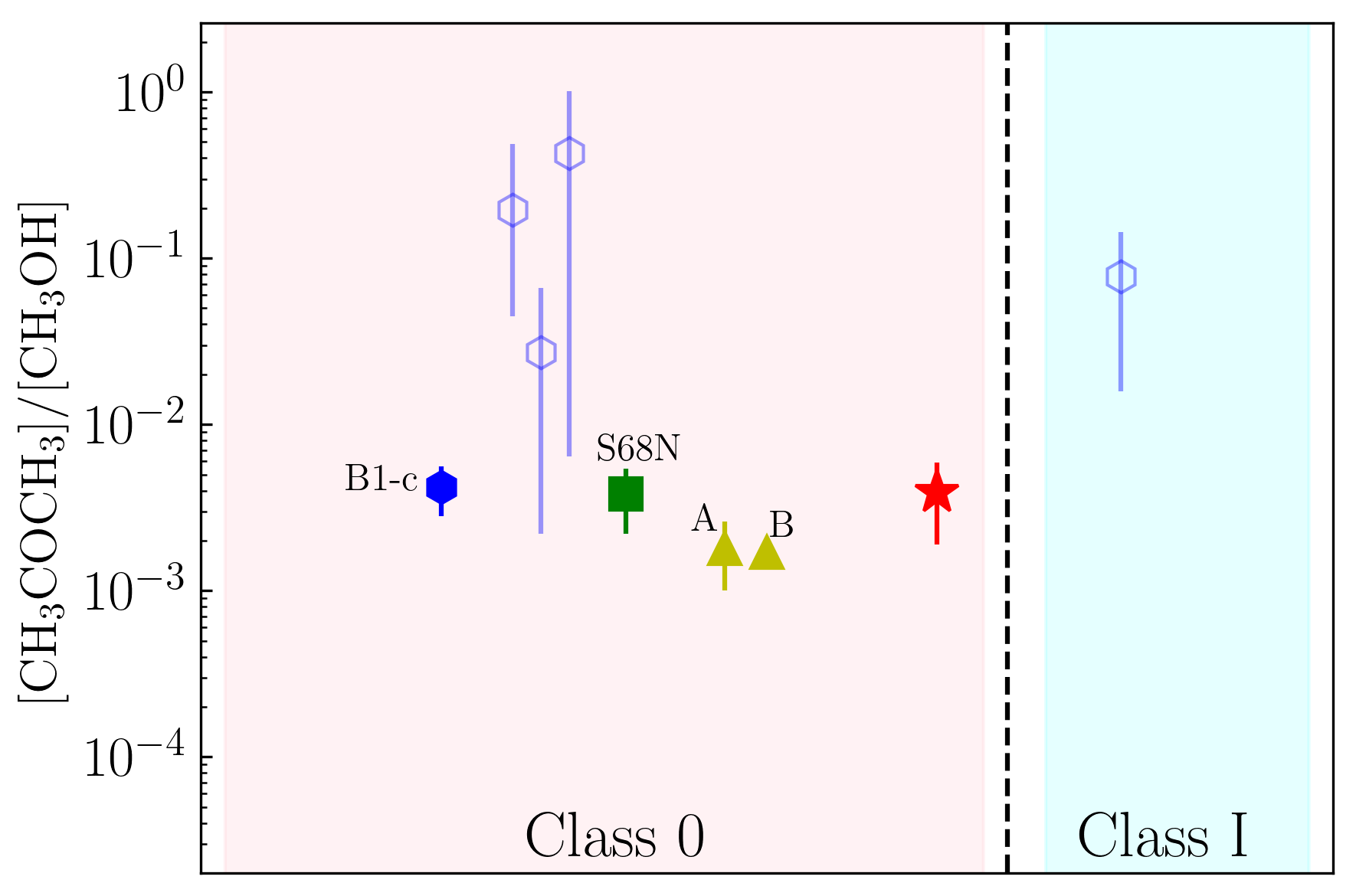} \\
        \includegraphics[width=0.12\textwidth]{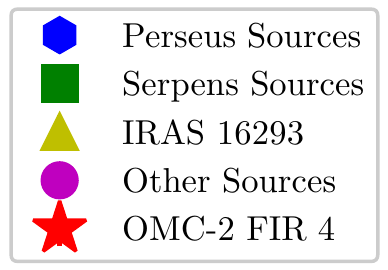} \\
        \includegraphics[width=0.32\textwidth]{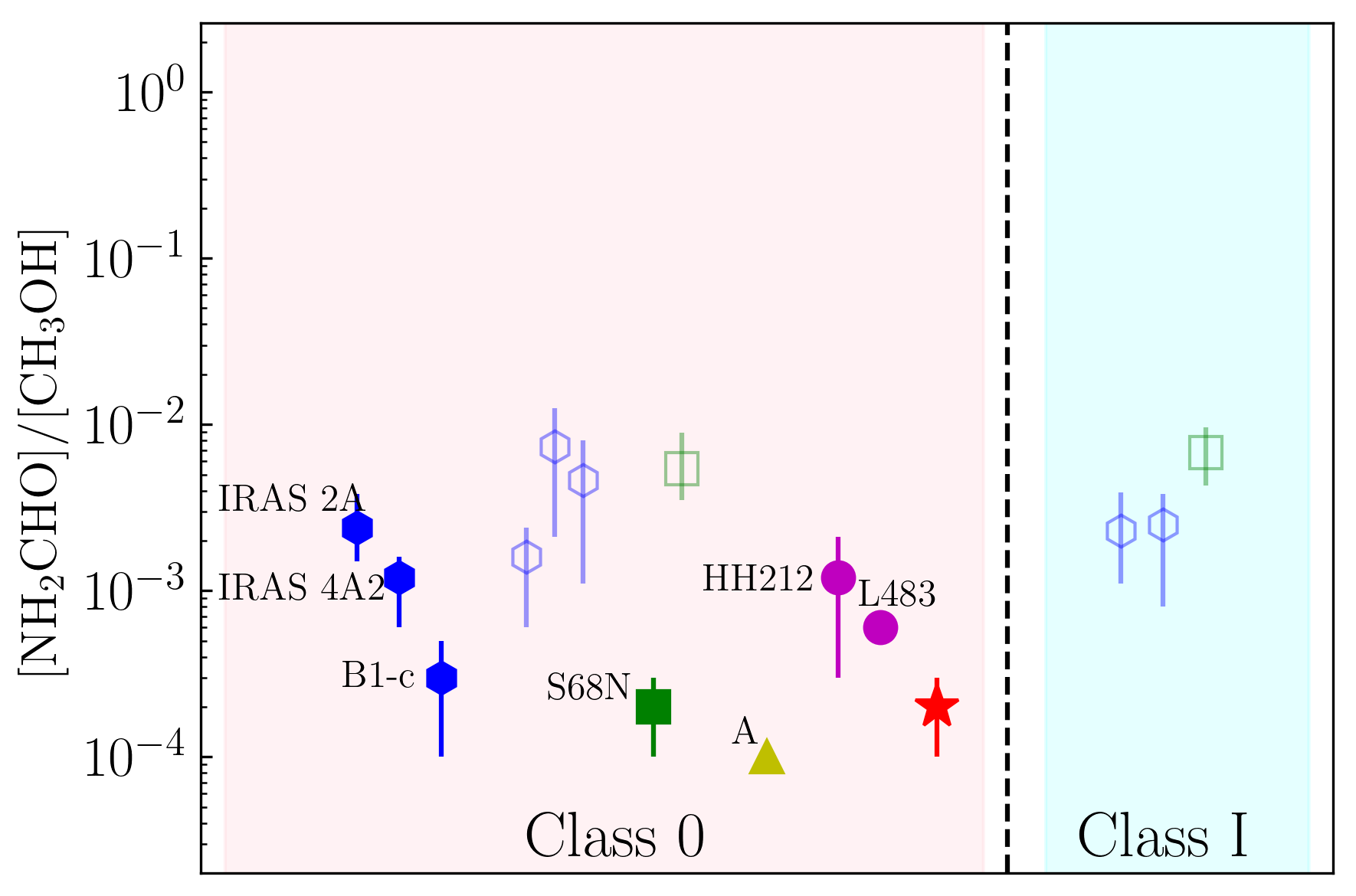} &
        \includegraphics[width=0.32\textwidth]{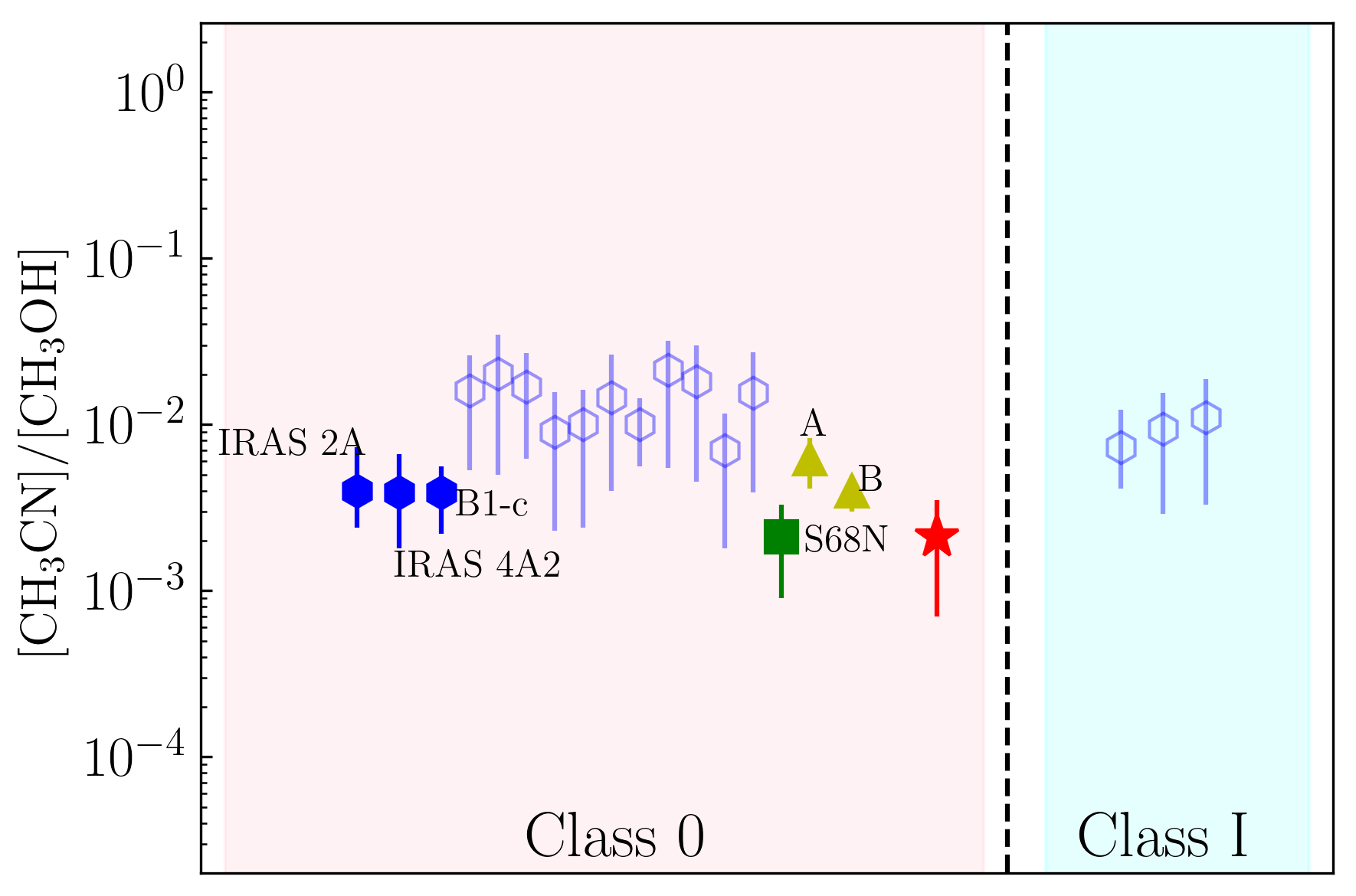} &
         \includegraphics[width=0.32\textwidth]{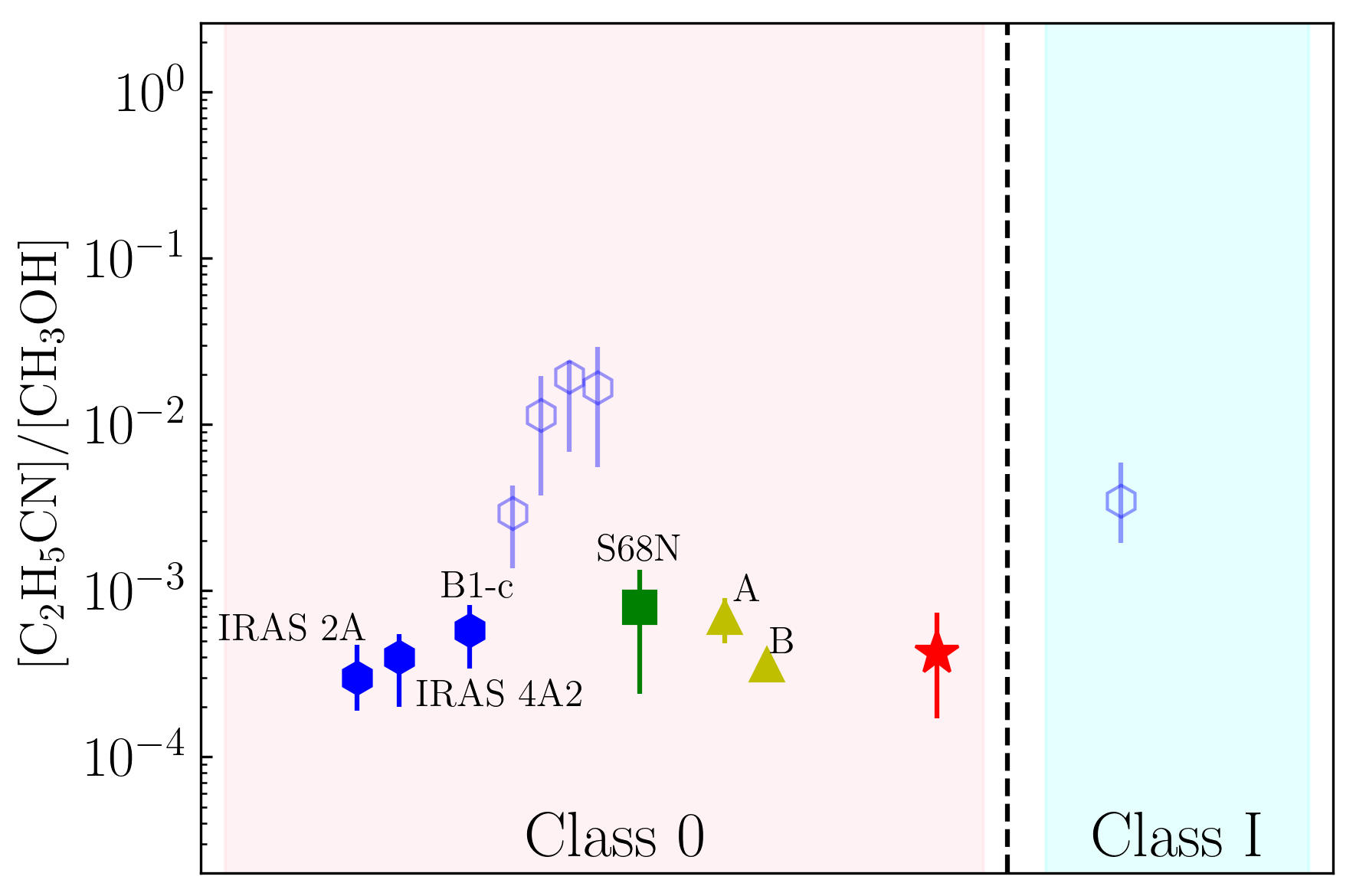} \\
    
    \end{tabular}
    
 \caption{Abundance ratio of iCOMs with respect to CH$_{3}$OH in HOPS-108 compared to other isolated hot corinos. The sources are separated according to their class, with no particular order within a given class. The open symbols correspond to the sources where the CH$_{3}$OH column density is most likely underestimated due to its high optical depth (See Sect. \ref{mf-discussion}). The Class 0 sources are: IRAS 2A, IRAS 4A2 \citep{Taquet2015,Lopez-Sepulcre2017}, B1-c, S68N \citep{vanGelder2020, Nazari2021}, IRAS19347 + 0727 in B335 \citep{Imai2016}, HH212 \citep{Bianchi2017, Lee2019}, L483 \citep{Jacobsen2019}, Ser-emb 1 \citep{Bergner2019}, IRAS 16293 A \citep{Manigand2020, Calcutt2018}, IRAS 16293 B \citep{Jorgensen2016, Jorgensen2018, Lykke2017, Calcutt2018}, HOPS-108 in OMC-2\,FIR\,4 (this work), and the sources from the PEACHES survey \citep{Peaches2021}, including Per-emb 1, Per-emb 2, Per-emb 5, Per-emb 10, Per-emb 11A, Per-emb 11C, Per-emb 12, Per-emb 13, Per-emb 18, Per-emb 20, Per-emb 21, Per-emb 22A, Per-emb 22B, Per-emb 26, Per-emb 27, Per-emb 29, Per-emb 33A, L1448 IRS3A, L1448-NW, SVS-13B, and B1-bS. The Class I sources are: Ser-emb 17 \citep{Bergner2019}, L1551 \citep{Bianchi2020}, and also sources from the PEACHES survey \citep{Peaches2021}, including Per-emb 35A, Per-emb 35B, Per-emb 42, Per-emb 44, Per-emb 53, and SVS13-A2. The same plots for HCOOCH$_{3}$ and CH$_{2}$DOH are represented in Figs. \ref{mf-meth-ratio} and \ref{deut-ratio}, respectively.}
   \label{mol-meth}
\end{figure}

\begin{figure}[hbt!]
    \centering
    \begin{tabular}{ccc}
         
         \includegraphics[width=0.32\textwidth]{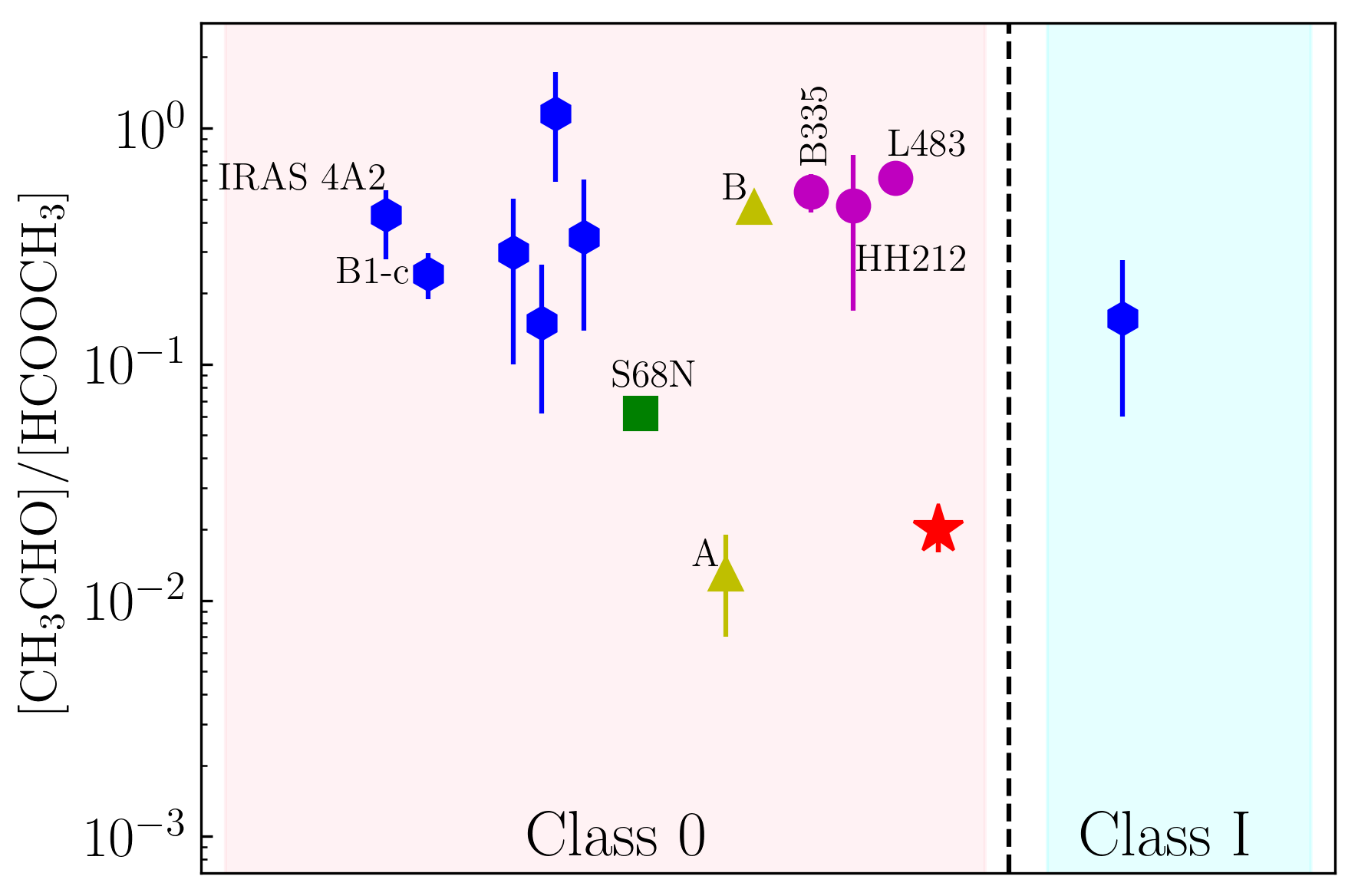} &
        \includegraphics[width=0.32\textwidth]{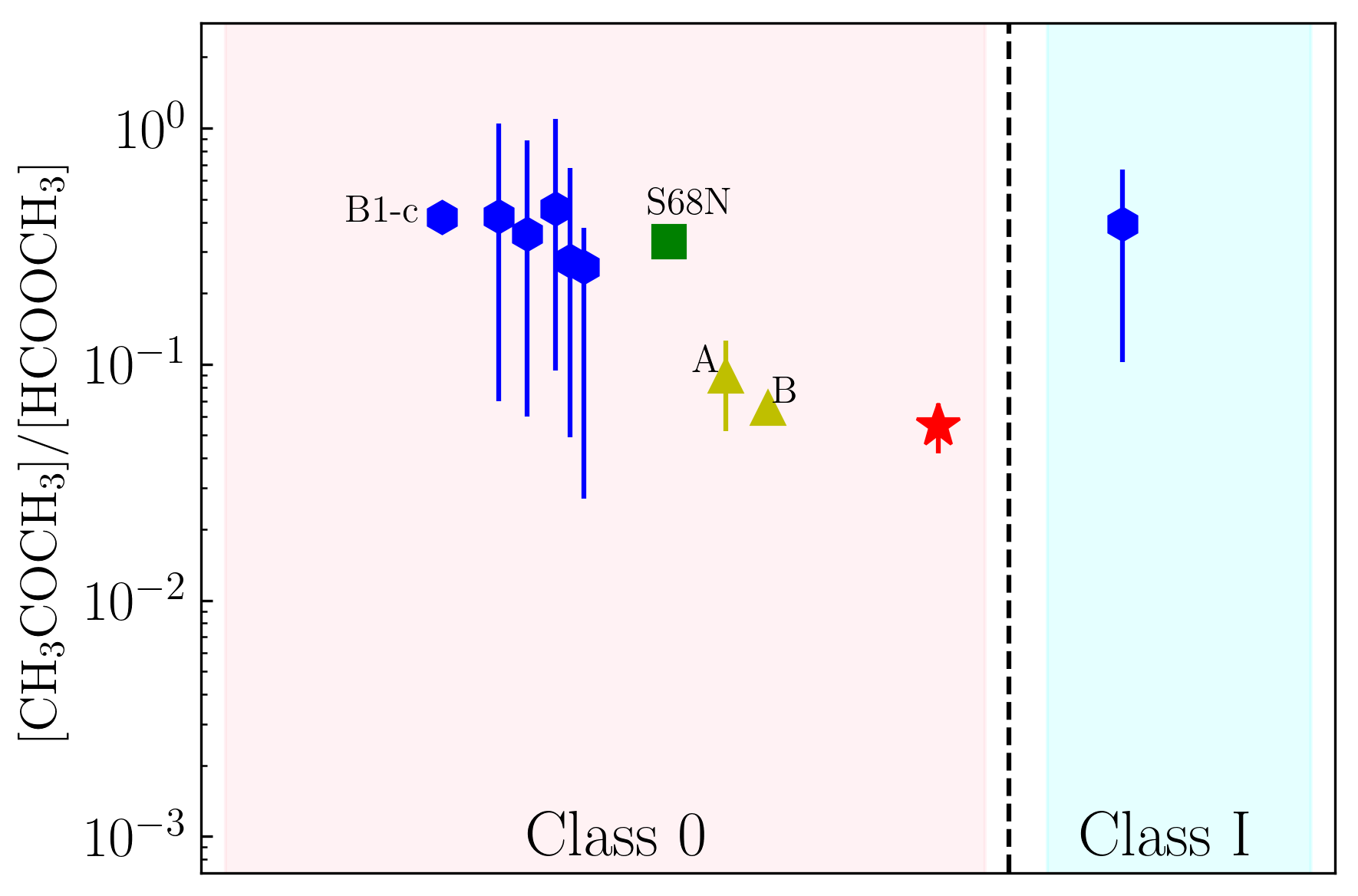} &
        \includegraphics[width=0.12\textwidth]{legend-2.pdf} \\
        \includegraphics[width=0.32\textwidth]{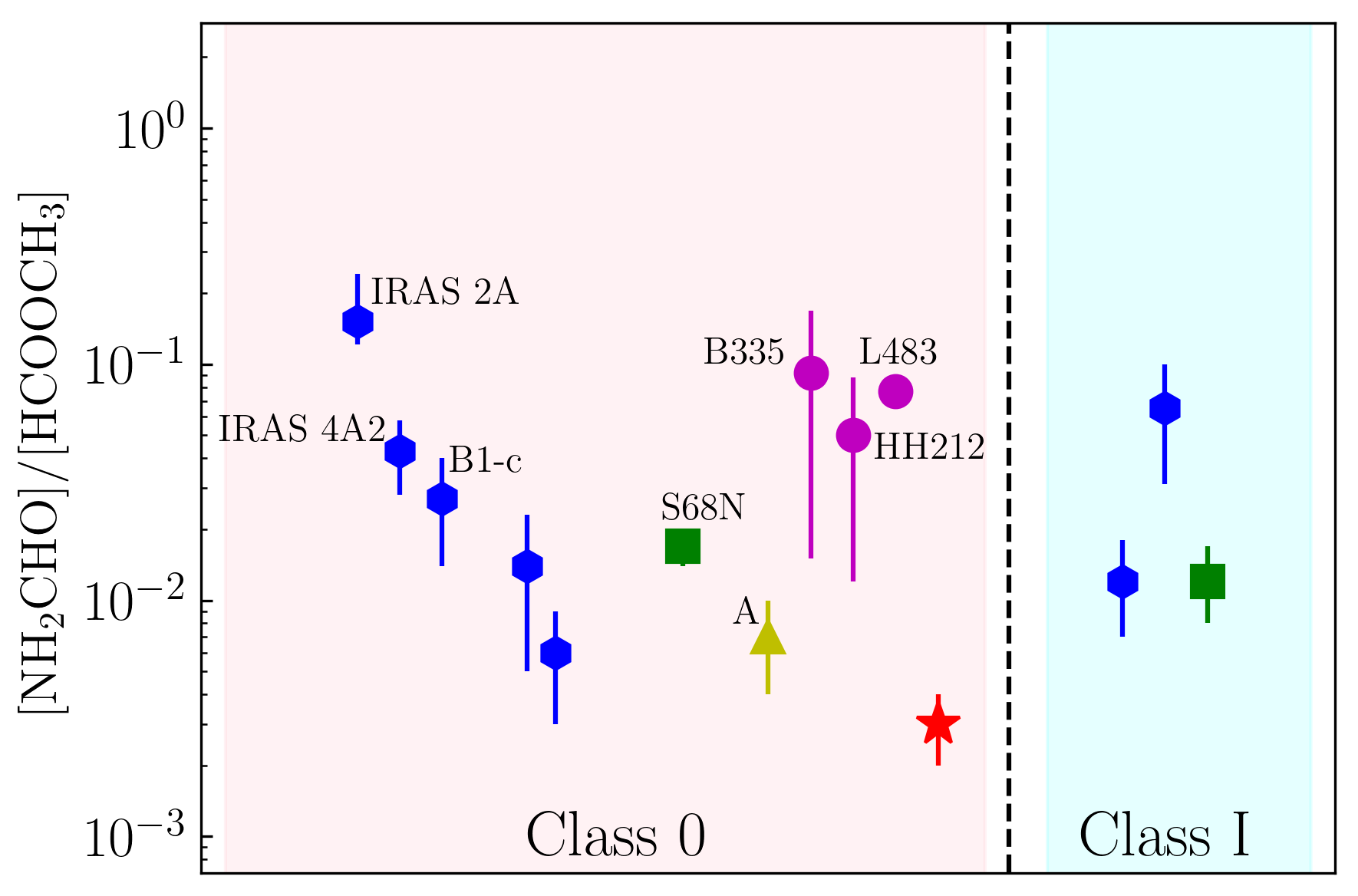} &
        \includegraphics[width=0.32\textwidth]{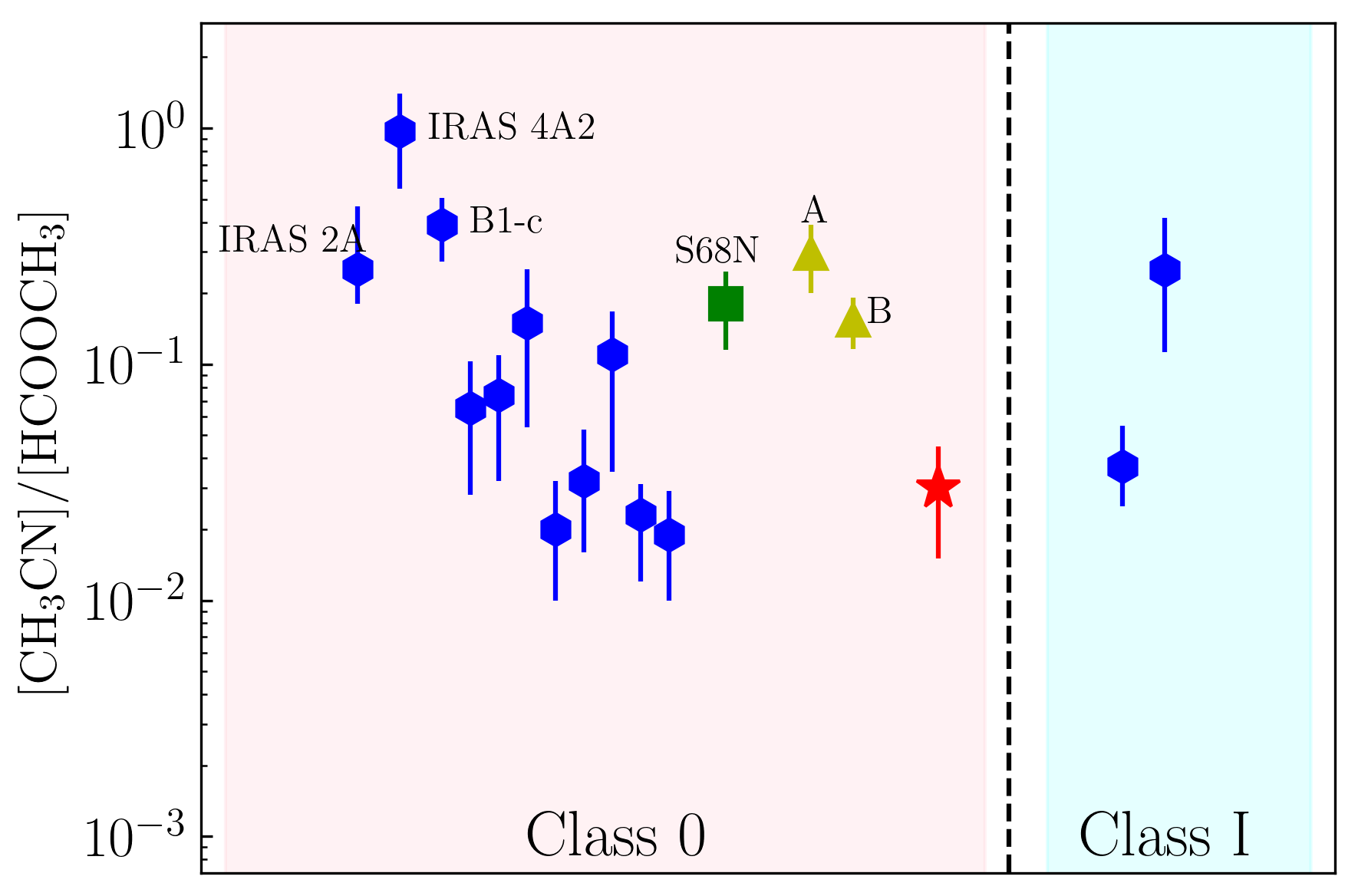} &
        \includegraphics[width=0.32\textwidth]{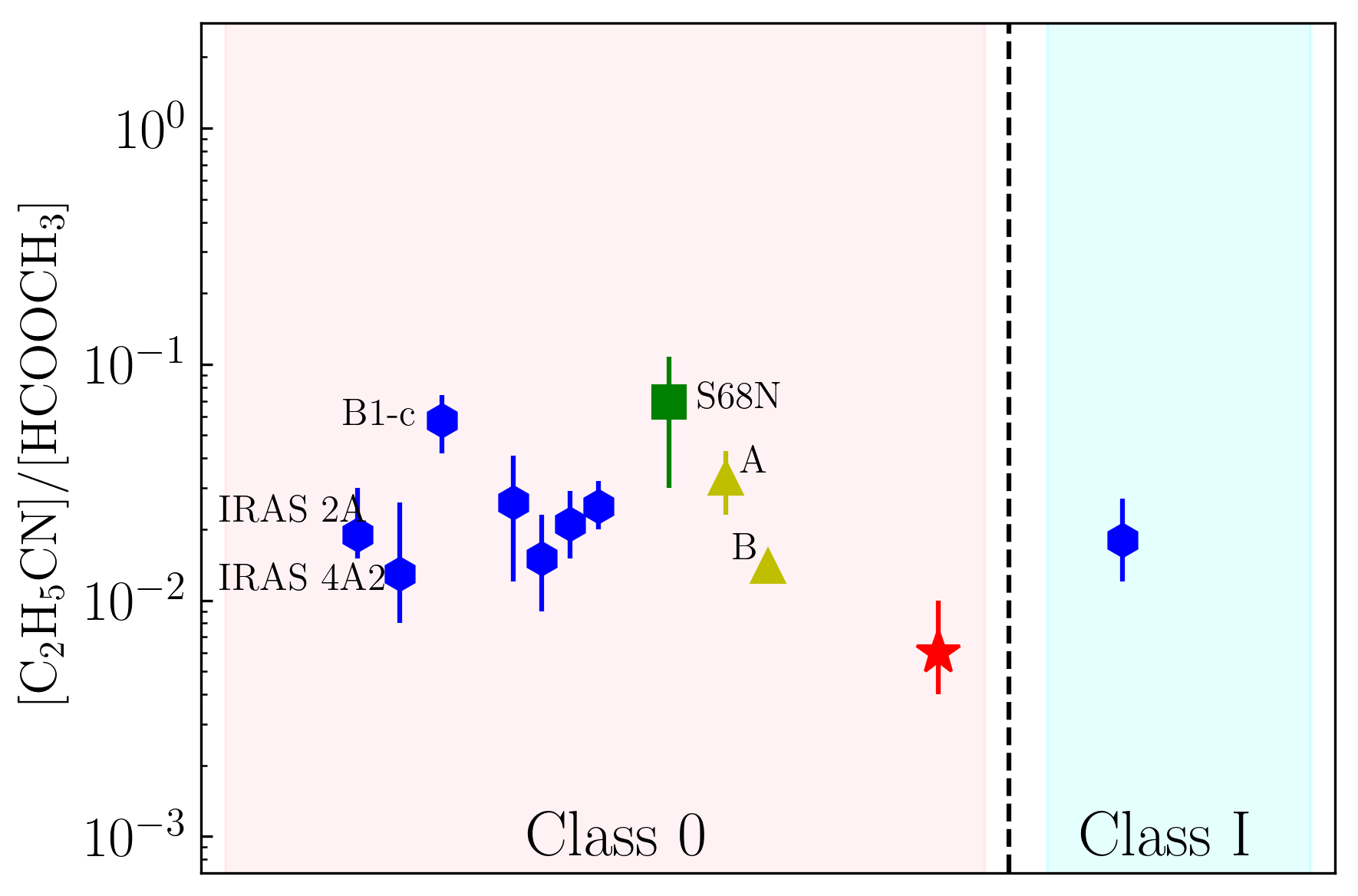} \\
    
    \end{tabular}
    
    \caption{Abundance ratio of iCOMs with respect to HCOOCH$_{3}$ in HOPS-108 compared to other isolated hot corinos. The sources are separated according to their class, with no particular order within a given class. The same sources from Fig. \ref{mol-meth} are used. The same plot for CH$_{3}$OCH$_{3}$ is represented in Fig. \ref{mf-dme-ratio}.}. 
    \label{mol-mf}
\end{figure}

\begin{figure}
    \centering
    \includegraphics[width=0.5\textwidth]{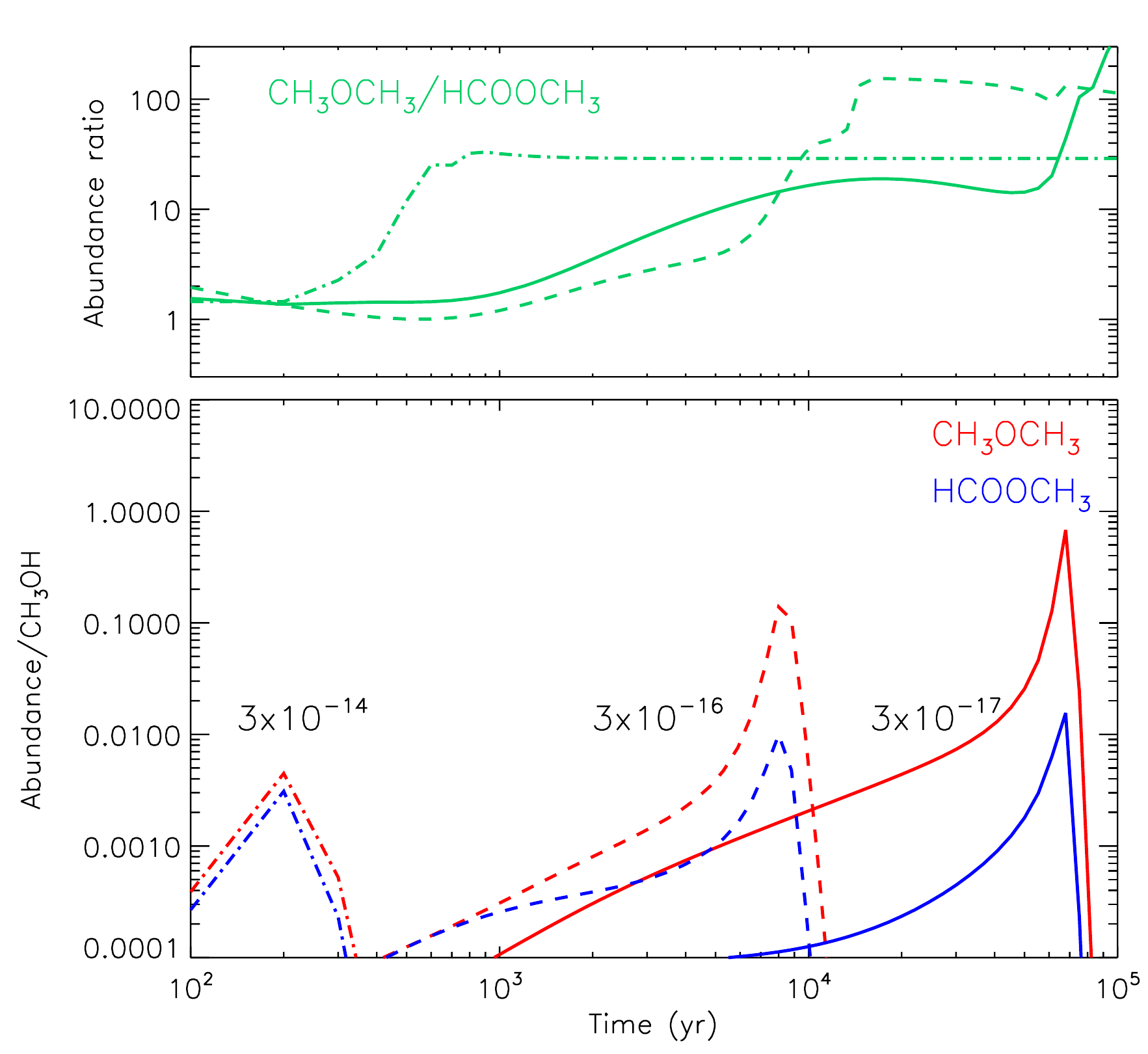}
\caption{Top: CH$_{3}$OCH$_{3}$/HCOOCH$_{3}$ ratio as a function of time for different ionisation rate values (in units of s$^{-1}$). Bottom: CH$_{3}$OCH$_{3}$ and HCOOCH$_{3}$ abundances with respect to CH$_{3}$OH as a function of time for different ionisation rate values $\zeta$. The ionisation rates in the upper and lower panel are the same: the solid lines correspond to $\zeta$=3$\times$10$^{-17}$ s$^{-1}$, the dashed lines correspond to $\zeta$=3$\times$10$^{-16}$ s$^{-1}$, and the dot-dashed lines correspond to $\zeta$=3$\times$10$^{-14}$ s$^{-1}$.}
    \label{crs-model}
\end{figure}

\end{document}